\pdfoutput=1
\documentclass[iop]{emulateapj}

\usepackage{amssymb,amsmath}
\usepackage{natbib}

\newcommand{\msun}{M_{\odot}}
\newcommand{\mbh}{M_{\bullet}}
\newcommand{\epsb}{\epsilon_{\rm B}}
\newcommand{\rs}{r_{\rm S}}
\newcommand{\mpia}
{
Max Planck Institute for Astronomy, K\"onigstuhl 17, 69117 Heidelberg, Germany
}
\newcommand{\cpa}
{
Centre for Plasma Astrophysics, K.U. Leuven, Celestijnenlaan 200B, 3001 Leuven, Belgium
}

\begin{document}
\title{Synchrotron radiation of self-collimating relativistic MHD jets}
\shorttitle{Jet base synchrotron radiation}
\author{Oliver Porth$^{1,}$\altaffilmark{3}}
\author{Christian Fendt$^{1}$}
\author{Zakaria Meliani$^{2}$}      
\author{Bhargav Vaidya$^{1,}$\altaffilmark{3}}
\affil{$^{1}$\mpia}
\affil{$^{2}$\cpa}
\email{Email: porth@mpia.de; fendt@mpia.de}
\altaffiltext{3}{Fellow of the 
                 \textit{International Max Planck Research School for Astronomy and Cosmic Physics at 
                         the University of Heidelberg} (IMPRS-HD), and the
                 \textit{Heidelberg Graduate School of Fundamental Physics} (HGSFP)}
\begin{abstract}
The goal of this paper is to derive signatures of synchrotron radiation from state-of-the-art simulation models of collimating relativistic magnetohydrodynamic (MHD) jets featuring a large-scale helical magnetic field. We perform axisymmetric special relativistic MHD simulations of the jet acceleration region using the PLUTO code. The computational domain extends from the slow magnetosonic launching surface of the disk up to $6000^{2}$ Schwarzschild radii allowing to reach highly relativistic Lorentz factors. The Poynting dominated disk wind develops into a jet with Lorentz factors of $\Gamma\simeq8$ and is collimated to $1^{\circ}$. In addition to the disk jet, we evolve a thermally driven spine jet, emanating from a hypothetical black hole corona. Solving the linearly polarized synchrotron radiation transport within the jet, we derive VLBI radio and (sub-) mm diagnostics such as core shift, polarization structure, intensity maps, spectra and Faraday rotation measure (RM), directly from the Stokes parameters. We also investigate depolarization and the detectability of a $\lambda^{2}$-law RM depending on beam resolution and observing frequency. We find non-monotonic intrinsic RM profiles which could be detected at a resolution of 100 Schwarzschild radii. In our collimating jet geometry, the strict bi-modality in polarization direction (as predicted by Pariev et al.) can be circumvented. 
Due to relativistic aberration, asymmetries in the polarization vectors across the jet can hint to the spin-direction of the central engine.  

\end{abstract}

\maketitle

\section{Introduction}
Relativistic jets are launched from accretion disks around compact objects and accelerated and
collimated by magnetohydrodynamic forces \citep{1982MNRAS.199..883B}.
The disk jet flow may be accompanied by a highly relativistic central spine jet resulting from electrodynamic effects within the black hole magnetosphere \citep{1977MNRAS.179..433B}.

The formation of relativistic MHD jets has been investigated by a number of authors
since the seminal paper of Blandford \& Znajek.
The early attempts were to look for stationary state MHD solutions for the asymptotic
jet structure
\citep{1991ApJ...377..462C, 1993A&A...270...71A,  1997A&A...323..999F}, or
the jet  formation domain of collimation and acceleration in case of self-similarity
\citep{1994ApJ...432..508C, 1995ApJ...446...67C},
or taking into account the 2.5D force-balance in case of special relativity
\citep{1987A&A...184..341C, 2001A&A...365..631F},
or general relativity \citep{1997A&A...319.1025F, 2000MNRAS.315...89G, 2001A&A...369..308F}.

Time-dependent simulations of jet formation from the disk surface have first been investigated
in the non-relativistic approximation \citep{1995ApJ...439L..39U, 1997ApJ...482..712O}.
General relativistic MHD simulations of accretion disks launching outflows
\citep{2005ApJ...625...60N, 2007MNRAS.375..513M, 2008ApJ...678.1180B, 2009MNRAS.394L.126M},
indicate that highly relativistic jets may not be launched by
the disk itself, but from the black hole magnetosphere.
These jets - so-called funnel flows - may reach Lorentz factors up to 10 however,
the mass loading is put in by hand (maybe corresponding to a pair-creation process)
and is not related to the accretion process itself.
\citet{2007MNRAS.380...51K} proposed special relativistic MHD simulations of AGN jets over a
large spatial scale and find asymptotic Lorentz factors of about 10. These jets were
pressure-confined by the outer boundary condition.  

In a recent paper we have treated another setup of relativistic jet formation
\citep{2010ApJ...709.1100P}.  
We have applied special relativistic MHD to launch outflows from a fixed-in-time hot 
surface of an accretion disk which is rotating in centrifugal equilibrium with the
central compact object.
This allowed us to pass through the slow-magnetosonic point, thus obtaining consistent mass and Poynting fluxes,
and to investigate the subsequent acceleration and collimation degree of a variety of jets.
Essentially, we find these relativistic MHD jets to be truly self-collimating - similarly
to their non-relativistic counterparts. 
In this paper, we will {\em prescribe} certain mass fluxes and explore the influence of the poloidal current distribution on the jet formation process.  
In contrast to the previous work, the jet energy and its partitioning is now a true parameter of the models, 
allowing us to investigate the regime of highly relativistic flows.  
More importantly, observational signatures of the jet formation models are discussed.

The existence of an ordered large-scale $\mu$G- mG magnetic field in extragalactic jets 
is well established by the detection of radio synchrotron emission, however the exact
geometry of the field structure cannot be derived.
The recent observational literature strongly suggests {\em helical magnetic fields}
(e.g. \cite{2009MNRAS.393..429O}),
a scenario which is consistent with theoretical models and numerical simulations of
MHD jet formation (see e.g.
\citep{1982MNRAS.199..883B, 1997ApJ...482..712O, 1999ApJ...526..631K, 2002A&A...395.1045F,                       2006ApJ...651..272F, 2010ApJ...709.1100P}).  
The wavelength dependent rotation of the polarization plane known as Faraday rotation provides a valuable diagnostic of magnetic field structure in astrophysical jets.  Consistent detections of $\Delta\chi\propto\lambda^{2}$ are found in resolved jets as well as in unresolved radio {\em cores} \citep{2003ApJ...589..126Z}.  
Helical magnetic fields are generally perceived to promote transversal Faraday rotation measure (RM)
{\em gradients} owing to the toroidal field component.  
Observationally, such gradients were first detected by \cite{asada2002} and \cite{2005ApJ...626L..73Z}
in the jet of $\rm 3C\,273$.
The RMs are generally found to follow a monotonic profile across jets
\citep{gabuzda2004, 2009MNRAS.393..429O, 2010MNRAS.402..259C}.  

While the magnetic field structure of pc and Kpc-scale jets can in principle derived
from radio observations, not much is known about the field structure within the jet
forming region very close to black hole - mainly because of two reasons.
Firstly, this very core region cannot be spatially resolved and, thus, cannot readily be
compared with expected RM profiles of helical jets.
This is somewhat unfortunate as close to the jet origin we expect the magnetic field helix
to be well preserved and not much affected by environmental effects.
Secondly, the observed rotation measures are so high that it is impossible to draw firm
conclusions about the intrinsic field geometry from the polarization vector with mere radio observations.

Only very few cases exist where this core of jet formation could be resolved observationally.
Among them is the close-by galaxy M87, where the VLBI/VLBA resolution of $\simeq 0.1$\,mas
is sufficient to resolve about 0.01\,pc within the central region, and
allows to trace the jet origin down to $\simeq 20$ Schwarzschild radii $\rs$ when the recent mass estimate of $6.6\times 10^{9} \msun$ by \cite{gebhardt2011} is adopted.  
This pinpoints the launching area within $\simeq 30 \rs$
\citep{junor1999,kovalev2007}.  
The radio maps clearly show limb brightening and indicate an initial jet opening angle of
about $60^{\circ}$.

Despite a vast amount of observational data spanning over a huge frequency scale,
and also time series of these multifrequency observations,
rather little is known about the dynamical status of relativistic jets.
There are no direct unambiguous observational tracers of jet velocity or density as only (if at all)
the pattern speed of radio knots is detected.
Kinematic modelling of knot ejection suggests pc-scale Lorentz factors of
typically $\Gamma \simeq 10$, while Kpc-jet velocities are believed to be definitely lower
and of the order of $0.1$c.
Kinematic modeling of jet propagation has been combined with synchrotron emission models
of nuclear flares, resulting in near perfect fits of the observed, time-dependent
radio pattern of jet sources such as 3C\,279 \citep{2006A&A...456..895L}.

However, what was missing until recently is a consistent combination of {\em dynamical
models} with {\em radiation models} of synchrotron emission resulting
in theoretical radiation maps which can then be compared with observations.  
\cite{Zakamska2008} and \cite{2009ApJ...695..503G} have taken a step into this direction by providing optically 
thin synchrotron and polarization maps from self-similar MHD solutions.  
\citet{broderick2010} presented synchrotron ray-castings from 3D general relativistic jet formation simulations 
that also include the evolution of a turbulent accretion disk performed by \citet{2009MNRAS.394L.126M}.  
In their approach, the MHD solution is extrapolated by means of an essentially self-similar scheme in order
to reach distances up to $10 \rm pc$.  
Their study focussed on the rotation measure provided by Faraday rotation in the disk wind \emph{external} to the emitting region in the Blandford-Znajek jet.

In comparison, using axisymmetric large scale simulations, we do not rely on an extrapolation within the AGN core
(up to $0.3\rm pc$) and treat the Faraday rotation also internal to the emitting region in the fast jet that gradually
transforms into a sub-relativistic disk wind.  
The observational signatures obtained in our work are derived entirely from the (beam convolved) Stokes parameters 
which allows us to also investigate the breaking of the $\lambda^{2}$ law due to opacity effects. \\
As the above studies, we also rely on post-hoc prescriptions for the relativistic particle content.  
Towards a more consistent modeling of the non-thermal particles, \citet{2010MNRAS.401..525M} have presented a method to follow the spectral evolution of a seed particle distribution due to synchrotron losses within a propagating relativistic hydrodynamic jet.  
The question of particle acceleration and cooling is in fact essential to close the loop for a fully self-consistent treatment of jet dynamics, jet internal heating, and jet radiation.

To derive signatures of synchrotron emission from
relativistic MHD jet formation is the main goal of the present paper.
We apply the numerically derived dynamical variables such as velocity, density,
temperature, and the magnetic field strength \& configuration to calculate
the synchrotron emission from these jets, taking into account proper beaming
and boosting effects for different inclination angles.
In particular, we apply the relativistically correct polarized radiation transfer
along the line of sight throughout the jet.

\section{The relativistic MHD jet}

We perform axisymmetric jet acceleration simulations with the PLUTO 3.01 code
\citep{2007ApJS..170..228M} solving the special relativistic magnetohydrodynamic equations.
As in \cite{2010ApJ...709.1100P}, the simulations are of the \textit{disk-as-boundary} type,
however in the present study the jet starts out with the slow magnetosonic velocity opening
the freedom to assign all energy channels as input parameters.  
The energy of the jet base is dominated by Poynting flux, driving a large-scale poloidal
current circuit.  
This current distribution is prescribed as a boundary condition in the toroidal magnetic field
component.
We investigate three cases for $B_{\phi}(r)\propto r^{-s}$ with $s\in\{1,1.25,1.5\}$, where $r$ is the cylindrical radius, 
resulting in asymptotic jets which are either in a current-carrying (one of them) or
current-free (two others) configuration.  
Injected initially with slow magnetosonic velocity, the jet material accelerates through
Alfv\`en and fast magnetosonic surfaces within the simulated domain.  

\subsection{Numerical grid setup}
To eradicate any artificial collimation effect from the outflow boundaries, we decided to move
the domain boundaries to such large distance that they are out of causal contact with the
region of interest.  
An inner equidistant grid of $20$ cells extends to the scale radius $r_1$ corresponding to the
inner radius of the accretion disk.  
Beyond $r_1$ the grid is linearly stretched to $(r,z)=(500,500)$ with a scaling factor of $1.0013$,
and for $(r,z)>(500,500)$ with a larger scaling factor of $1.0047$.
So far, we apply a square box of $2000^{2}$ scale radii corresponding to $2555^{2}$ grid cells.
Magnetic fields are advanced on a staggered grid using the method of constrained transport \citep{1999JCoPh.149..270B} supplied by PLUTO.  

For the calculation of the radiation maps, we consider only a subdomain of $1417\times2356$ grid
cells corresponding to $200\times 1000$ scale radii.  

The first possible contamination by boundary effects takes place when the bow shock reaches
the upper outflow boundary after $t_{\rm c1}< z_{\rm end}/c = 2000$.  
A typical simulation is terminated after maximally $t_{\rm c2}=3000$ time units when a
signal traveling at the speed of light could have returned from the $z_{\rm end}$ boundary
to the subdomain.  This conservative treatment ensures that no spurious boundary effects can occur.  

In order to test this 
we have re-run one
of our simulations on a five times larger computational domain with comparable resolution.
This simulation provided the same result as the lower grid size simulation and acquired a
nearly stationary state when terminated.
Thus, the region of interest is not at all affected by any outflow boundary effects on account of an increased computational overhead.  

\subsubsection{Inflow boundary conditions}
In a well posed MHD boundary, the number of outgoing waves (i.e. seven minus the downstream critical
points) must equal the number of boundary constraints provided.  
Thus, in addition to the $B_{\phi}(r)$ and $v_{p}(r)$ profiles,
we choose to prescribe the thermal pressure $p(r)$ and density $\rho(r)$ as boundary conditions for the jet injection (the jet inlet).  
The fifth condition sets $v_{p}||B_{p}$ and thus constrains the toroidal electric field $E_{\phi}(r)\equiv0$.  
This is a necessary condition for a stationary state to be reached by the axisymmetric simulation.  

The remaining primitive MHD variables $v_{\phi}$ and $B_{r}$ are extrapolated linearly from the computational domain, while the component $B_{z}$ (which determines the magnetic flux) follows from the solenoidal condition.  

Specifically, the fixed profiles read
\begin{align}
\rho(R)        &= \rho_{1}\left[(1-\theta) R + \theta  R^{-1.5} \right]\label{eq:rho}\\
p(R)            &= p_{1}\left[(1-\theta) (1 - \rho_{1} \ln R) + \theta R^{-2.5}\right]\label{eq:p} \\
B_{\phi}(r) &= B_{\phi,1} \left[ (1-\theta) r + \theta r^{-s}\right] \\
v_{p}(r)     &= v_{\rm sm}(r) \label{eq:vpin} 
\end{align}
where
\begin{align}
\theta = \left\{
\begin{array}{cc}
0;\ r<1\\
1;\ r\ge1
\end{array}
\right.
\end{align}
is the step function and $r$ and $R$ denote the cylindrical and spherical radius, respectively. 
Here and in the following, a subscript 1 indicates the quantity to be evaluated at $(r,z)=(1,0)$.
 
The slow magnetosonic velocity profile along the disk $v_{\rm sm}(r)$ of
relation (\ref{eq:vpin}) is updated every time step to account for the variables which are
extrapolated from the domain.  
Within $R<1$, an inner ``black hole corona'' with relativistic plasma temperatures is modeled.  
The physical processes responsible for the formation of an inner hot corona could be an accretion shock 
or a so-called CENBOL shock (CENtrifugal pressure supported BOundary Layer shock) \cite[see also][]{kazanas1986, das1999}.  
Also \cite{blandford1994} has proposed a mechanism of dissipation near the ergosphere as a consequence of 
the Lense-Thirring effect.  
Due to the large enthalpy and decreasing Poynting flux of the inner heat bath, a thermally driven
outflow is anticipated from this region \citep[e.g.][]{2010arXiv1006.2073M}.  
We apply the causal equation of state introduced by \cite{mignone2005} to smoothly join the relativistically hot central region to the comparatively cold disk jet. This choice permits a physical solution for both regions of the flow \citep[see also][]{mignone2007}.  

\subsubsection{Outflow boundary conditions}

At the outflow boundaries we apply power-law extrapolation for $p,\rho$ and the parallel magnetic field component, while we apply the solenoidal condition to determine the normal magnetic field vector $B_{z}$ respectively $B_{r}$.
For the velocities and the toroidal field, zero-gradient conditions are applied.  
This choice is particularly suited to preserve the initial condition that is 
well approximated by power laws.  A more sophisticated treatment such as force-free \citep[introduced by][]{Romanova1997} or zero-current boundaries as in \cite{2010ApJ...709.1100P} is rendered unnecessary by the increased computational grid as detailed before.    

\subsection{Initial conditions}
As initial setup we prescribe a non-rotating hydrostatic corona threaded by a force-free magnetic
field.
For the initial poloidal field distribution we adopt
$B_{r} =  1/r-z/r\left(r^2+z^2\right)^{-1/2}$ and
$B_{z} = \left(r^2+z^2\right)^{-1/2}$
(\cite{1997ApJ...482..712O}, see also \cite{2010ApJ...709.1100P}).
The hydrostatic corona is balanced by a point-mass gravity in a Newtonian approximation.
In order to avoid singularities in the density or pressure distribution, equations 
(\ref{eq:rho}), (\ref{eq:p}),
we slightly offset the computational domain from the origin
by $(r_{\rm 0},z_{\rm 0})=(0,1/3)$.

\subsection{Parametrization}
In the present setup we focus on effects of the poloidal current distribution and
parametrize accordingly.  
Thus, we set the Kepler speed at $r=1$ to $v_{\rm K}=0.5 c$ for convenience, yielding a sound speed $c_{\rm s,1}=(\gamma p_{1}/\rho_{1})^{1/2}\simeq0.4 c$ from the hydrostatic condition.  
To further minimize the number of free parameters, we tie the toroidal field strength
given by $B_{\phi,1}$ to the poloidal field strength via $B_{\phi,1}=0.5 B_{p,1}$.  
This Ansatz is consistent with the sub-Alfv\`eninc nature of the flow, since at the Alfv\`en point $B_{\phi,1} \simeq B_{p,1}$ \cite[e.g.][]{1999ApJ...526..631K} is valid.  
The two remaining parameters are the poloidal magnetic field strength measured by
the plasma-beta $\beta_{1}^{2} = 2 p_{1} /B_{p,1}^{2}$
and the toroidal field profile power law index $s$.  

Note that with the choice of a fixed in time toroidal field, the injected Poynting flux is 
controlled via the $\beta_{1}$ parameter, implying the toroidal field being induced  
in a non-ideal MHD disk below the domain.  
The conserved total jet energy flux and its partitioning between Poynting and kinetic energy given by the $\sigma$-parameter essentially become boundary conditions and determine the terminal Lorentz factor.  

\subsection{Physical scaling}\label{sec:scaling}
In order to convert code units to physical units we need to define the two scaling parameters
of length and density, while the velocity is naturally normalized to the speed of light $c$.    
To obtain an approximate radial scale, we assume that the transition between the inner corona
to the
disk-driven jet at $r=1$ corresponds to the innermost stable circular orbit at $3 R_{\rm S}$
(for a Schwarzschild black hole).  
Hence the physical length scale is given by
\begin{align}
r_{\rm cgs} = 8.9 \times 10^{14} \frac{M_{\bullet}}{10^9 M_{\odot}} \ r\ \rm cm.
\end{align}
The physical density is then obtained by assuming a total jet power $\dot{E}_{\rm 43}$ in
units of $10^{43}\rm erg\ s^{-1}$,
\begin{align}
\rho_{\rm cgs} = 4.7 \times 10^{-17}
        \frac{\dot{E}_{43}}{\dot{E}}
        \left(\frac{M_{\bullet}}{10^9 M_{\odot}}\right)^{-2} \rho\rm\ g\ cm^{-3}, 
\end{align}
where $\dot{E}$ is the corresponding power in code units for a certain simulation run (of order $\sim 10$).  
With this, the magnetic field strength follows to 
\begin{align}
B_{\rm cgs} = 7.3\times10^{2}
  \left(\frac{\dot{E}_{43}}{\dot{E}}\right)^{1/2}
  \left(\frac{M_{\bullet}}{10^9 M_{\odot}}\right)^{-1}\ B\ \rm Gauss
\end{align}
and the physical time scale becomes 
\begin{align}
t_{\rm cgs} = 3.1 \times 10^{4} \frac{M_{\bullet}}{10^9 M_{\odot}} \ t\ \rm s\ .
\end{align}

Unless stated otherwise, we adopt a black hole mass of
$M_{\bullet}=10^{9}M_{\odot}$ and a total jet power of
$\dot{E}_{43}$.
In order to calculate the observable radiation fluxes, we assume a photometric distance of $D=100\ \rm Mpc$.  
The angular scale of the Schwarzschild radius then becomes $\alpha_{\rm rS}=0.2~ \mu\rm as$.  
For the case of M87's supermassive black hole with $M_{\bullet}=6.6\times 10^{9}M_{\odot}$ and $D=~16~ \rm Mpc$ we would have $\alpha_{\rm rS}=8~ \mu\rm as$ yielding an increase in resolution by a factor of $40$ compared to our fiducial scaling.   

\section{Jet dynamics: Acceleration and Collimation}

The large intrinsic scales of the relativistic MHD jet acceleration process require
substantial numerical effort when simulated with a dynamical code.  
Codes optimized for such tasks were developed by \cite{2007MNRAS.380...51K} or
\cite{2008MNRAS.388..551T}, involving particular grid-extension techniques which can
speed up the simulations of a causally de-coupled flow.
In both seminal papers the flow acceleration could be followed substantially beyond the
equipartition regime to establish tight links to analytical calculations.  
However, it can be argued that by using a rigid, reflecting boundary of certain shape as done by
\citet{2007MNRAS.380...51K}, or a force-free approach as applied by
\citet{2008MNRAS.388..551T},
the rate of jet collimation resulting from those simulations could be altered.  

Here we aim at studying MHD \textit{self-}collimation including inertial forces.
We therefore solve the full MHD equations omitting the outer fixed funnel around the jet
and replace it with a stratified atmosphere which may dynamically evolve due to the
interaction with the outflow.
By placing the outer boundaries out of causal contact with the solution of interest,
we can be certain to observe the intricate balance between jet self-collimation and
acceleration that is inaccessible otherwise.  
We like to emphasize that the density has to be considered in the flow equations for 
two reasons - one is to take into account the inertial forces which are important
for collimation and de-collimation, the other is our aim to consistently treat the
Faraday rotation which is given by cold electrons.  The latter could in principle be taken into account in magnetodynamic simulations by introducing electron tracer particles as additional degree of freedom, but is not immediately satisfied by applying the force-free limit of ultrarelativistic MHD.  

\begin{figure*}[htbp]
\begin{minipage}{0.64\textwidth}
\includegraphics[width=0.45\textwidth]{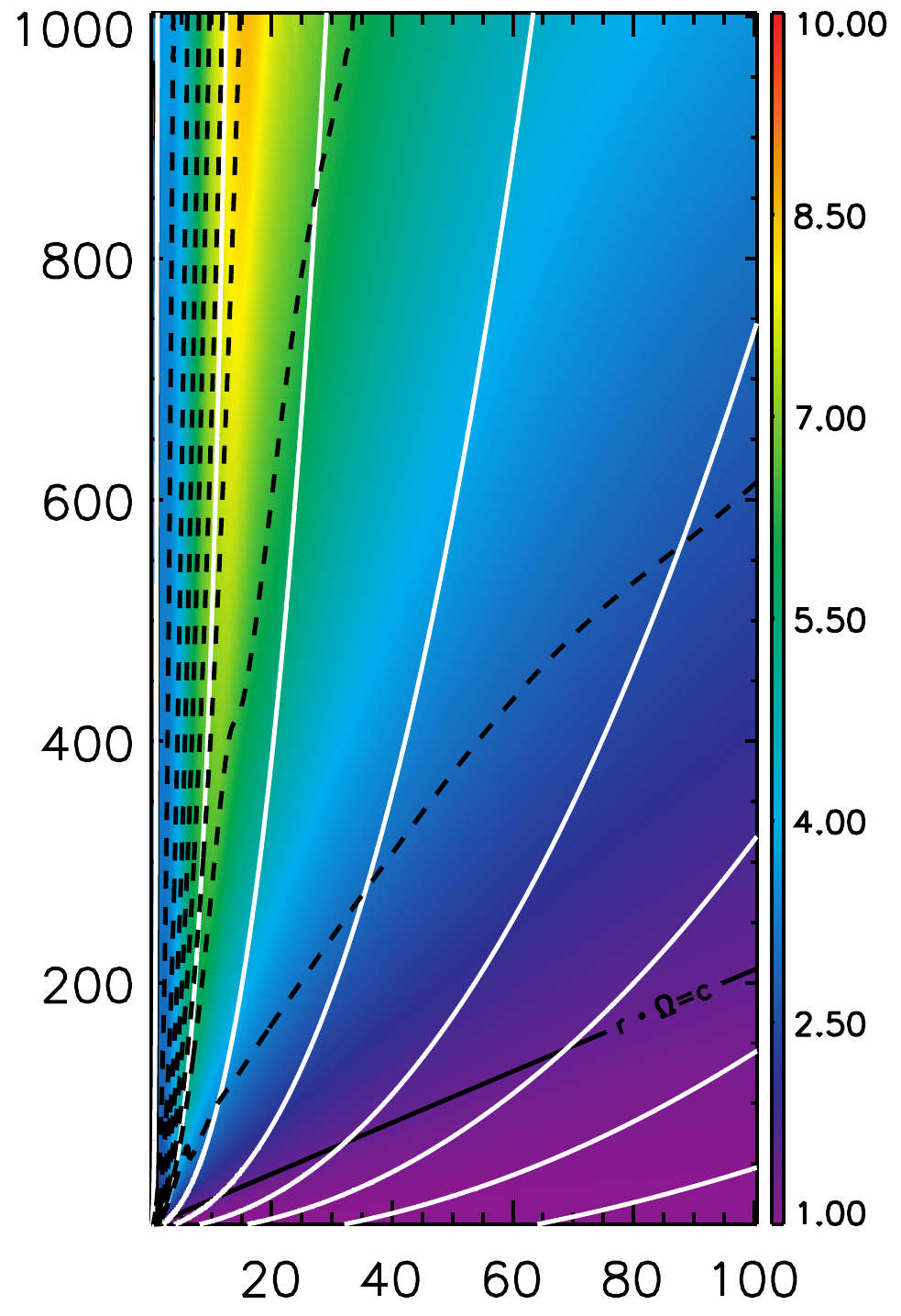}
\includegraphics[width=0.51\textwidth]{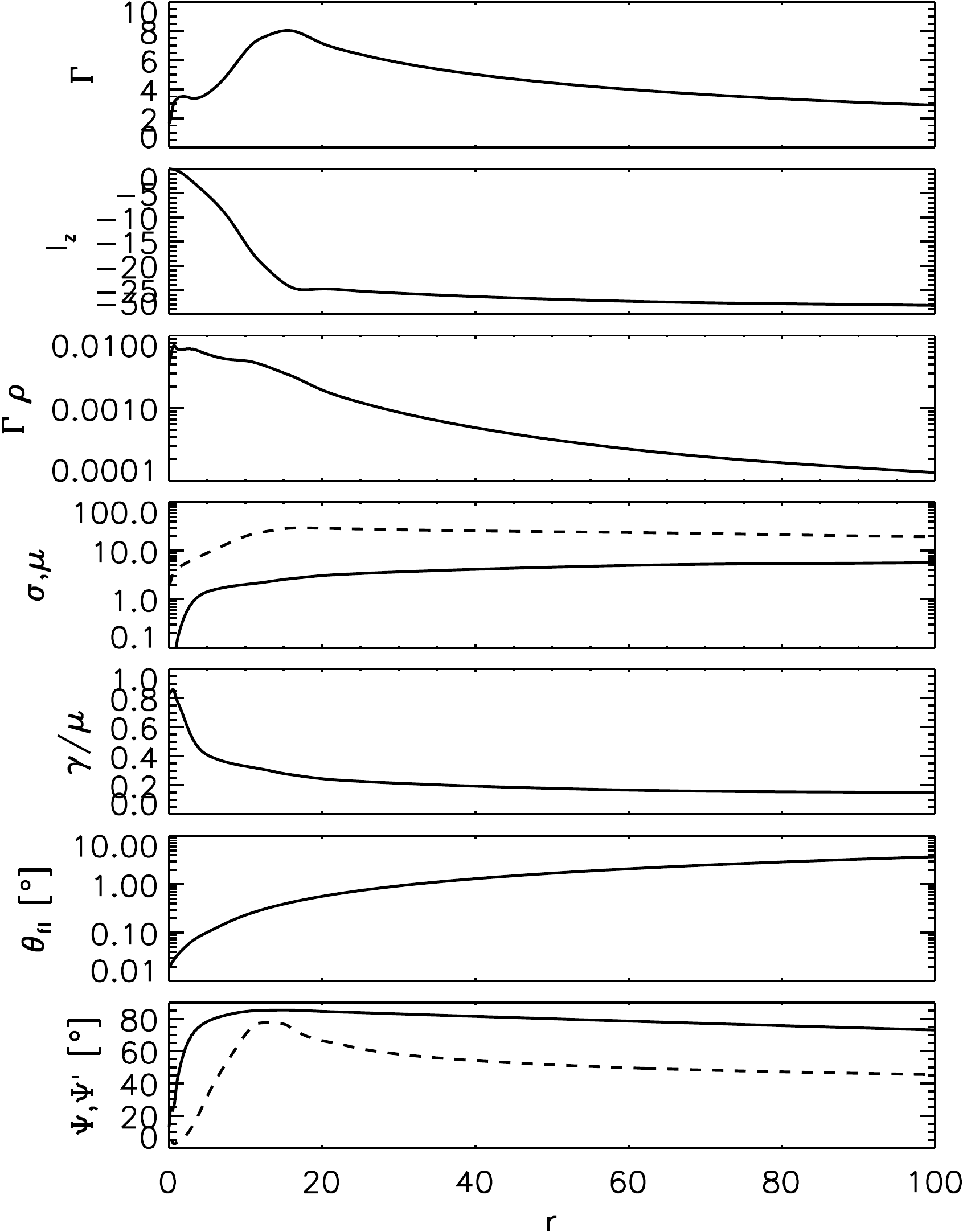}
\end{minipage}
\begin{minipage}{0.31\textwidth}
\centering
\includegraphics[width=.96\textwidth]{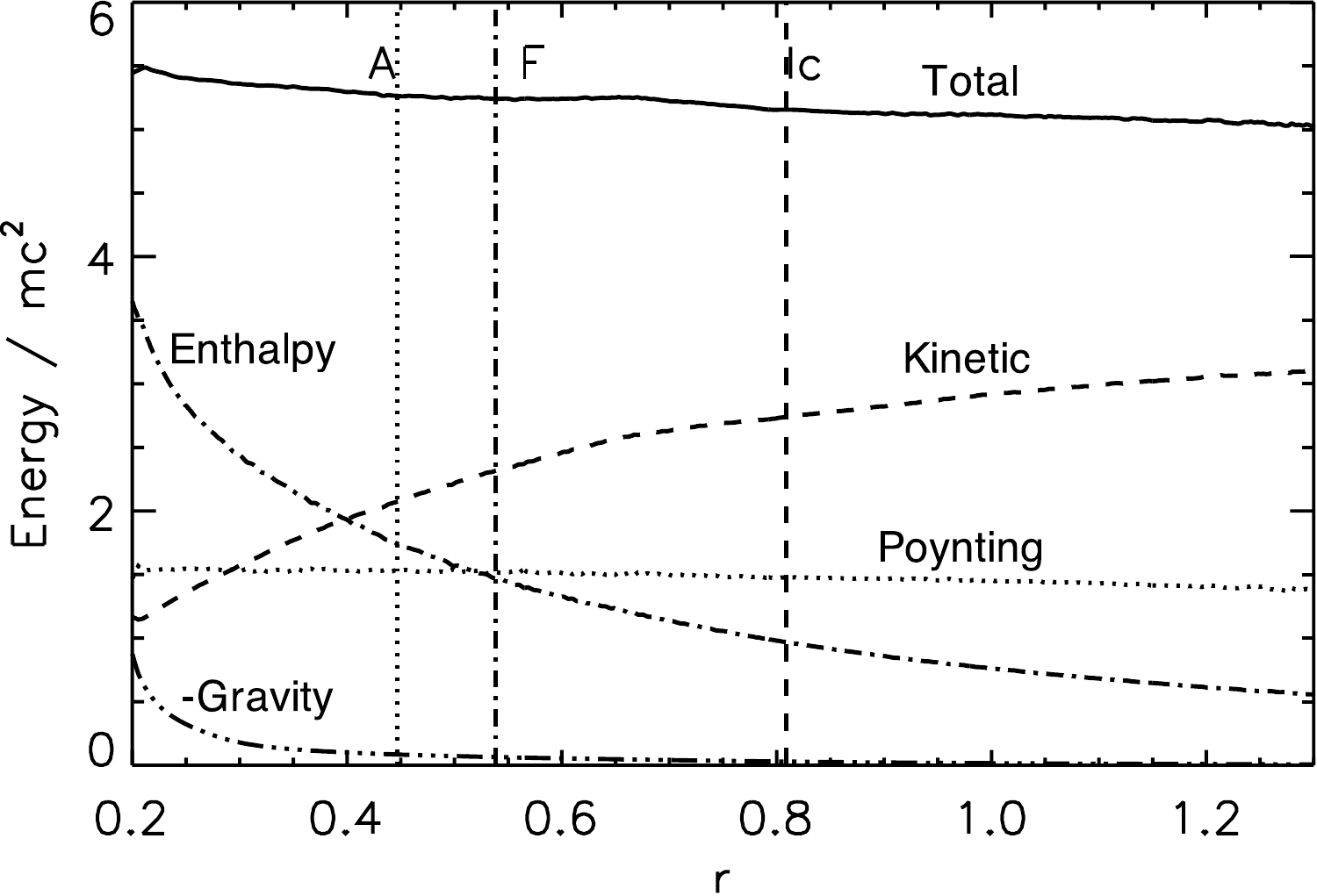}
\includegraphics[width=\textwidth]{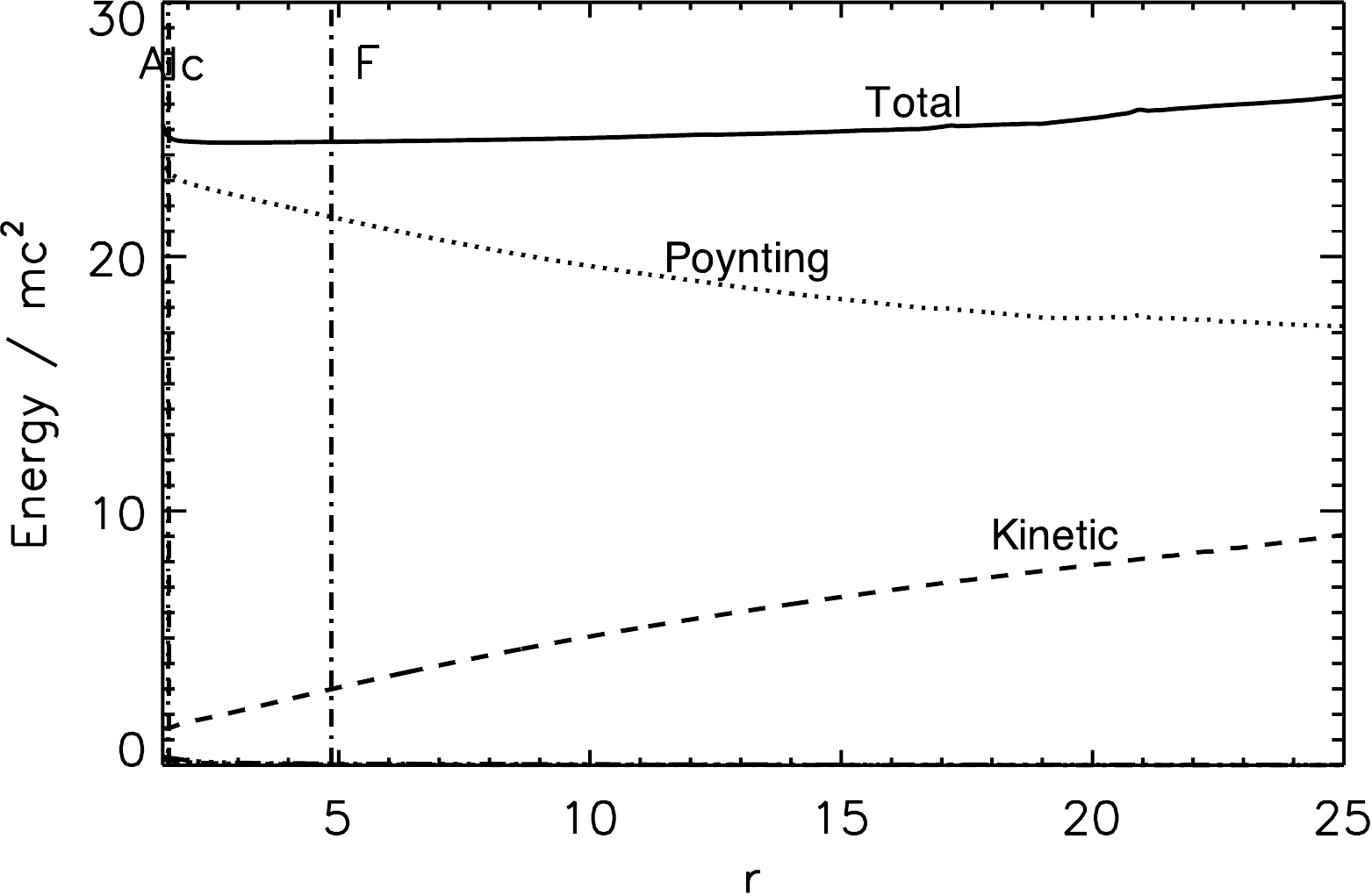}
\end{minipage}
\caption{Jet model 1h allowing no outgoing current in the disk.  \textit{Left:} Current lines (dashed) shown on Lorentz factor color-contours in the (r,z) plane (note the extreme aspect ratio).  Field lines are given in solid white and the light-cylinder is indicated by the solid black line.  
\textit{Center:}
Cuts through $z=750$ for Lorentz-factor $\Gamma$, integral current $I_{z}$, lab-frame density $\Gamma \rho$, Poynting-to-kinetic energy flux ratio $\sigma$ (solid) and total normalized energy $\mu$ (dashed), acceleration efficiency $\Gamma/\mu$, field line collimation angle $\theta_{\rm fl}$ and the pitch angles of the co-moving system $\Psi'$ (dashed), respectively the lab-frame $\Psi$ (solid).  
\textit{Right:}
Acceleration along selected field-lines against the cylindrical radius r showing thermal acceleration for a field line in the spine (footpoint $r_{\rm fp}=0.2$, above) and magnetic acceleration in the jet ($r_{\rm fp} =1.5$, below).  Vertical lines indicate the crossing of the Alfv\`en (A) and fast (F) critical point as well as the light cylinder (lc).  
\label{fig:models1}}
\end{figure*}

\begin{figure*}[htbp]
\begin{minipage}{0.64\textwidth}
\includegraphics[width=0.45\textwidth]{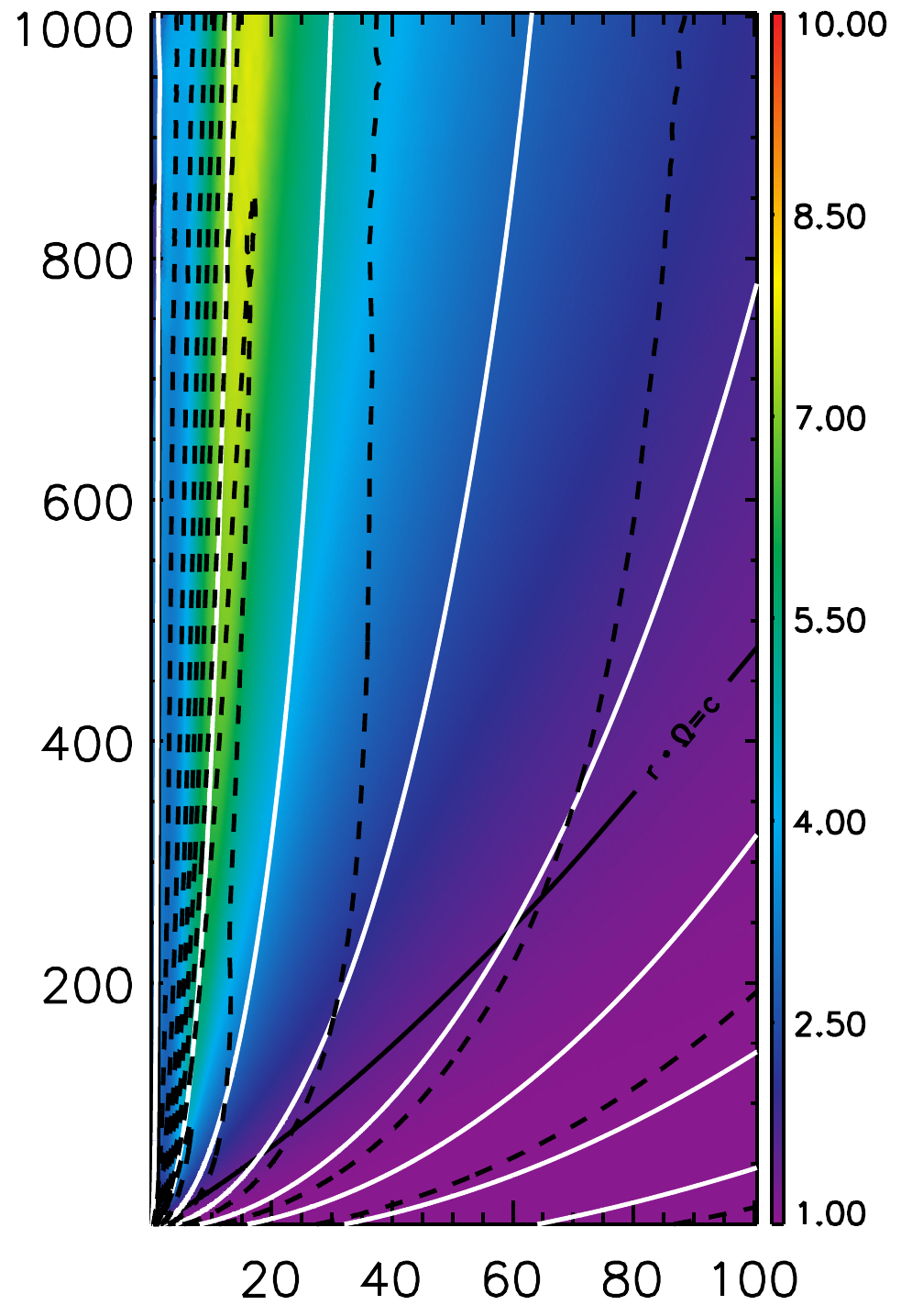}
\includegraphics[width=0.51\textwidth]{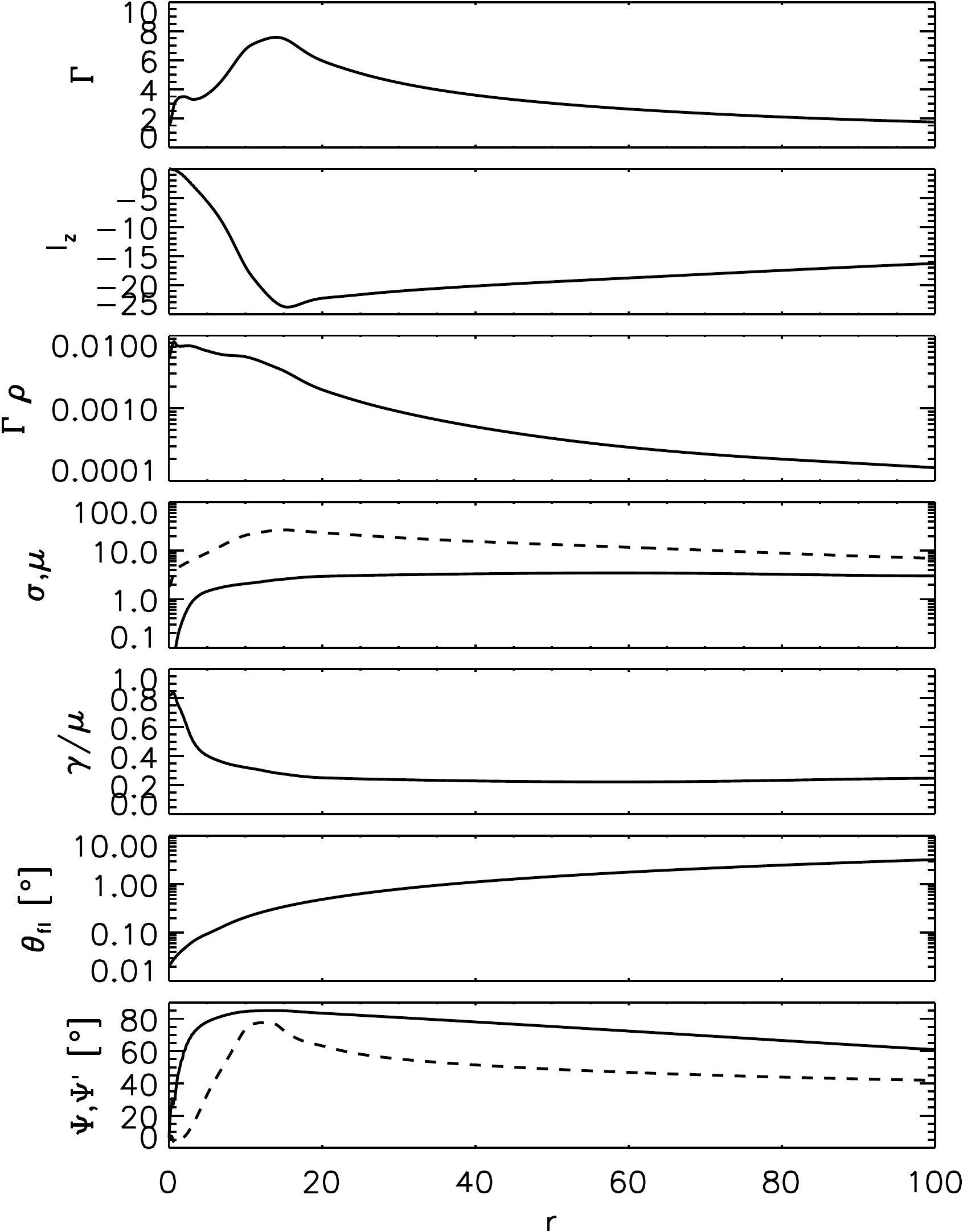}
\end{minipage}
\begin{minipage}{0.31\textwidth}
\centering
\includegraphics[width=.96\textwidth]{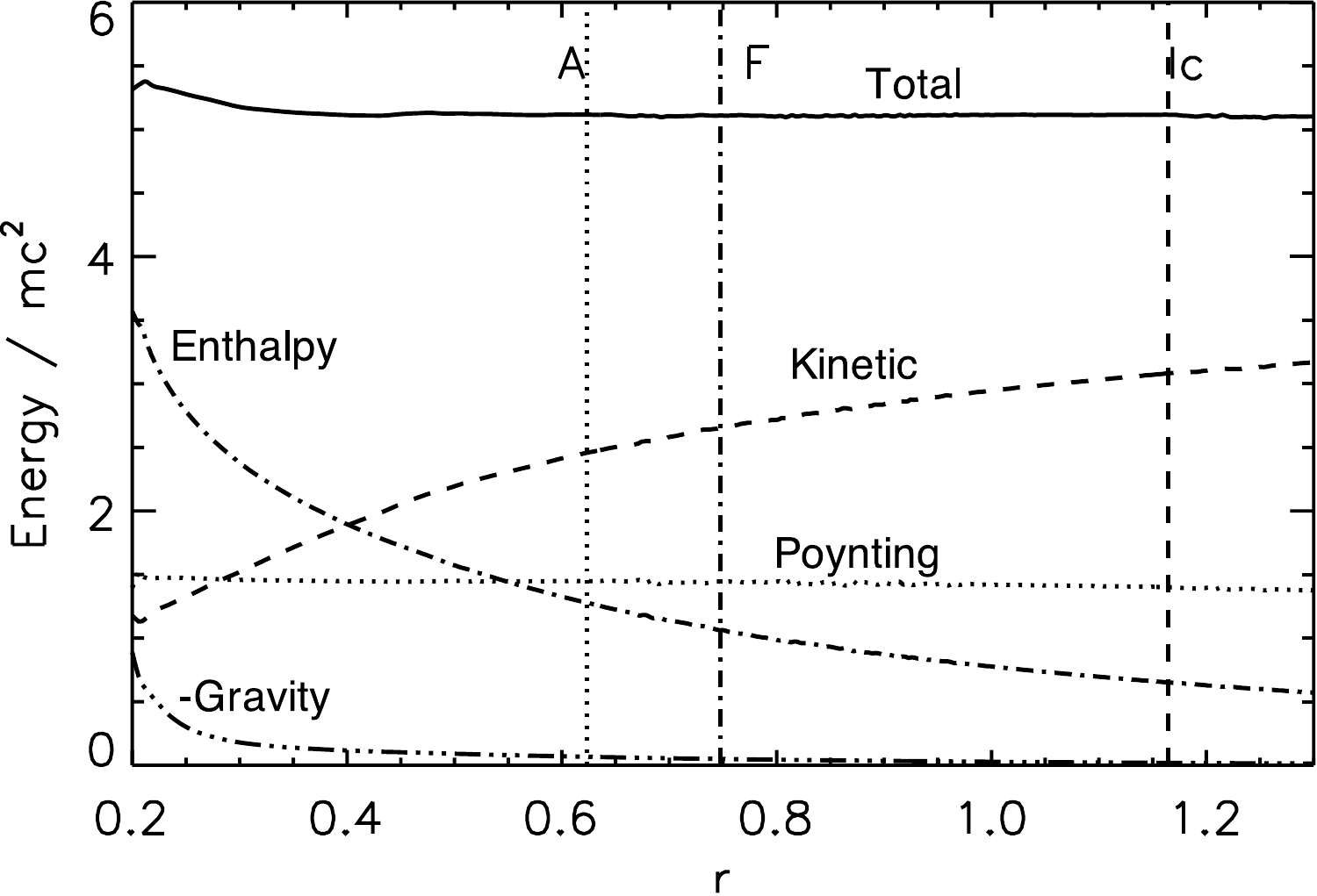}
\includegraphics[width=\textwidth]{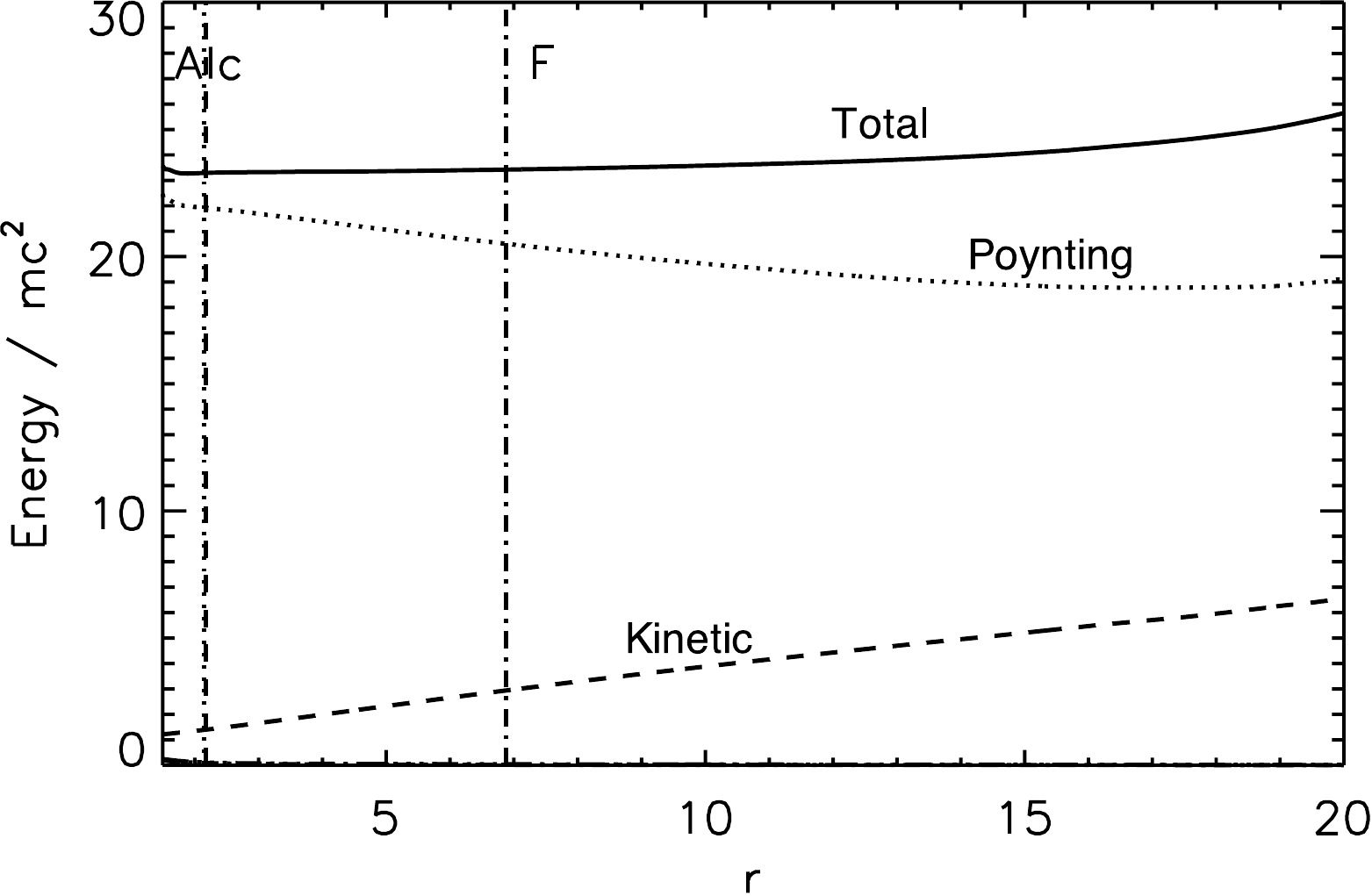}
\end{minipage}
\caption{As figure \ref{fig:models1} for model 2h.
\label{fig:models2}}
\end{figure*}

\begin{figure*}[htbp]
\begin{minipage}{0.64\textwidth}
\includegraphics[width=0.45\textwidth]{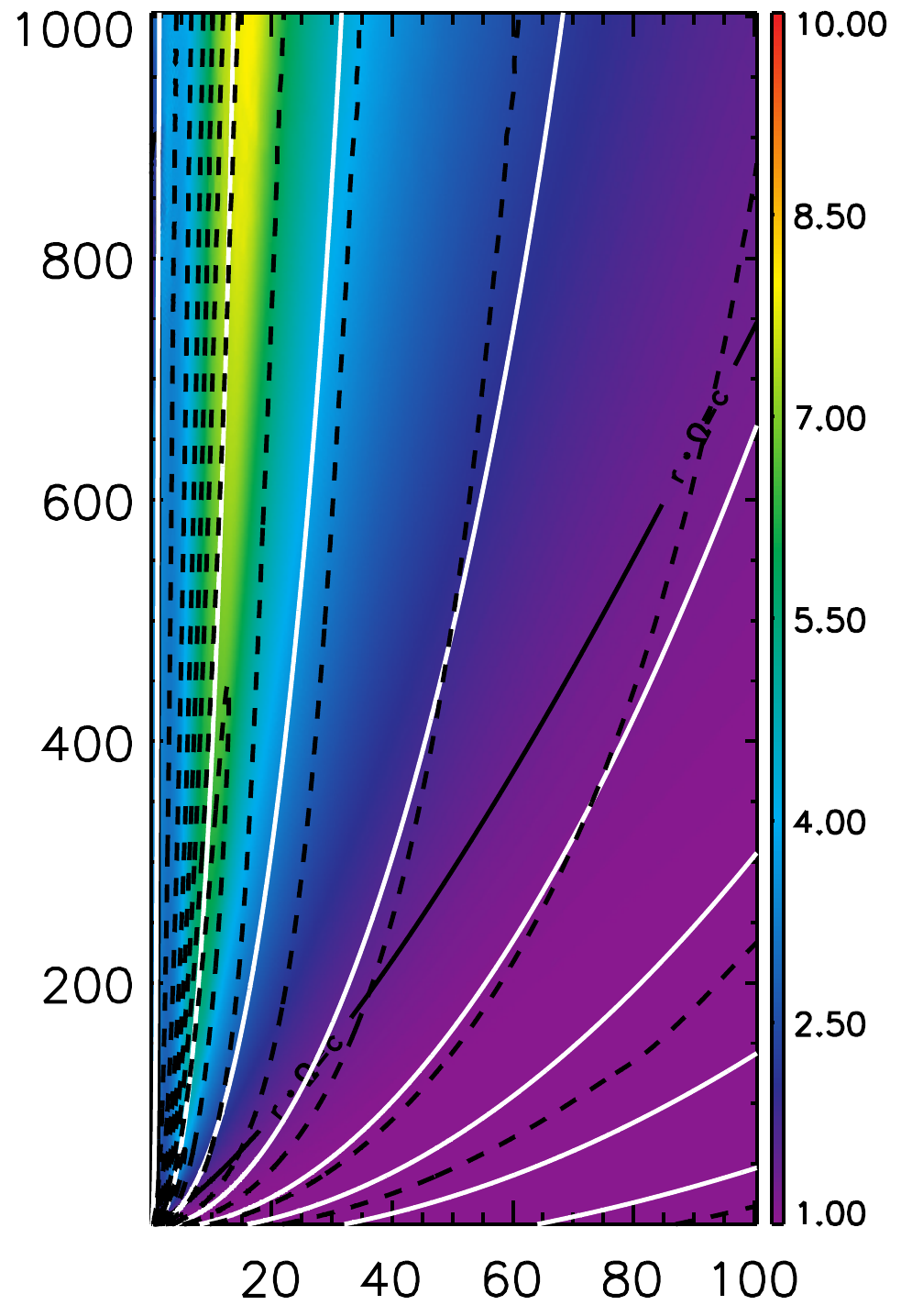}
\includegraphics[width=0.51\textwidth]{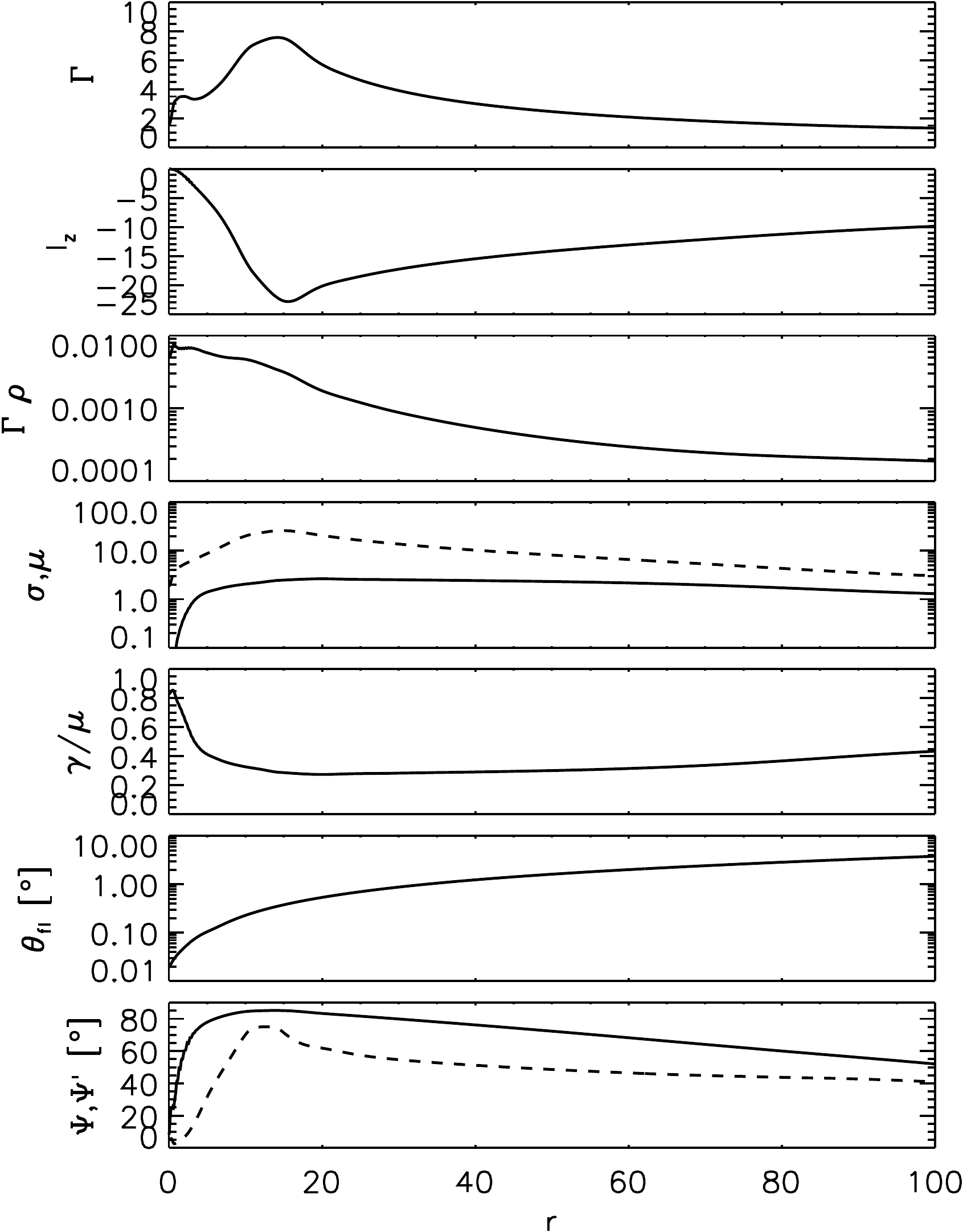}
\end{minipage}
\begin{minipage}{0.31\textwidth}
\centering
\includegraphics[width=.96\textwidth]{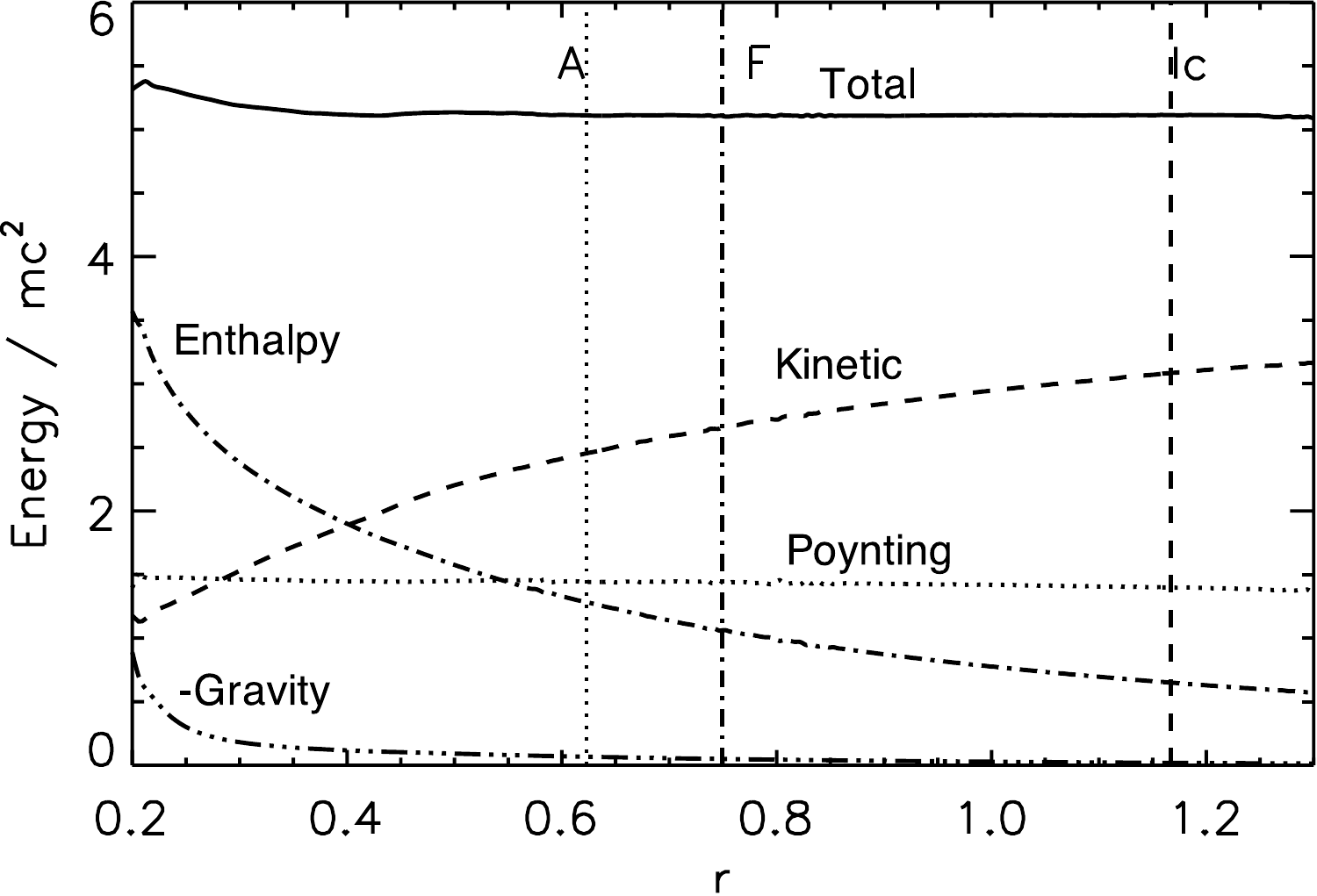}
\includegraphics[width=\textwidth]{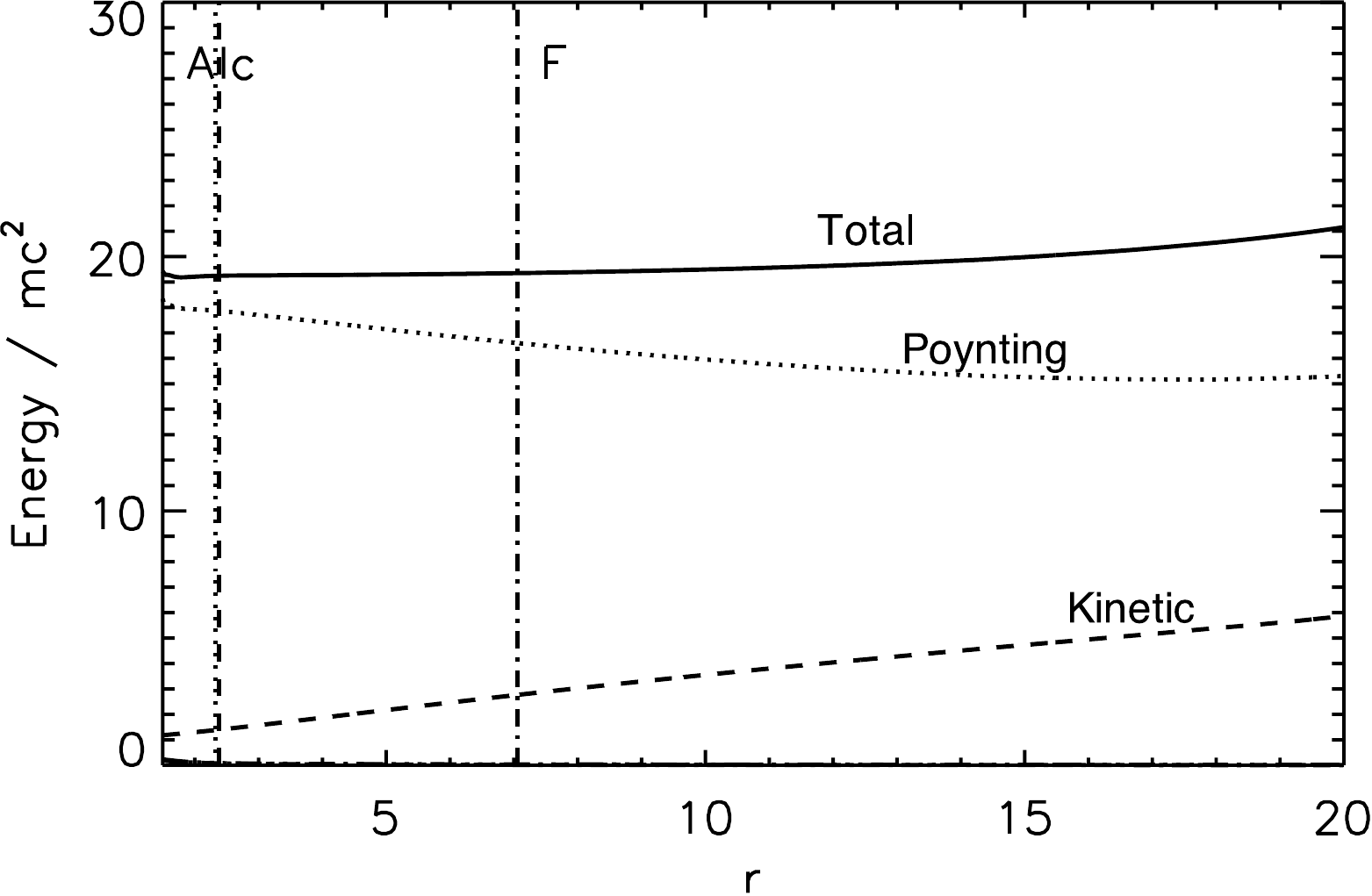}
\end{minipage}
\caption{As figure \ref{fig:models1} for model 3h.
\label{fig:models3}}
\end{figure*}

We summarize our parameter runs of different jet models in table \ref{tab:sim},
indicating the maximum Lorentz factor attained, $\Gamma_{\rm max}$ and other dynamical quantities of interest on which our results discussed in the following are based.  
\begin{table}[htbp]
\caption{Simulation runs}              
\centering
\label{tab:sim}      
  \begin{tabular}{l l l | l l l}
\hline\hline                        
run ID & $s$ & $\beta_{1}$ & $\Gamma_{\rm max}$ & $\Gamma_{\infty}$ & $\theta_{\rm fl,1}$\\
\hline                                   
1h & 1    & 0.005 & 8.5  & 25  & 0.21$^{\circ}$\\ 
2h & 1.25 & 0.005 & 7.9  & 24  & 0.16$^{\circ}$\\  
3h & 1.5  & 0.005 & 7.9  & 23  & 0.17$^{\circ}$\\   
1m & 1    & 0.01  & 5.9  & 13  & 0.14$^{\circ}$\\   
2m & 1.25 & 0.01  & 5.6  & 13  & 0.16$^{\circ}$\\   
3m & 1.5  & 0.01 & 5.9   & 13  & 0.24$^{\circ}$\\   
\hline                                             
\end{tabular}
\tablecomments{
Columns denote: 
Simulation ID; 
Radial power law slope of the toroidal field; 
Plasma beta at (r,z)=(1,0); 
Maximal Lorentz factor obtained in the Raycasting domain; 
Maximal attainable Lorentz factor assuming complete conversion into kinetic energy $\Gamma_{\infty}\equiv \mu$.  
Collimation angle of the field line rooted at the inner disk radius evaluated at z=1000.  
}
\end{table}
Although the different electric current distributions applied in the inflow boundary condition promote
a distinct jet dynamics as seen for example in the position of the light cylinder shown in figures \ref{fig:models1} to \ref{fig:models3},
the geometry of the field lines turns out to be quite similar.  
At a height $z=750$ (corresponsing to $\approx 2200 r_{\rm S}$), the fast jet component is
collimated into an opening angle less than $1^{\circ}$ in all our models.  
More obvious differences are found in the radial distribution of the Lorentz factor which
is peaked at the maximum of vertical current density $j_{z}$.
In the case of closed-current models, the fast jet component becomes narrower as the integral
electric current levels off more steeply. 
We find an acceleration efficiency in terms of the total energy per rest mass energy, $\mu$ of $\Gamma/\mu\approx 80\%$ for the axial spine
and acceleration efficiencies varying between $20\% < \Gamma/\mu < 40\%$
for the outer parts of the jet (see the outer field lines in the middle panels).  
Since the flow has not reached equipartition within the considered domain,
the acceleration efficiencies we obtain represent only a lower limit to the total
efficiency $\Gamma_{\infty}/\mu$.  
Energy conversion is depicted in the right panels of figures \ref{fig:models1} to \ref{fig:models3} showing the individual energy channels normalized to the conserved rest mass energy flux $\rho u_{p} c^{2}$ along selected field lines.  
The terms are defined: $E_{\rm Enthalpy}\equiv \Gamma (h-1)$, $E_{\rm Kinetic}\equiv \Gamma$, $E_{\rm Gravity}\equiv\varphi$ and the Poynting flux $E_{\rm Poynting} \equiv - r \Omega B_{p} B_{\phi}/(4\pi\rho u_{p} c^{2})$.
\footnote{Where $\rho$ denotes the co-moving density, $u_{p}=\Gamma v_{p}$ is the poloidal part of the four-velocity and h signifies the specific enthalpy defined through the equation of state.  }
In the latter relation, we have introduced Ferraro's iso-rotation parameter $\Omega$ defining the ``angular velocity of the field line'' given by 
\begin{align}
\Omega r\equiv v_{\phi}-\frac{B_{\phi}}{B_{p}}v_{p}.
\end{align}

\subsection{Poynting dominated flow}
We first consider the MHD acceleration of the disk component of the jet flow.  
Here, the bulk of the acceleration takes place in the relativistic regime beyond the
light surface $r_{L}(r,z) \equiv c/\Omega$,
and can therefore be approximated asymptotically $x\equiv r/r_{L}\gg 1$, $\Gamma\gg1$.
For a cold wind initially dominated by Poynting flux, the total energy flux
per rest mass energy flux $\mu$ can be expressed as
\begin{align}
\mu=\Gamma + E_{\rm Poynting}=\Gamma-\frac{\Omega r B_{p} B_{\phi}}{4\pi \rho u_{p}c^{2}},\label{eq:energy}
\end{align}

\begin{figure}[htbp]
\begin{center}
\centering
\hfill
\includegraphics[height=0.22\textwidth]{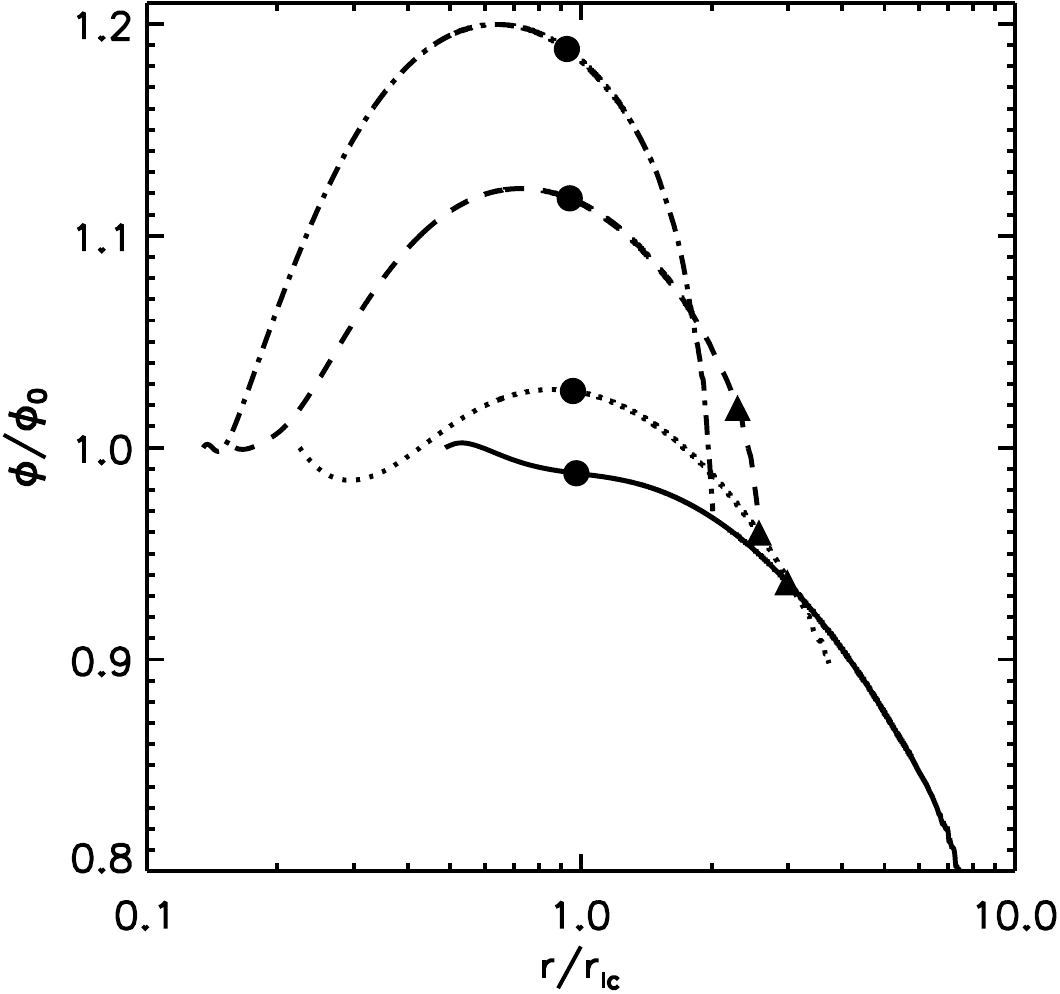}
\hfill
\includegraphics[height=0.22\textwidth]{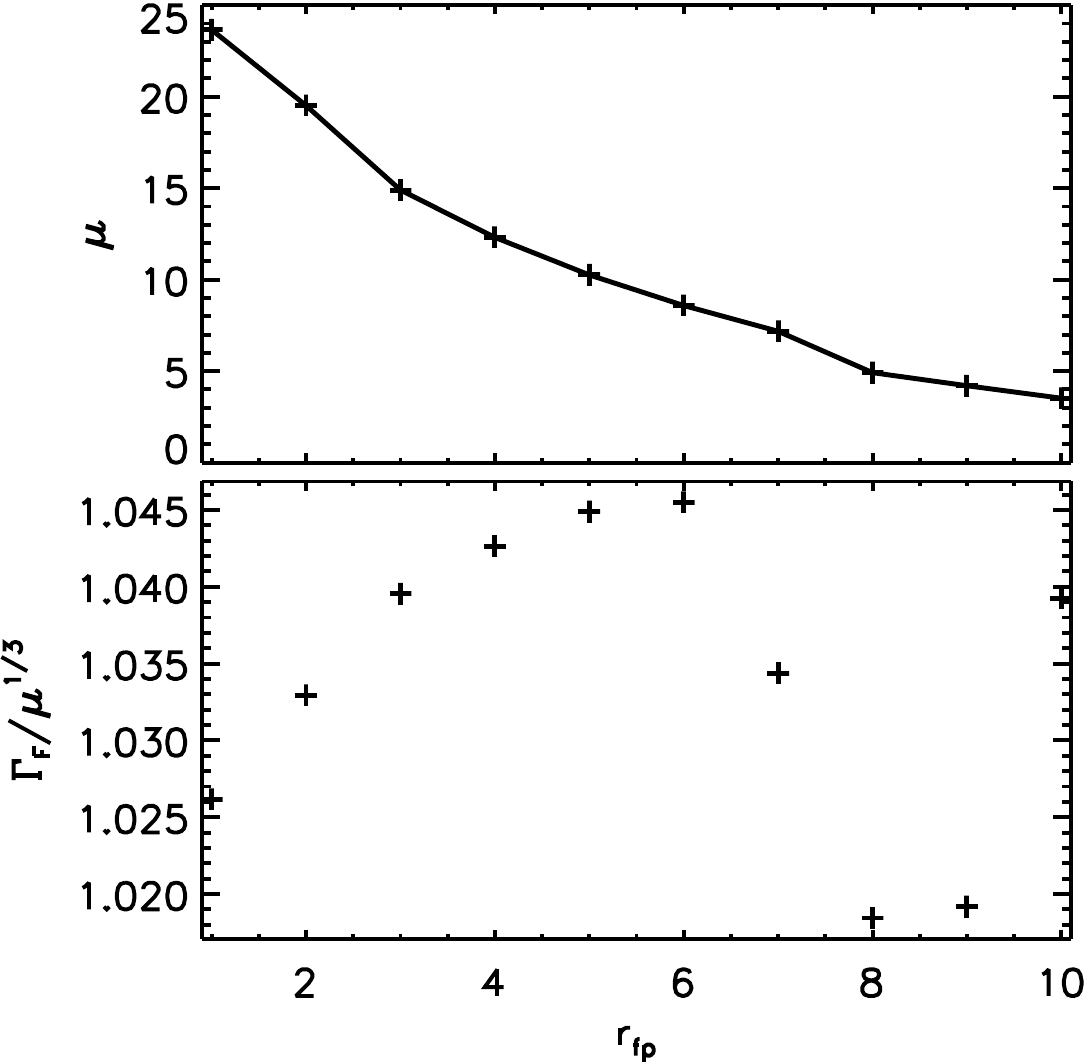}
\hfill
\caption{Acceleration along field lines $r_{\rm fp}\in\{2,4,6,8\}$ shown respectively as \{solid,dotted,dashed,dash-dotted\} lines in model 2h.  
\textit{Left:} The $\phi=B_{p}r^{2}$ function of the expanding flux tube normalized by footpoint value against radius in terms of light cylinder radii $r/r_{\rm lc}$.
Alfv\`en ($\bullet$) and fast ($\blacktriangle$)
 critical point transitions are marked accordingly.
\textit{Right:} Total energy flux ratio $\mu$ \textit{(top)} and Lorentz factor at the fast point $\Gamma_{F}$ compared to the expected value of $\mu^{1/3}$ for various field line footpoints \textit{(bottom)}.  We find Michel's scaling to be satisfied within $5\%$.  
}
\label{fig:phimichel}
\end{center}
\end{figure}

Following the asymptotic relations by \cite{Camenzind1986} we have $v_{\phi}\to 0\, (x\gg1)$,
and hence 
$
\Omega r\simeq -B_{\phi}B_{p}^{-1}v_{p}
$
can be used to eliminate the toroidal field from equation \ref{eq:energy}.  
With $\Gamma \gg 1$ we can write
\begin{align}
\mu=\Gamma-\frac{\Omega^{2} r^{2}B_{p}}{4\pi k c^{3}}, \label{eq:mumichel}
\end{align}
where $\mu$, $k\equiv\rho u_{p}/B_{p}$, and $\Omega$ are conserved quantities along the stationary streamline 
(for details see e.g. \cite{2010ApJ...709.1100P}).   
Thus, the asymptotic flow acceleration depends solely on the decrease of 
\begin{align}
\phi\equiv r^{2}B_{p}
\end{align}
along the flow line by differential fanning out of the field lines.\footnote{Sometimes denoted as ``field line bunching'' in the recent literature.}
We show the evolution of the $\phi$ function along selected field lines for the intermediate
model 2h in figure \ref{fig:phimichel} (left panel).  
In the non-asymptotic regime, $\phi$ increases until the $x=1$ surface,
while it is decreasing for $x\gg 1$ as expected.  

The second term of (\ref{eq:mumichel}) corresponds to the Michel magnetization parameter
$\sigma_{\rm M}$ \citep{1969ApJ...158..727M}
\footnote{Where we added the subscript ``M'' in order to avoid confusion with the parameter
$\sigma=\sigma_{\rm M}/\Gamma$ defined previously.}  
For a critical solution in a monopole field geometry where the fast magnetosonic velocity
is reached at infinity ($x_{\rm F}\to \infty$), the terminal Lorentz-factor becomes
\begin{align}
\Gamma(x_{\rm F})=\mu^{1/3}.  \label{eq:sig13}  
\end{align}
Different derivations of this fundamental result are given by \cite{Camenzind1986,2009ApJ...699.1789T}.  
For small perturbations from the monopole field geometry \cite{begelman1994, 1998MNRAS.299..341B} could show that $x_{F}$ can be crossed at a finite distance, where again (\ref{eq:sig13}) is satisfied.
This general scaling was also found by \cite{1996A&A...313..591F,2001A&A...369..308F,2004ApJ...608..378F} for collimating relativistic jets.  
Our jet solutions quickly accelerate to the fast magnetosonic point and, despite the departure from the monopolar shape, follow Michel's scaling there remarkably well.
As illustrated in figure \ref{fig:phimichel} (right panel), the deviation from the expectation of $\mu^{1/3}$ is less than $5\%$.  
\begin{figure}[htbp]
\begin{center}
\centering
\hfill
\includegraphics[height=0.22\textwidth]{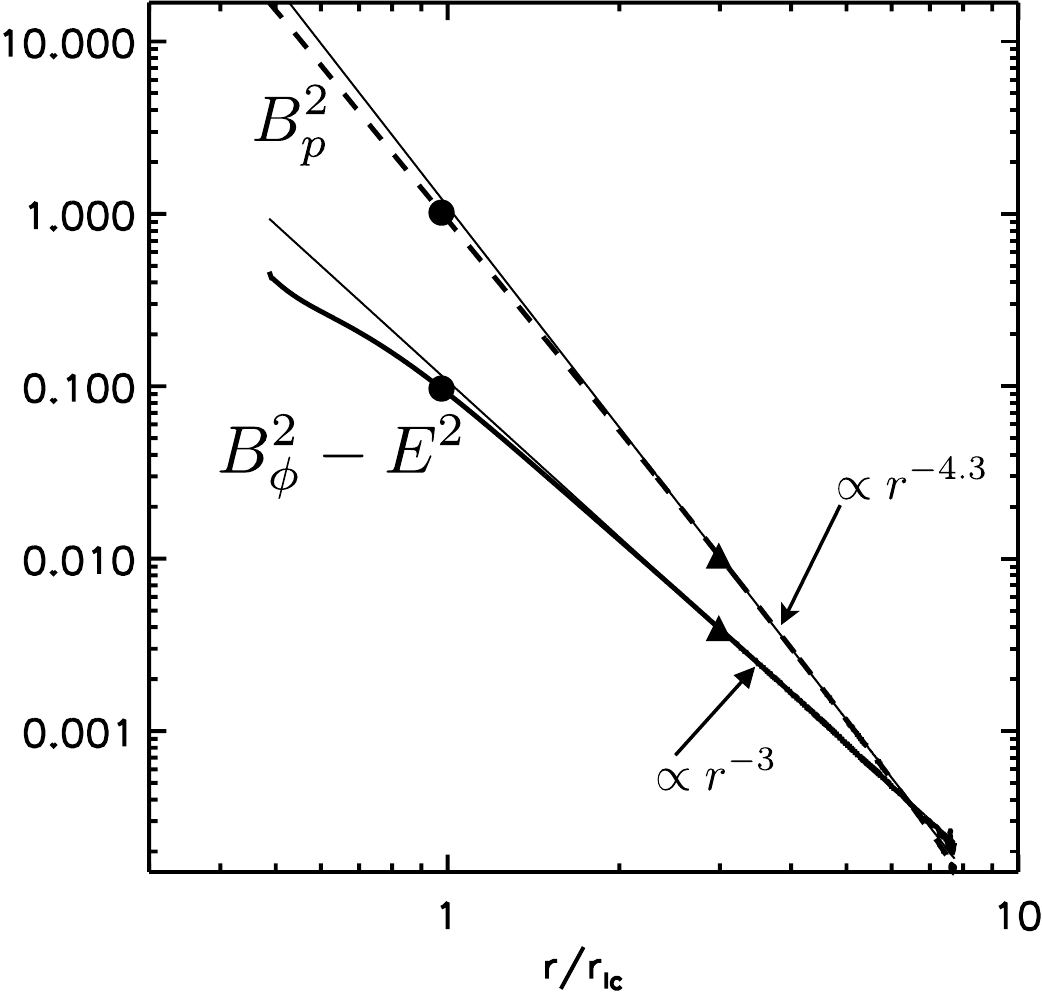}
\hfill
\includegraphics[height=0.22\textwidth]{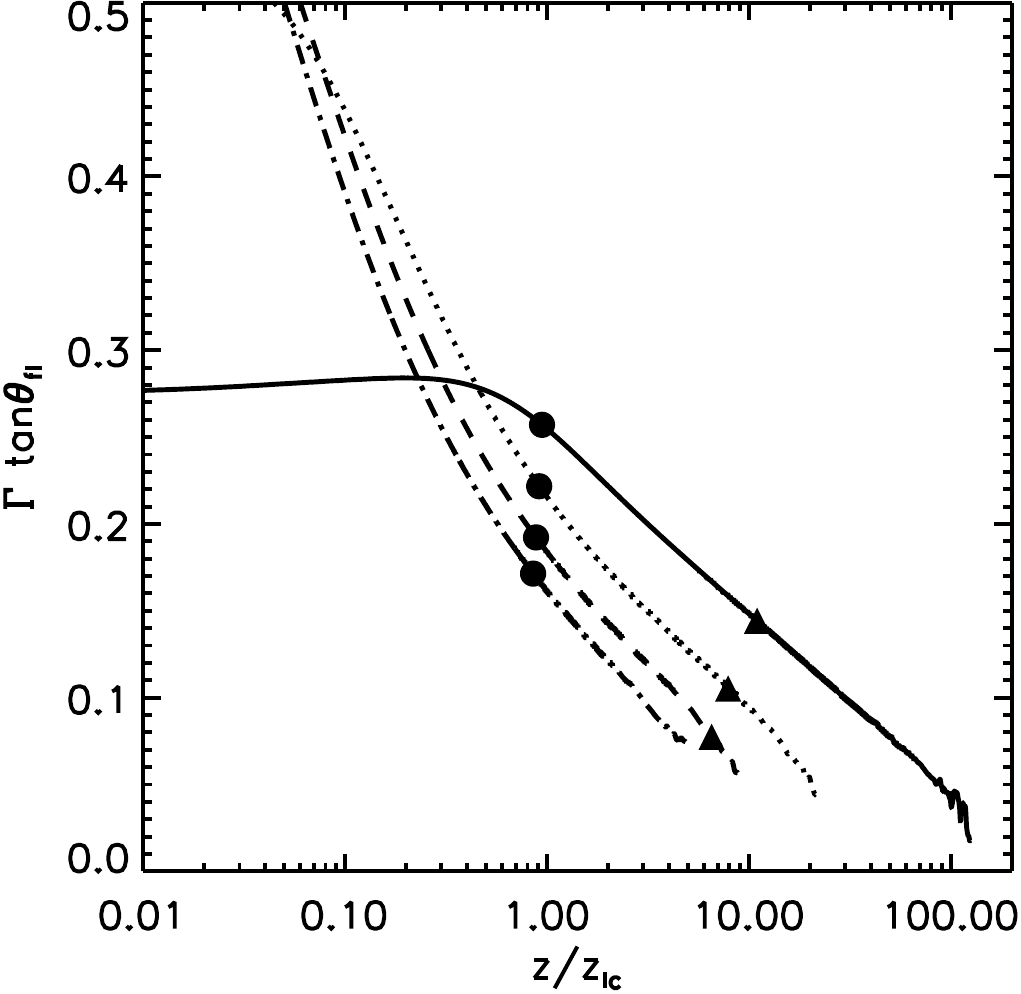}
\hfill
\caption{
Characterization of the acceleration in model 2h.  
\textit{Left:} Comparison of the field strengths for $r_{\rm fp}=2$ along the flux tube.  We find $B_{p}^{2}> B_{\phi}^{2}-E^{2}$ for the most part of the domain yielding the linear acceleration regime.  Also shown are power-law fits to the super-fast regime (thin solid lines).
\textit{Right:} The quantity $\Gamma \tan\theta_{\rm fl}$ along the same field lines of figure \ref{fig:phimichel}.  Efficient acceleration in the power-law regime would yield $\Gamma \tan\theta_{\rm fl}\simeq const$.
Alfv\`en ($\bullet$) and fast ($\blacktriangle$) 
critical point transitions are marked accordingly.
}
\label{fig:regimes}
\end{center}
\end{figure}

Insight into the ongoing acceleration process can be gained by an analysis of the trans-field force equilibrium as performed for example by \citet{1991ApJ...377..462C, vlahakis2004}.   
The asymptotic relativistic force balance can conveniently be decomposed into ``curvature'', ``electromagnetic'' and ``centrifugal'' contributions.  
Depending on the dominating terms, at least two regimes are possible \citep[see also the discussion by][]{2009MNRAS.394.1182K}: 
When the curvature term is negligible, the equilibrium is maintained by balancing of the centrifugal force with the electromagnetic contribution. This constitutes the first or linear acceleration regime.  
The transition to the second regime occurs when field line tension begins to dominate over the centrifugal force, maintaining the equilibrium between purely electro-magnetic forces.
The occurrence of curvature in the force equilibrium leads to a tight correlation between collimation and acceleration since the tension force also becomes the governing accelerating force\footnote{The latter was demonstrated using the parallel field force-balance in application to relativistic disk wind simulations by \cite{2010ApJ...709.1100P}.}.  

As far as a stationary state is reached, we find that the acceleration is well described by the
linear acceleration regime $\Gamma\propto r$, or  
\begin{align}
\Gamma^{2}\approx\frac{B_{\phi}^{2}}{B_{p}^{2}}\label{eq:gamma2b}
\end{align}
as suggested for the initial acceleration of rotating flows by various authors (e.g. \cite{1998MNRAS.299..341B}, \cite{2007MNRAS.375..548N}, \cite{2008MNRAS.388..551T} and \cite{2009MNRAS.394.1182K}).  
This corresponds to 
\begin{align}
B_{p}^{2}\gg B_{\phi}^{2}-E^{2} \label{eq:critfields}
\end{align}
which is satisfied for the most part of the flow in our simulation domain.  
Figure \ref{fig:regimes} (left panel) shows $B_{p}^{2}$ and $B_{\phi}^{2}-E^{2}$ for a sample field line in the fast jet.  We find that the critical field strengths are fairly well approximated by power-laws in the asymptotic super fast-magnetosonic regime. For the particular case shown, we have $B_{p}^{2}\propto r^{-4.3}$ and $B_{\phi}^{2}-E^{2}\propto r^{-3}$ such that the flow experiences a transition to the second, or power-law acceleration regime where the inverse of relation \ref{eq:critfields} becomes true.  

For the power-law regime, a correlation  
between Lorentz factor $\Gamma$ and half-opening angle of the jet $\theta_{\rm fl}$, 
\begin{align}
\Gamma \tan \theta_{\rm fl}\simeq 1 \label{eq:gammatheta}
\end{align}
was discovered by \cite{2009MNRAS.394.1182K} in the context of ultra-relativistic gamma-ray bursts.  
Figure \ref{fig:regimes} (right panel) illustrates the run of $\Gamma \tan \theta_{\rm fl}$
along a set of field lines in our fiducial model.  
Compared to the suggestion of equation \ref{eq:gammatheta}, our simulation setup shows efficient MHD self-collimation, but appears less efficient in terms of acceleration.  

We note that only when a substantial part of flow acceleration takes place in the power-law regime, relation \ref{eq:gammatheta} will hold.  
Our AGN jet models are however collimated to $\simeq1^{\circ}$ and 
accelerated with efficiencies of $40\%$ ($\Gamma\simeq 8$) already in the linear regime.  Even if the flow acceleration is followed indefinitely, $\Gamma \theta_{\rm lf}\simeq 1$ can not be recovered as this would require terminal Lorentz factors of $\Gamma_{\infty}>60$ and thus violate energy conservation.

It could be argued that the low acceleration efficiency is due to the loss of causal connection
for the relativistic flow.  
In this case, the bunching of field-line can not be communicated across the jet anymore, thus stalling 
the acceleration process.  
This should in fact occur when the fast Mach-cone half opening-angle
$\theta_{\rm MF}\simeq\pi/2\sqrt{\mu/\Gamma^{3}}$ does not comprise the jet axis, hence $\theta_{\rm MF}<\theta_{\rm fl}$ \citep[see also][]{Zakamska2008,2009MNRAS.394.1182K}.  
We have checked this conjecture by comparing both angles and found our still
moderate Lorentz factor, highly collimated, jet models to be in causal connection throughout the whole acceleration domain.  

\subsection{Thermal spine acceleration}

In this work, the very inner jet spine is modeled as a thermal wind.  An alternative approach would be to prescribe a Poynting dominated flow originating in the \cite{1977MNRAS.179..433B} process.  In this case, the toroidal field would be generated by the frame dragging in the black hole ergosphere below our computational domain similar to the induction in the disk.  
However, our attempts to increase the central magnetization $\sigma$ by further decreasing the coronal density failed at the inability of the numerical scheme to handle the steep density gradients emerging at the boundary.  Due to the vanishing toroidal field at the axis, also the \cite{1977MNRAS.179..433B} mechanism is not able to provide acceleration of the axial region \cite[see also][]{2006MNRAS.368.1561M}.  

In principle, it would be possible to convert the thermal enthalpy first into Poynting flux when
the jet is expanding, and then back into kinetic energy via the Lorentz force as observed by 
\cite{2009MNRAS.394.1182K}.  
However, as we see in figures \ref{fig:models1} to \ref{fig:models3} (right top panels), this does
in fact not occur in our simulations since the Poynting flux is approximately conserved along
the inner flux lines that show little expansion.  
It is the magnetic field distribution and the collimated structure of the outer (disk-jet)
component which merely provide the shape of the trans-sonic nozzle for the thermal wind.  
We can thus understand the acceleration in the jet spine by using the relativistic Bernoulli
equation, which we cast in the form
\begin{align}
h^{2}\left[1+2\varphi+(u/c)^{2}\right] = c^{4}\Gamma_{\infty}^{2}={\rm const}. \label{eq:bernoulli}
\end{align}
An order of magnitude estimate sufficiently far from the compact object yields $h\Gamma\simeq const$. Using mass conservation $\Gamma\rho r^{2}= const\ (v\to c)$ and a polytropic equation of state with
the enthalpy $h=c^{2}+\gamma/(\gamma-1)\,p/\rho$, we obtain a scaling relation
$\Gamma\propto r^{-2+2/(2-\gamma)}\ (p/\rho\gg c^{2})$.  
For a relativistic polytropic index of $\gamma=4/3$ this results in $\Gamma\propto r$.  

Applying a non-relativistic index of $\gamma=5/3$, the latter relation would yield $\Gamma\propto r^{4}$, however, the non-relativistic limit also implies $p/\rho\ll c^{2}$, and thus $h\to c^{2}\ (\gamma\to 5/3)$ and the acceleration ceases.  

\begin{figure}[htbp]
\begin{center}
\includegraphics[height=0.4\textwidth]{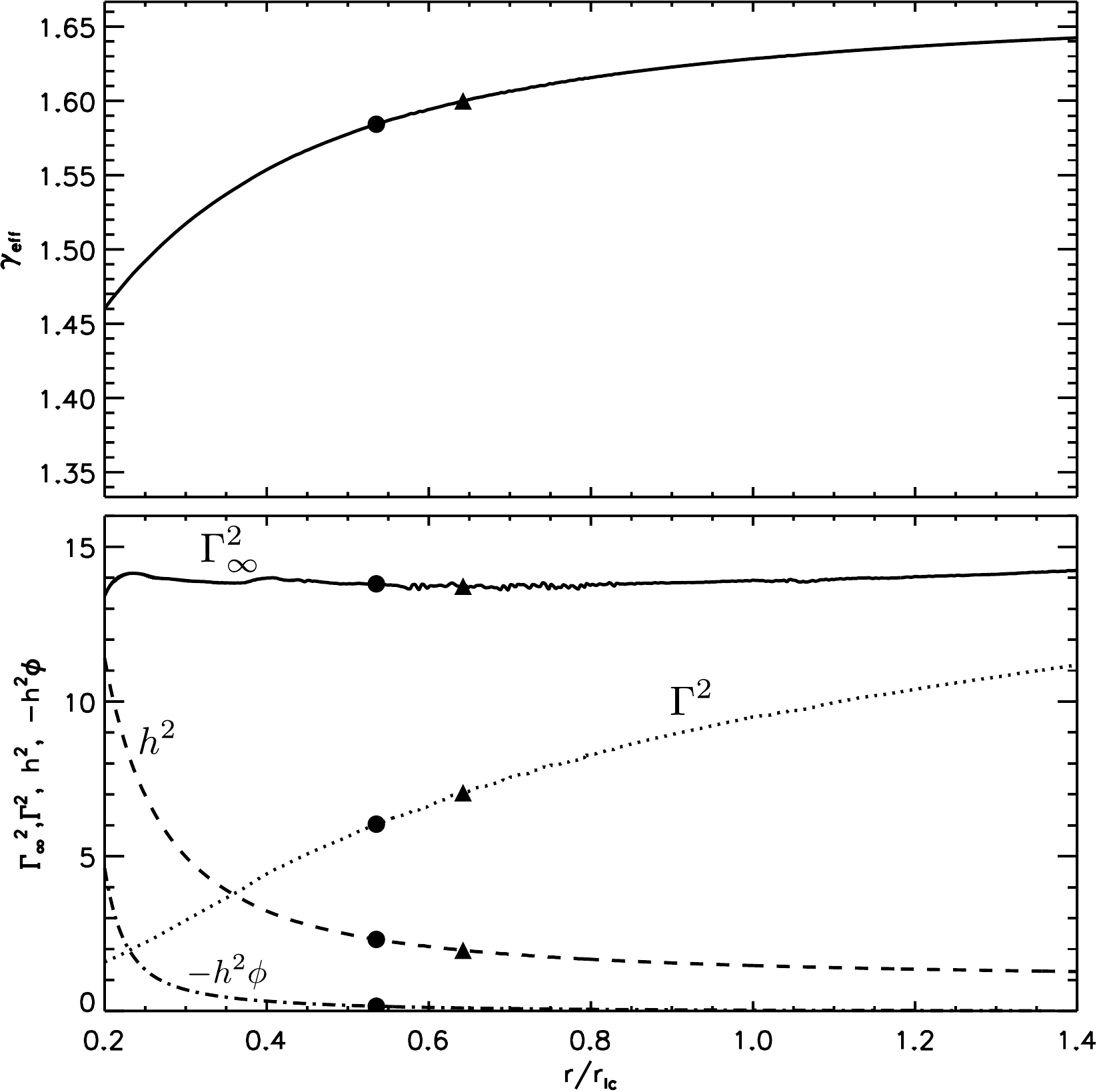}
\caption{
Thermal acceleration along the field line $r_{\rm fp}=0.2$ in model 2h.  
\textit{Top:} Effective polytropic index $\gamma_{\rm eff}$ along the flow.  
\textit{Bottom:} Individual terms of equation \ref{eq:bernoulli} showing thermal energy conversion and the conservation of $\Gamma_{\infty}\simeq 3.7$. In this plot, we normalized to c=1.   
Alfv\`en ($\bullet$) and fast ($\blacktriangle$) 
critical point transitions are marked accordingly.
}
\label{fig:bernoulli}
\end{center}
\end{figure}

Our simulations are performed employing the causal equation of state (obeying the
\cite{1948PhRv...74..328T} inequality) introduced by \cite{mignone2005}.  
Thus we obtain a variable effective polytropic index
\begin{align}
\gamma_{\rm eff}\equiv\frac{d \ln p}{d \ln \rho} = \frac{(h-1)\rho/p}{(h-1)\rho/p-1}
\end{align}
between $4/3$ and $5/3$.  
Figure \ref{fig:bernoulli} (top panel) shows the effective polytropic index along a
stream line / flux surface.  
In the sample stream line we find $\gamma_{\rm eff}$ to vary between
$1.45<\gamma_{\rm eff}<1.65$ as the plasma adiabatically cools from relativistic to
non-relativistic temperatures.  
Thermal acceleration saturates for $\gamma_{\rm eff}\to 5/3$ as the enthalpy approaches the specific rest mass energy $c^{2}$ (see also Fig.~\ref{fig:bernoulli}, bottom panel).  
The maximum attainable Lorentz factor $\Gamma_{\infty}$ is given by the footpoint values at
the sonic point to $\Gamma_{\infty}=h_{0}(\Gamma_{0}+2\phi_{0})^{1/2}$ and depends on the
detailed modeling of the inner corona.  
In our approach the jet spine Lorentz factor is thus limited to values of $\Gamma_{\infty}<4$.  

\section{Synchrotron radiation and Faraday rotation}\label{sec:radiation}

  The numerical MHD simulations discussed above provide an intrinsic dynamical model
  for the parsec-scale AGN core.
  In the following we will use this information - kinematics, magnetic field
  distribution, plasma density and temperature - to calculate consistent synchrotron
  emission maps.
  What is still missing for a fully self-consistent approach is the acceleration model
  for the highly relativistic particles which actually produce the synchrotron radiation.
  However, we have compared a few acceleration models and discuss differences in the ideal 
  resolution synchrotron maps (see below).

\begin{figure}[htbp]
\centering
\includegraphics[width=0.45\textwidth]{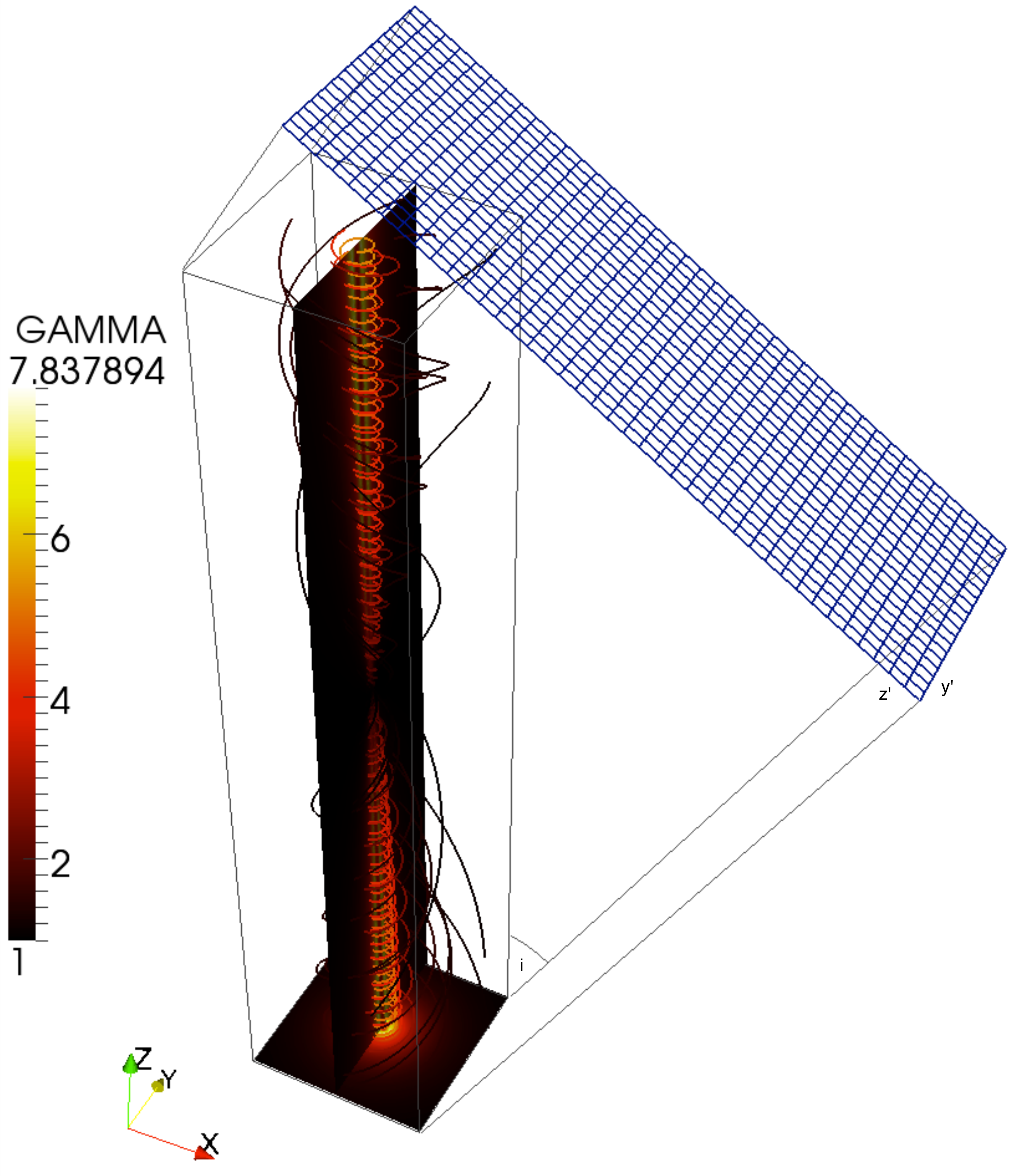}
\caption{Illustration of the ray-casting geometry on an exemplary
  solution.  Color-coding in slices ($x=0$ and $z=-1000$) and field-lines represents the
  bulk Lorentz-factor.  An inclination angle $i=0$
  corresponds to looking directly into the jet. \label{fig:view}}
\end{figure}

Radio observations of nearby AGN-cores show optically thick and thin emission
features with a high degree of Faraday rotation \citep[e.g.][]{zavala2002}.
The nature of the Faraday rotation could either be internal, thus directly produced
in the emitting volume, or due to an external Faraday sheet, possibly comprised of a
magnetized disk-wind as ventured e.g. by \cite{broderick2010}, or an ambient
jet cocoon.
On these scales, even with global VLBI experiments, the radio emission is barely
resolved for most of the known sources.
In order to confront the existing observations, we perform linearly polarized
synchrotron radiation transport in the relativistically moving gas, taking into
account self-absorption and internal Faraday rotation.  We apply beam averaging to examine the resolution dependence of the results.  
An illustration of our ray-tracing procedure with a rendering of an exemplary MHD
solution of a collimating jet is shown in Fig.~\ref{fig:view}. 
For a grid of lines of sight, each corresponding to one pixel in the final image,
we solve for the parameters of linear polarization
$\mathbf{I}=\{I^l,I^r,U^{lr}\}$ as defined e.g. by \cite{Pacholczyk:1970}.  This treatment provides the equivalent information as the Stokes parameters $\{I,Q,U\}$.  
Within the aforementioned notation, the transport equation is a linear system of equations
\begin{align}
    \frac{d\mathbf{I}}{dl} = \mathbf{\boldsymbol{\mathcal{E}}-\underline{A}\ I}\label{eq:radtransp}
\end{align}
where $\boldsymbol{\mathcal{E}}$ denotes the emissivity vector and $\underline{A}$ the opacity matrix, taking into account relativistic beaming, boosting and swing of the
polarization as defined in Appendix
\ref{sec:stokest}.  Faraday rotation of the relativistically moving plasma has first been considered by 
\cite{2009ApJ...703L.104B} and is directly incorporated into the previous relation via the observer system Faraday rotation angle 
\begin{align}
    \frac{d\chi_{\rm F}}{dl} = \frac{e^3}{2\pi m_e^2c^2}\frac{f(\gamma_t)n_e
      D^2}{\nu^2}  \mathbf{(\hat{n} - \boldsymbol{\beta})\cdot B'}\label{eq:Frot}
\end{align}
in cgs units, 
where $e, m_e, n_e$ denote the electron charge, mass and number
density, $\nu, \mathbf{\hat{n}}$ the
observed photon frequency and direction, $D
= \left(\Gamma\left(1-\mathbf{\hat{n}\cdot \boldsymbol{\beta}}\right)\right)^{-1}$ the
Doppler factor and $\mathbf{B'} = \mathbf{B}/\Gamma+\Gamma\left(
\mathbf{\beta B}\right)\cdot \boldsymbol{\beta}$ the co-moving field.  
The dimensionless function $f(\gamma_t)$ takes into account that for
high plasma temperatures the natural wave modes do not remain circular \citep{CambridgeJournals:18267}, suppressing Faraday rotation in favor of conversion between linear and circular polarization.  
We follow \cite{2008ApJ...676L.119H} and \cite{2008ApJ...688..695S} in defining
\begin{align}
  f(\gamma_{\rm t}) =
\gamma_{\rm t}^{-1}\left(\gamma_{\rm t}^{-1}\left(1-\frac{\ln \gamma_{\rm t}}{2\gamma_{\rm t}}\right)+\frac{\ln\gamma_{\rm t}}{2\gamma_{\rm t}}\right)
; \ \gamma_{\rm t} = 1 + \frac{k_B T_e}{m_e c^2}
\end{align}
in terms of the thermal electron Lorentz-factor $\gamma_{\rm t}$ to
interpolate between the cold and relativistic limits.  Especially for
the hot axial flow, Faraday rotation is thus substantially suppressed.  
Assuming an electron-proton plasma, the electron number density follows from the mass density of the simulations.  
We assume further that a small subset of these ``thermal'' electrons is accelerated to a power-law distribution and thus responsible for the non-thermal emission of synchrotron radiation.  The modeling of particle acceleration is detailed further in section \ref{sec:particles}.  

A fraction of the Faraday rotation thus takes place already in the emitting region of the relativistic
jet, such that the radiation undergoes depolarization due to {\em internal} Faraday rotation.
In this case, the angular difference $\Delta \chi_{\rm obs}$ between the observed polarization angle $\chi_{\rm obs}$ and the ($\lambda\to 0$) case can depart from the integral
\begin{align}
\Delta \Psi \propto \int n_e \nu^{-2} \mathbf{B\cdot dl}, \label{eq:dpsi}
\end{align}
which is customarily used in the diagnostics of jet observations.  
For example, in a uniform optically thin medium with internal Faraday rotation, the value of $\Delta \chi_{\rm obs}$ is just half of relation \ref{eq:dpsi}.  Non-uniform optically thin media will break the $\lambda^{2}$-law and exhibit depolarization once $\Delta \chi_{\rm obs}$ exceeds $\sim 45^{\circ}$ \citep[e.g.][]{burn1966}.  For optically thin media with $\Delta \chi_{\rm obs}<45^{\circ}$, $\lambda^{2}$-law rotation measures can be recovered also in the non-uniform case, however the observed rotation angle is always less than $\Delta \Psi$.  

\begin{figure}[htbp]
   \centering
   \includegraphics[width=.45\textwidth]{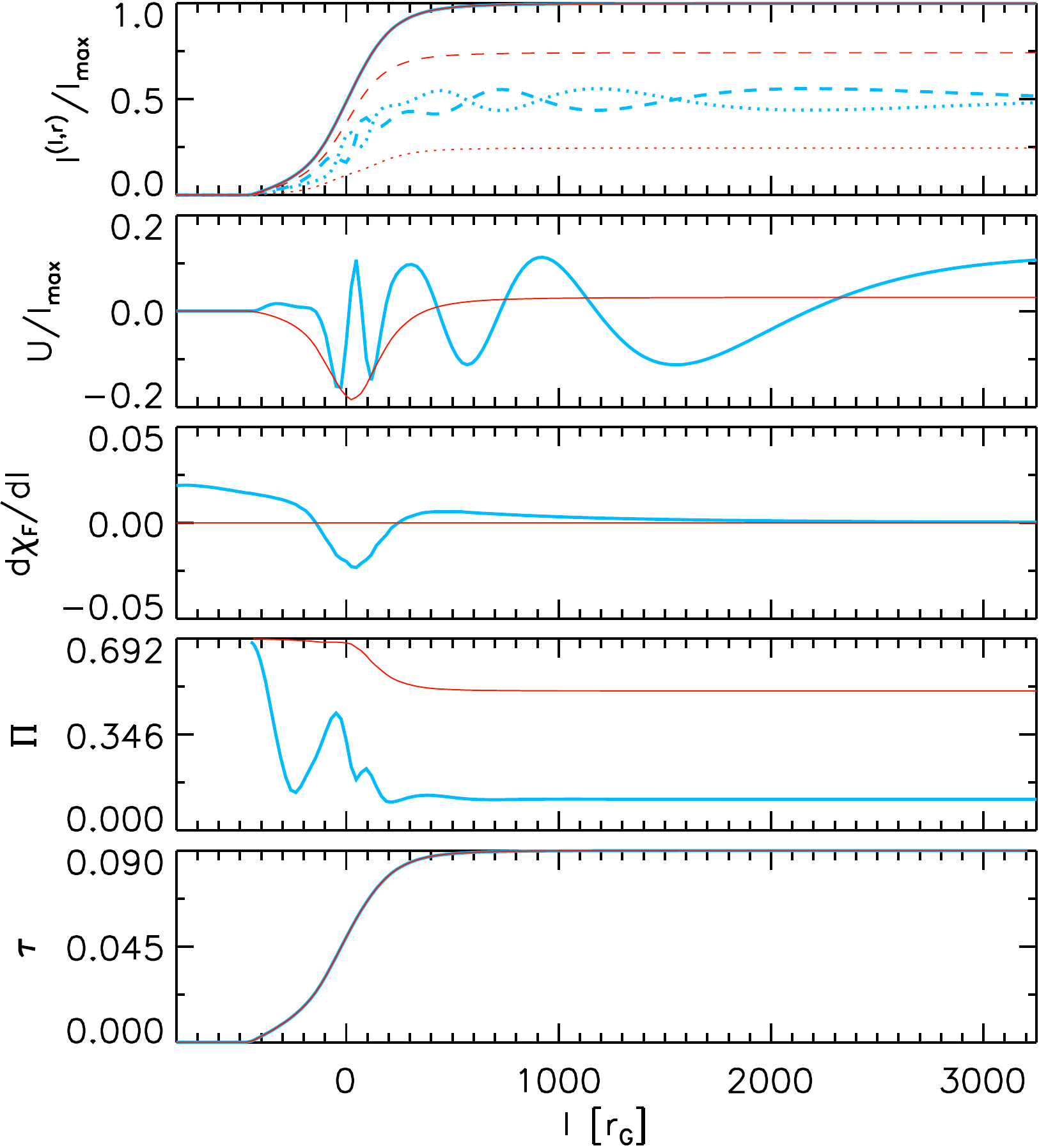}
   \caption{Raytracing for an individual line of sight.  Thick blue lines including Faraday rotation compared to a case where the latter was neglected - illustrated by thin red lines.  The upper panels show intensity ($I^{tot}$) and the Stokes parameters $I^{l}$ (dotted), $I^{r}$ (dashed) and $U$ (second panel).  Faraday depth $d\chi_{F}/dl\ \rm [rad /r_{G}]$, polarization degree $\Pi$ and the optical depth $\tau$ is shown in the subsequent panels.  Internal Faraday rotation and the accompanying depolarization is observed in the emitting region near the $x=0$ plane ($l=0$).}
   \label{fig:los}
\end{figure}

We show the effect of internal Faraday rotation along an individual ray compared to the case 
with no Faraday rotation in Fig.\ref{fig:los}. 
In the emitting volume the polarization degree oscillates as expected for an optically thin medium with Faraday rotation.
As a consequence, the observed polarization degree is lowered.
Following the density and magnetic field strength, the differential Faraday depth $d\chi_{\rm F}/dl$
decreases fast enough towards the observer, so we can be confident not to miss a substantial part
of the Faraday screen in the ray-casting domain.

To speed up the computation in cases of high optical or Faraday depths, we
(i) limit the integration to $\tau<100$, and
(ii) solve the polarized transport only for the last $200$ internal Faraday rotations
$\tau_{\rm F} < 200 \pi$.
Both optimizations do not at all affect the resulting emission maps, as the observed
radiation typically originates in the photosphere of $\tau=1$, and only a few internal Faraday
rotations suffice to depolarize the radiation in the models under consideration.

Once $\mathbf{I}$ is recovered, we obtain beam-averaged quantities via the convolution
\begin{align}
\langle \mathbf{I} \rangle (\mathbf{x}) = \int d^{2}\mathbf{x'}\, \mathcal{G}(\mathbf{x-x'}) \mathbf{I}(\mathbf{x'})
\end{align}
with a Gaussian beam $\mathcal{G}$.  
The beam-averaged Stokes parameters are then used for mock observations providing spectral indices, polarization maps, rotation measure maps, and spectra to be compared to the model parameters.

\subsection{Particle acceleration recipes}\label{sec:particles}

Within the MHD description of the jet plasma, knowledge about the relativistic particle
distribution, which is needed as input for the synchrotron emission model, is not
available.
To recover the information from the velocity-space averaged quantities of MHD, we have to rely
on further assumptions.
To mention other approaches, \cite{2009ApJ...696.1142M} were able to follow the spectral evolution
of an ensemble of relativistic particles embedded in a hydrodynamic jet simulation.
Their treatment includes synchrotron losses, assuming a power-law seed distribution derived from the gas thermal pressure and density at the jet inlet. 
  
Alas, for our purposes a consistent prescription for in-situ acceleration and cooling would be
required - which seems unfeasible at the time.
We therefore take a step back and assume that relativistic electrons are distributed following a global power law with index $p$ as
$dn_{\rm e} = N_0 E^{-p} dE$ for $E_{\rm l}\le E\le E_{\rm u}$ where $N_{0}$ signifies the overall normalization of the distribution and $E_{\rm l}$, $E_{\rm u}$  denote the lower and upper cutoffs.  
The optically thin flux density for the synchrotron process then reads
$S_{\nu}\propto \nu^{-\alpha}$ with $\alpha=(p-1)/2$.
Optically thick regions radiate according to the source function $S_{\nu}= \epsilon_{\nu}/\kappa_{\nu}\propto\nu^{2.5}$. 

This choice of particle distribution is justified by observations as well as theoretical
expectations for the particle acceleration.
The major physical mechanisms capable of producing non-thermal relativistic electrons are
(internal) shock acceleration of relativistic seed electrons \cite[e.g.][]{2000ApJ...542..235K}
and MHD processes like magnetic reconnection \citep{2005MNRAS.358..113L} or hydromagnetic turbulence \citep{1971Ap&SS..12..302K}.  
Considering differential rotation in relativistic jets \cite{2002A&A...396..833R,2004ApJ...617..155R,aloy2008} suggested particle
acceleration by shear or centrifugal effects. 

In addition to $E_l$ and $E_u$, the normalization $N_0$ depends highly on the mechanism
under consideration.
A straight-forward recipe is to connect the particle energy to the overall mass density
\citep[similar to][]{2009ApJ...695..503G}
  
\begin{align}
  \rho = m_p\int_{E_l}^{E_u} N_0 E^{-p}dE,\label{eq:density}
\end{align}
where an ionic plasma consisting of equal amounts of protons and relativistic electrons is assumed.  Thus, all available electrons are distributed following this relation and the Faraday effect is maintained by the ``equivalent density of cold electrons'' $\propto n_{e}E_{l}^{-2}$ \citep[e.g.][]{1977ApJ...214..522J} in contrast to relation (\ref{eq:Frot}) where we assumed that the most part of electrons is non-relativistic.  

An alternative to (\ref{eq:density}) is to specify the integral particle energy density and thus the first moment of the distribution function.  
For their leptonic jet models, \citet[][]{Zakamska2008} have assumed that the internal energy is carried by relativistic particles, 
hence the relation
  \begin{align}
  \epsilon = 3p = \int_{E_l}^{E_u}N_0 E^{1-p}dE\label{eq:pressure}
\end{align}
can be used to provide $N_{0}$ from the gas pressure resulting from the simulations. 

In contrast to relativistic shock acceleration where the energy reservoir for the particles
is the bulk kinetic energy of the flow,
MHD processes directly tap into the co-moving magnetic energy density, and can effectively
accelerate the particles up to equipartition.
Accordingly, for the equipartition fraction $\epsilon_{\rm B}$,
  
\begin{align}
\epsilon_{\rm B}
\frac{B'^2}{8\pi} = n_{\rm e} \langle E\rangle = \int_{E_l}^{E_u} N_0 E^{1-p}dE\label{eq:minimal e}
\end{align}
is customarily used to estimate jet magnetic field strength from the observed
emission \citep[e.g.][]{Blandford1979} or vice versa
\citep{2005MNRAS.360..869L, broderick2010}.
To obtain peak fluxes in the Jy range, we have applied $\epsilon_{\rm B}=0.1$ for our fiducial model.
For the following discussion we have adopted $\alpha=0.5$ ($p=2$), $E_{\rm u}=10^{6}E_{\rm l}$ and specified $E_{\rm l}=\gamma_{\rm t}m_{e}c^{2}$ for application with relation (\ref{eq:density}).
For $\alpha=0.5$ and applying the recipes (\ref{eq:pressure},\ref{eq:minimal e}), only the cutoff
\emph{energy ratio} $E_{\rm u}/E_{\rm l}$ enters logarithmically into the determination of $N_{0}$. 
Within these assumptions, the influence of the cutoff values on resulting jet radiation is marginal.  
The magnitude of $E_{\rm l}$ then merely determines the number density of relativistic particles, to be chosen consistent with the number of particles available for acceleration.  
In this first study, we neglect the spectral changes introduced by the cutoff energies as this would require a more detailed modeling of the particle content which is beyond the scope of the current paper.

The observed morphology of the intensity maps 
is mainly given by the various prescriptions of $N_{0}$ mentioned above.
In the following we briefly compare the resulting radio maps.
  
\par
\subsection{Radio maps for different particle acceleration models}
\begin{figure}
\centering
\hfill
\includegraphics[height=0.35\textwidth]{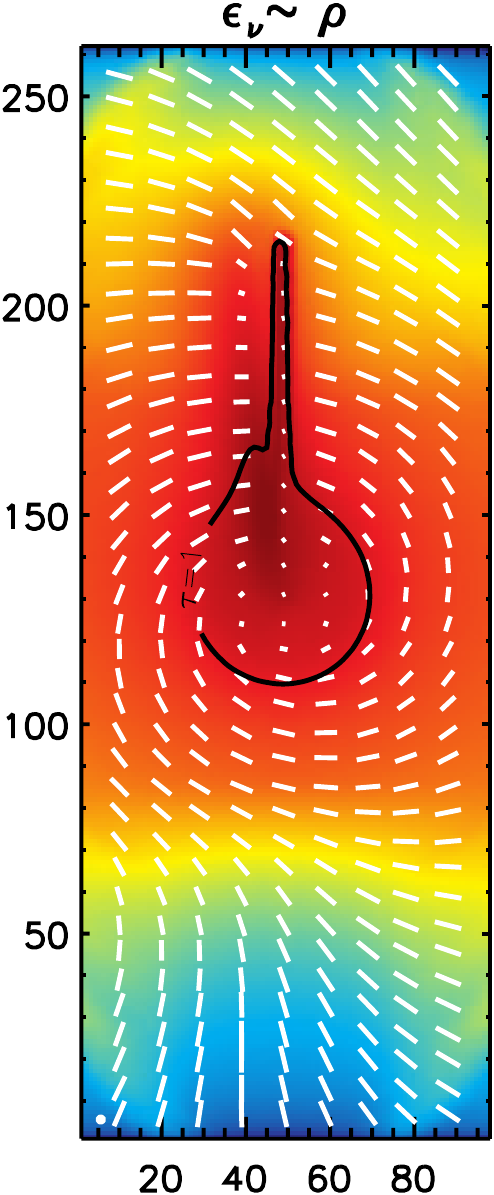}
\hfill
\includegraphics[height=0.35\textwidth]{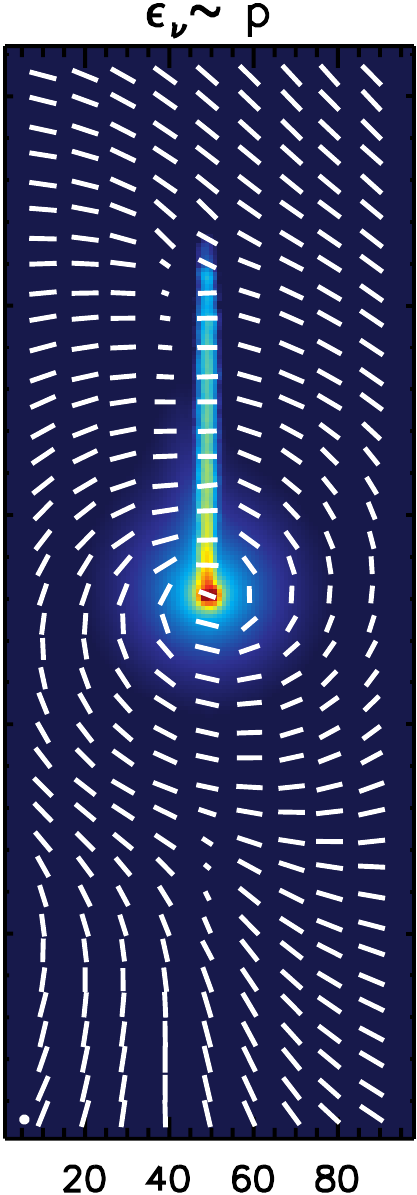}
\hfill
\includegraphics[height=0.355\textwidth]{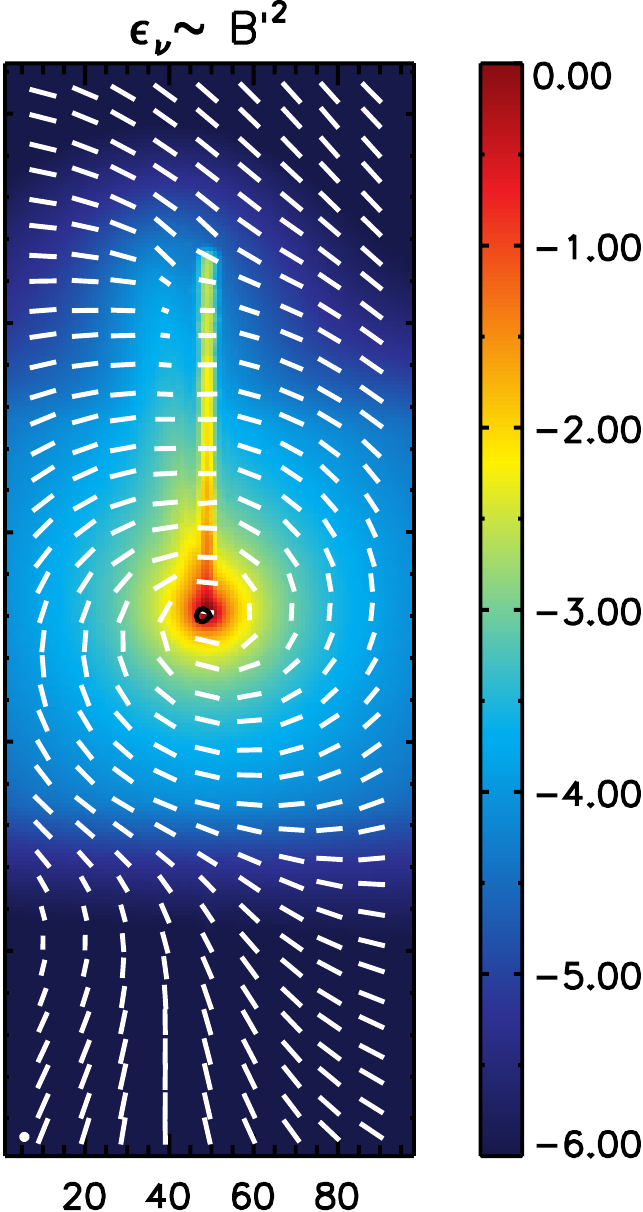}
\hfill
\caption{Ideal resolution logarithmic $I_{\nu}/I_{\nu,\rm max}$ maps for model 2h at $\nu\!=\!43\ \rm GHz$, $i= 30^\circ$ using various tracers for the relativistic particles: Density \textit{(left)}, thermal pressure \textit{(center)} and magnetic energy density \textit{(right)}.  Linear ($\mathbf{\hat{e}}$) polarization vectors are overlaid as white sticks.  The (x-) scale is given in terms of ray-casting footpoint and corresponds to a physical extent of $1200 r_{\rm S}$.  
\label{fig:maps}}
\end{figure}

Figure \ref{fig:maps} shows ideal resolution maps for the aforementioned particle acceleration
tracers.
For the sake of comparison, Faraday rotation is neglected and with $43\rm GHz$ we choose
a high radio frequency to penetrate through the opaque jet base. 

All tracers show an almost identical polarization structure, and highlight a thin ``needle'' owing
to the cylindrically collimated axial flow with high density, and high magnetic and thermal pressure.
The axial flow is slower than the Poynting dominated disk wind which is de-beamed and, thus, not
visible at this inclination.
Since the axial flow features $\mathbf{\boldsymbol{\beta} || B}$ (cf. \ref{eq:ehat}), the resulting
$\mathbf{\hat{e}}$ polarization vector reduces to the classical case, and points in direction
perpendicular to the projected vertical field of the axial spine.
In the case of the density tracer, the emission becomes optically thick, as indicated by the
$\tau=1$ contour.
Correspondingly, the polarization degree is lowered and the direction of the spine polarization
turns inside the $\tau=1$ surface.  
Depending on the radial density and pressure profiles $\rho\propto r^{-3/2}$, $B^{2}\propto r^{-2}$
and the pressure distribution in the disk corona $p\propto r^{-5/2}$,
the emission at the base of the jet is more or less extended and dominates the flux in all three
cases.
Relativistic beaming cannot overcome the energy density which is present in the disc corona,
and therefore necessitates a more elaborate modeling of the accelerated particles in the jet.
This will be provided in section \ref{sec:radiationinjet}.

\subsection{Relativistic swing and beaming}

In optically thin, non-relativistic synchrotron sources, the observed $\mathbf{\hat{b}}$ polarization
vector directly corresponds to the projected magnetic field direction of the emitting region and thus
carries geometric information about the jet.
This allows us to interpret parallel $\mathbf{\hat{e}}$ vectors in terms of toroidal fields, while perpendicular $\mathbf{\hat{e}}$ vectors indicate a poloidal field
\citep[e.g.][]{1980MNRAS.193..439L}.
Similar to all realistic models of MHD jet formation, our simulations feature a helical field structure
that is tightly wound within the fast jet, but increasingly poloidal further out.
Hence, the resulting polarization structure is that of the telltale \textit{spine and sheath} geometry -
across the jet, the polarization $\mathbf{\hat{e}}$ direction flips from being perpendicular to parallel
and eventually returns to a perpendicular orientation.
  
Due to aberration and the accompanying swing of the polarization \citep[e.g.][]{Blandford1979},
an interpretation in terms of pure geometrical effects is not longer applicable in flows with relativistic
velocities, instead a kinematic jet model is required.
For cylindrical (i.e. $(z,\phi)$-symmetric) relativistic jets, \cite{Pariev2003} have
demonstrated how the optically thin polarization follows a strictly bimodal distribution,
since the inclined polarization vector from the front of each annulus cancels with the
corresponding polarization vector from the back side.
This remains also valid for differentially rotating jets.

\begin{figure}[htbp]
\begin{center}
\includegraphics[width=0.4\textwidth]{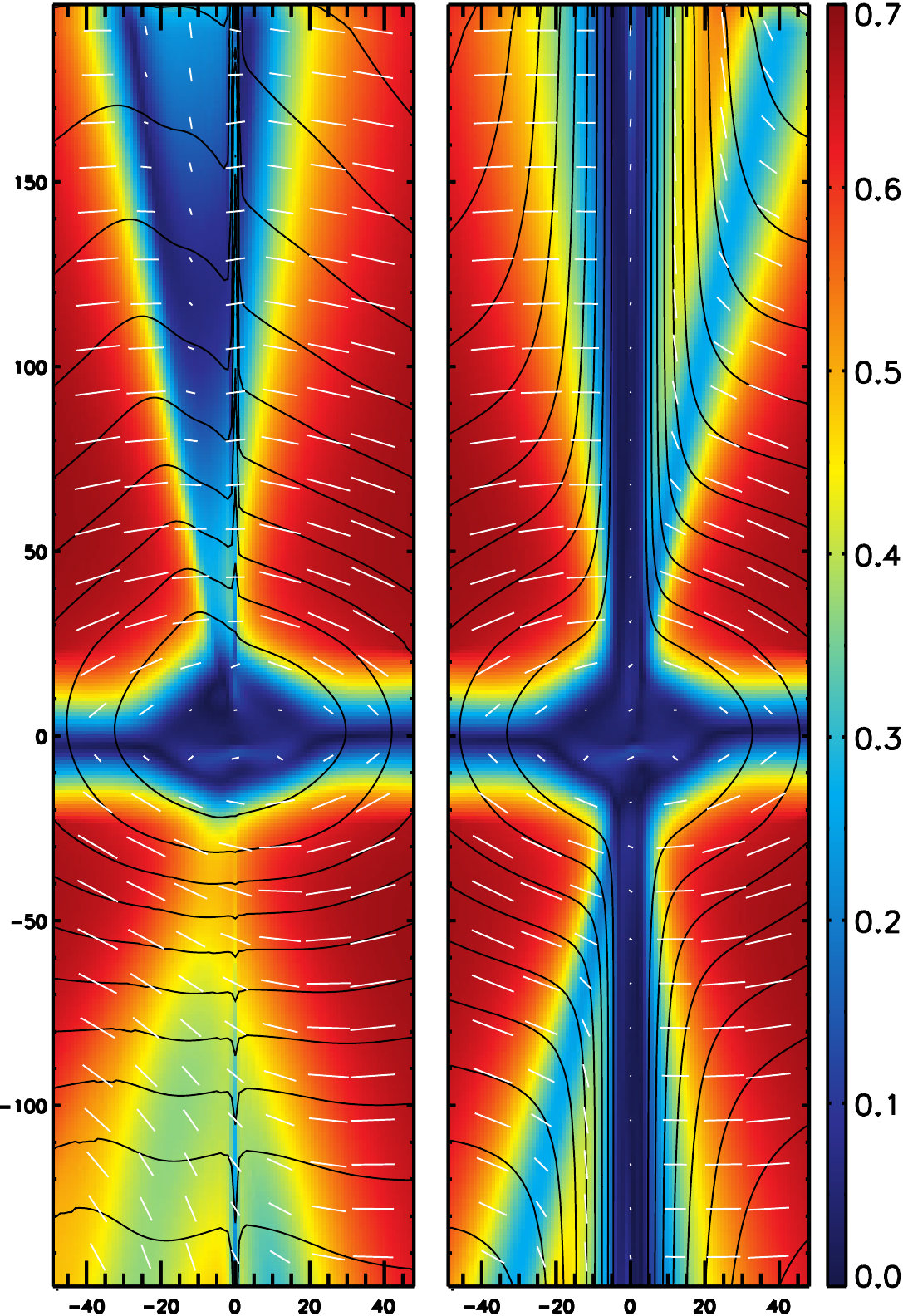}
\end{center}
\caption{Relativistic effect on optically thin polarization (run 2h) at $i=60^{\circ}$.  The polarization degree is indicated by the background coloring, black contours show total intensity levels spaced by factors of two.  \textit{Left:} Including relativistic aberration. At this high inclination, the jet is de-beamed and rotation is apparent in the asymmetry of the intensity contours.  \textit{Right:} In the absence of relativistic effects, the polarization pattern is point symmetric about the origin and the intensity clearly promotes the jet.  }
\label{fig:picomp}
\end{figure}

For the case of a collimating and accelerating jet as shown here, we loosen the constraint of
cylindrical symmetry to mere axisymmetry in the $\phi$ direction.
Additionally, our simulations feature a non-constant pitch of the magnetic field and a small
degree of rotation.
Together, this results in inclined polarization vectors which deviate from the strict bimodality
observed in $(z,\phi)$-symmetry.
Figure \ref{fig:picomp} shows the optically thin polarization structure in the presence of
relativistic effects (left panel), and in absence thereof (right panel).
To produce the non-relativistic map, we had simply set $\mathbf{v}\equiv0$ before conducting
the radiation transport.
At the base of the outflow where the velocities are only mildly relativistic, the polarization vectors are found predominantly perpendicular to the collimating poloidal magnetic
field in both cases.  
Further downstream, the bimodal spine-and-sheath polarization structure prevails as the jet
dynamics becomes increasingly cylindrical.
The figure clearly demonstrates how the relativistic swing skews the spine towards the approaching
side of the jet.
Note that the jet rotation is also apparent in the beamed asymmetric intensity contours.
At high inclination $i=60^{\circ}$, the main emission from the high-speed jet is de-beamed and
only the low-velocity ``needle'' of the thermal spine along the axis can be recognized.
It is worthwhile to note that both intensity and polarization of axisymmetric non-relativistic
synchrotron sources exhibit a point symmetry about the origin as illustrated in Fig.~\ref{fig:picomp}.

\subsection{Pitch-angle dependence}\label{sec:pitchangle}
\begin{figure*}
\begin{minipage}{8.5cm}
\includegraphics[height=4.5cm]{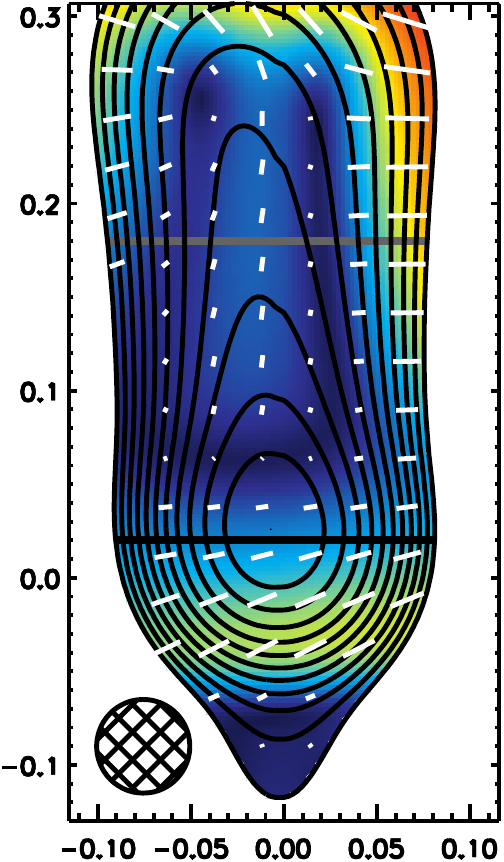}
\hfill
\includegraphics[height=4.5cm]{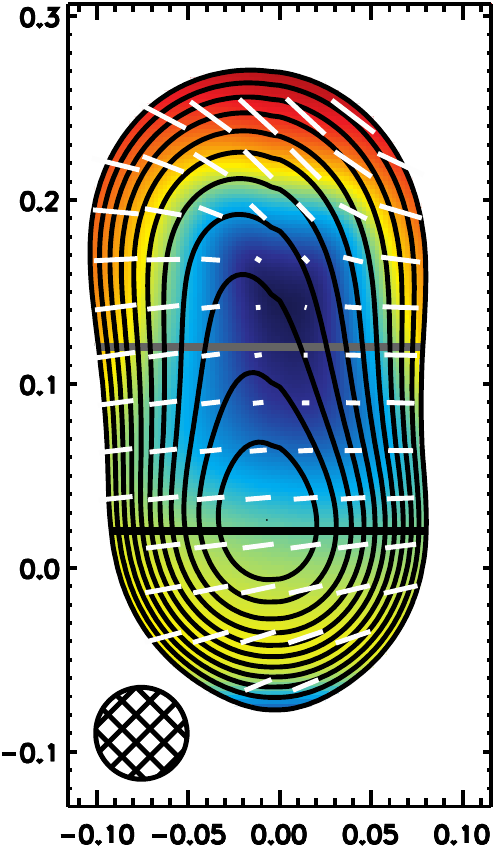}
\hfill
\includegraphics[height=4.5cm]{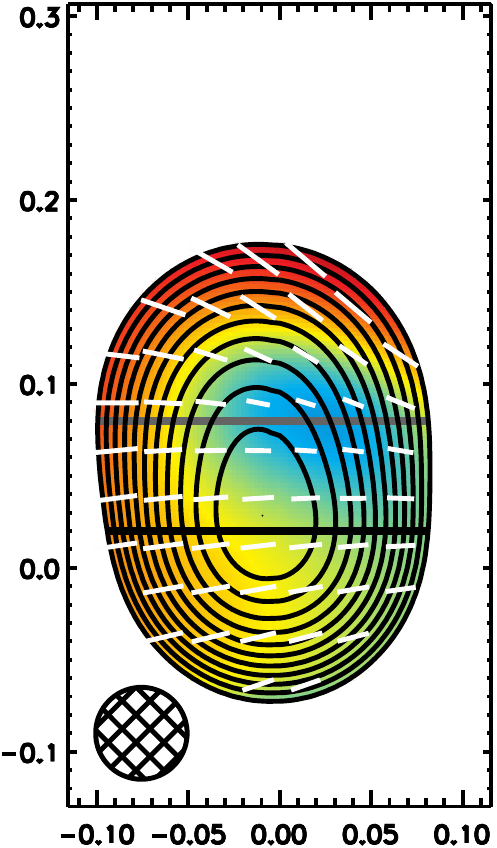}
\vfill
\includegraphics[height=4.5cm]{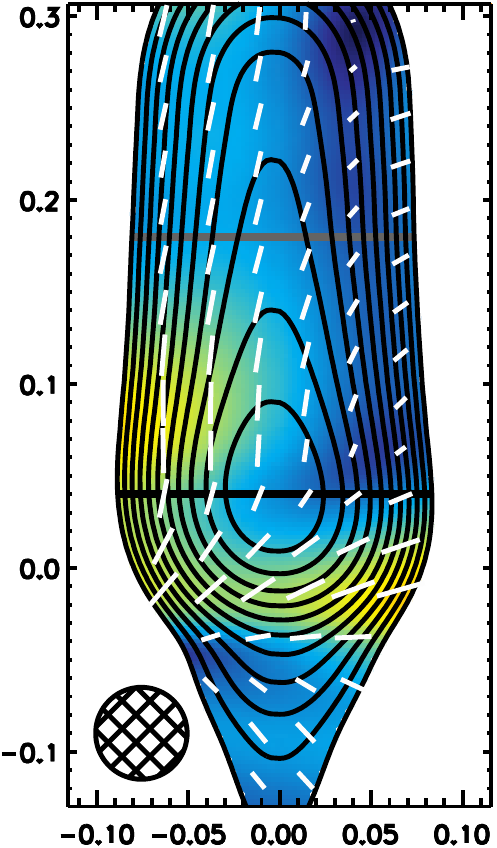}
\hfill
\includegraphics[height=4.5cm]{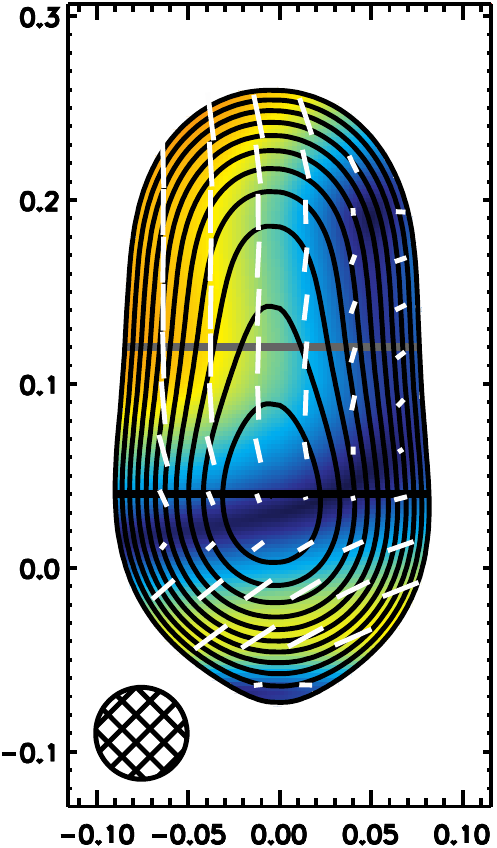}
\hfill
\includegraphics[height=4.5cm]{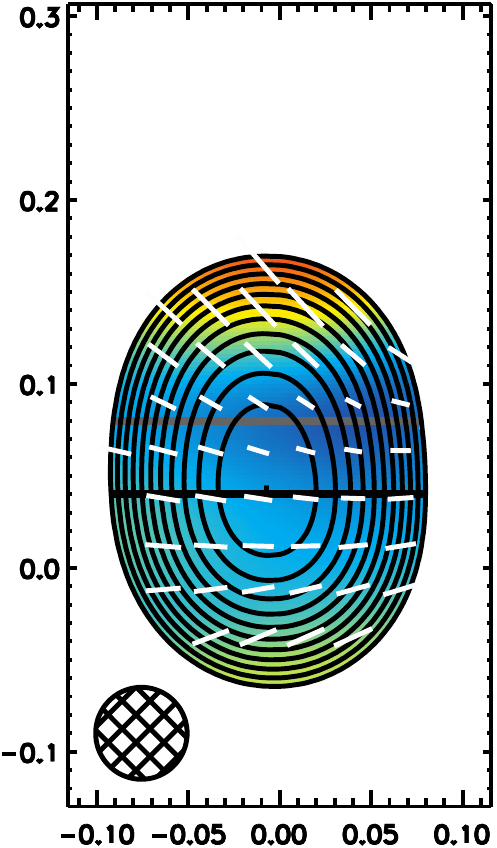}
\end{minipage}
\begin{minipage}{1cm}
\includegraphics[height=9.4cm]{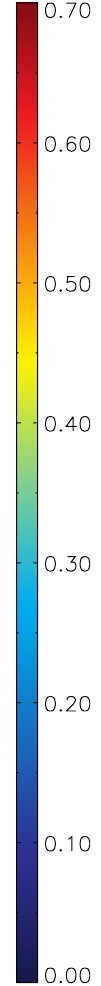}
\end{minipage}
\hspace{0.5cm}
\begin{minipage}{8cm}
\includegraphics[height=4.5cm]{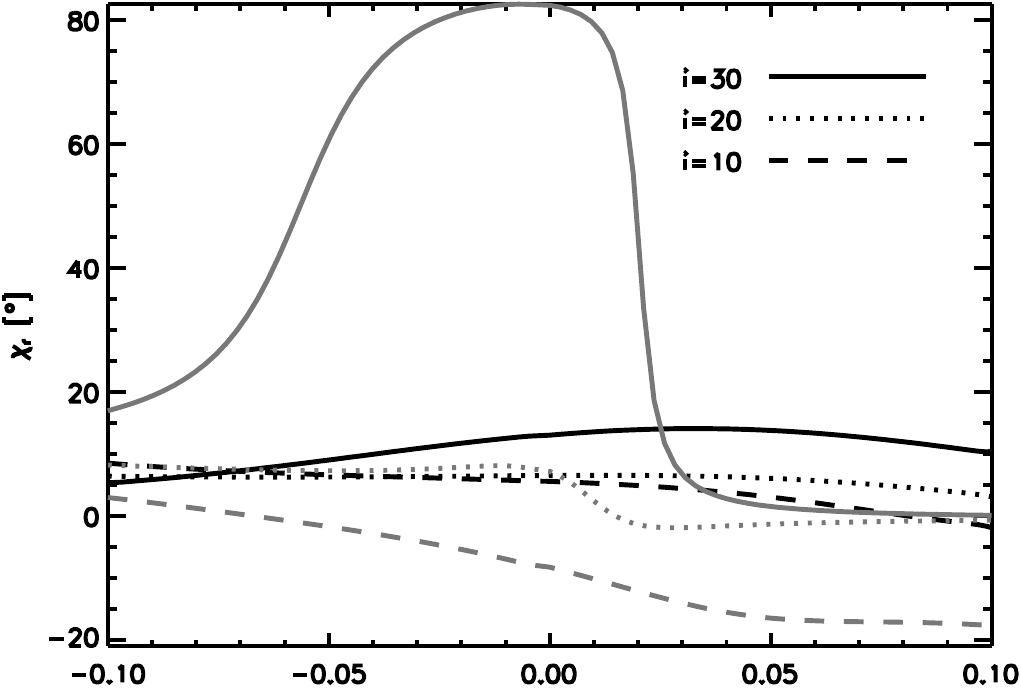}
\\
\vfill
\includegraphics[height=4.5cm]{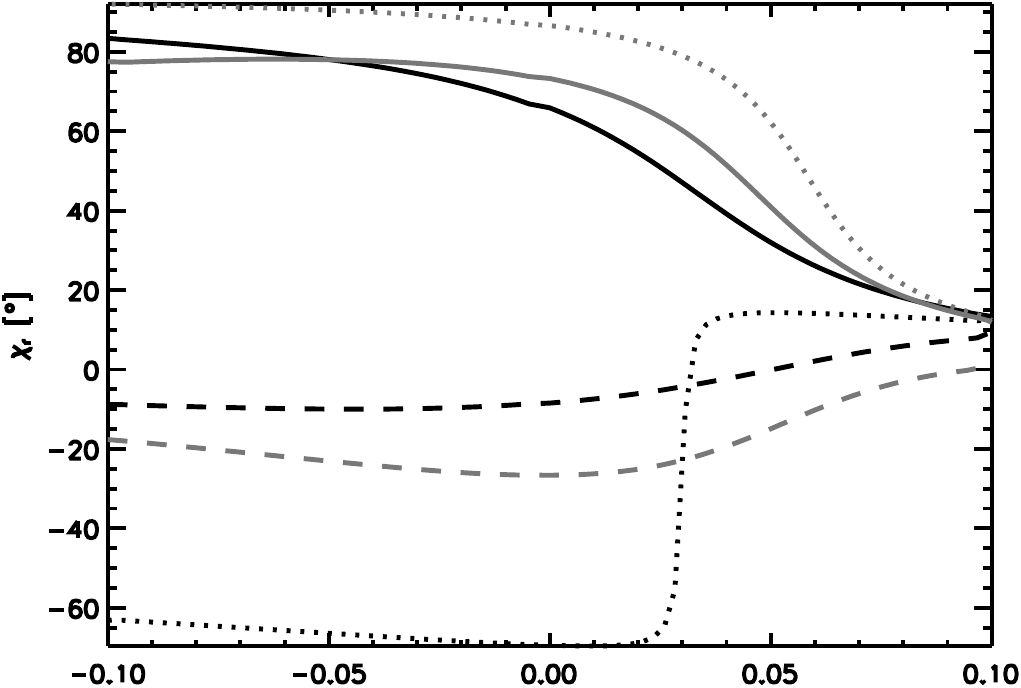}
\end{minipage}
\hfill
\caption{Polarizations for $i\in\{30^{\circ},20^{\circ},10^{\circ}\}$ \textit{(from left to right)} emitted from regions with co-moving pitches $B'_{\phi}/B'_{p}>1$ \textit{(above)} and $B'_{\phi}/B'_{p}>2$ \textit{(below)}.   The polarization degree $\Pi_{43 \rm GHz}$ is color-coded and $I_{43 \rm GHz}$ contours are shown.   Contours are spaced by a factor of $2$ out to $\simeq5\cdot 10^{-4} I_{\nu, \rm peak}$ where the image is cropped.   Spatial scale is given in milli arcseconds and a restoring beam with  FWHM=$0.05\, \rm mas$ was used.  
 The right-hand panel shows polarization angles along the cuts along core \textit{(black)} and jet \textit{(gray)}.  
\label{fig:polariz}}
\end{figure*}

Several Authors \citep{2002ApJ...577...85M, 2005AJ....130.1389L} found indications for a bimodal distribution of the electric vector position angles (EVPA) of quasars and BL-Lac objects either aligned or perpendicular to the jet direction.  It was also supposed that BL-Lacs tend to aligned $\mathbf{\hat{e}}$ vectors and overall higher degree of polarization.  
In a recent $86\, \rm GHz$ polarimetric survey however, \cite{agudo2010} found no such correlation in their flat-radio-spectrum AGN sample, rather are their data consistent with an isotropic (mis-)alignment.  
Alignment is customarily attributed to oblique shocks, while perpendicular EVPAs are then interpreted in terms of a shearing of the magnetic field with the surrounding.  
Another possible explanation for the bimodality is due to large-scale helical fields as shown in the previous section.  
By varying the emitting region within the collimating jet volume, we investigate to what degree the polarization still conveys the geometric information of the emitting region.  As before, Faraday rotation is neglected and we introduce the co-moving pitch angle $\tan \Psi' \equiv B'_{\phi}/B'_{p}$ in analogy to \cite{2005MNRAS.360..869L}.  In a non-rotating, cylindrical jet, the pitch-angle transforms as 
\begin{align}
\tan\Psi = \Gamma \tan\Psi'\label{eq:pitchtransform}
\end{align}
and hence the co-moving fields appear much less twisted than their laboratory frame counterparts.  
Relation \ref{eq:pitchtransform} together with equation \ref{eq:gamma2b} for the linear acceleration regime yields the co-moving fields 
\begin{align}
B'_{\phi}/B'_{p}\simeq 1.
\end{align}
Substantially higher pitches are not realized within the simulations.  
An impression on the pitch angle distribution throughout the jet can be obtained with the cuts shown in the middle panels of figures \ref{fig:models1} to \ref{fig:models3}. 
We restrict the emission to originate from $B'_{\phi}/B'_{p} \ge \{1,2\}$, corresponding to $\Psi' \ge \{45^{\circ}, 63^{\circ}\}, (\Psi > \{83^{\circ},86^{\circ}\})$ where equipartition particle energy density (eq. \ref{eq:minimal e}) is assumed.   
This way, only the regions of the current driven jet contribute to the emission and no radiation is observed from the spurious axis where $B_{\phi}$ must vanish.  
Figure \ref{fig:polariz} shows the polarization for the two cases and various viewing angles.  

In the high pitch-angle case $B'_{\phi}/B'_{p} > 2$, the resulting EVPA become parallel for viewing angles $i>10^{\circ}$ while for the standard case $B'_{\phi}/B'_{p} > 1$, this happens only at viewing angles $i>30^{\circ}$.  
Most structure is observed at moderate viewing angles where we can find the core polarization perpendicular to the ridge line polarization for the case $i=30^{\circ}, B'_{\phi}/B'_{p} > 1$ as observed in some sources \citep[e.g.][]{2005MNRAS.356..859P}.  Here we also see a spine and sheath polarization profile across the jet.  In the adopted parametrization of the emission region, the spine and sheath structure is only observed for $B'_{\phi}/B'_{p} \sim 1$ as higher pitches tend to alignment and lower pitches tend to counter align.  

Beam depolarization is apparent in regions where the polarization turns and, consistent with most observations, the degree of polarization increases towards the boundary of the jet.  
In the high-pitch case, a left-right asymmetry is most significant with parallel vectors on the approaching side of the rotating jet and perpendicular ones at the receding one.  
\cite{2005MNRAS.360..869L} have proposed that based on the asymmetric polarization signal, the handedness of the magnetic field and thus the spin direction of the black hole / accretion disk can be inferred.  Our results support this finding.  
Clearly, when increasing the pitch, the EVPAs turns from the perpendicular to the parallel direction as a general trend. 
However, pitches $B'_{\phi}/B'_{p} > 4$ were not realized by our simulations, such that the emitting region would vanish.  
Thus for the models under consideration, when approaching the Blazar case, the intrinsic polarization always appears perpendicular.  In the following we will mostly characterize the emission region by $B'_{\phi}/B'_{p} \ge 1$ as this best selects the relativistic jet contribution.

\section{Radiation in the jet models}\label{sec:radiationinjet}
We have now introduced all parameters to build models for the synthetic observations.  A radiation model comprises a MHD simulation run, a particular physical scaling given by $M_{\bullet}, \dot{E}$ and the specific parameters of the radiation transport $\Psi',\epsilon_{\rm B}$ and $i$.  To normalize the flux, a photometric distance $D=100\, \rm Mpc$ is assumed in all models.   
Table \ref{tab:radmod} summarizes the parameters adopted in this work.  We will mostly report results for the fiducial model A and consult the other models only for comparison with the standard case.  
\begin{table*}[htbp]
\caption{Jet radiation models}                
\label{tab:radmod}                  
\centering                                    
\begin{tabular}{c c c c c c c | c}
\hline\hline                              
model ID & run ID & $M_{\bullet}[M_{\odot}]$ & $\dot{E} [\rm erg/s]$ & $\epsilon_{\rm B}$ & $\tan \Psi'$ & $i [\rm deg]$ & $\left.\frac{\dot{M}}{\dot{M}_{\rm Edd}}\right.$ \\
\hline                                       
A & 2h & $10^{9}$ & $10^{43}$ & 0.1    & 1  & 20 &   $\rm 1.3\times 10^{-6}$\\
B & 2h & $10^{9}$ & $10^{44}$ & 0.001  & 0.5  & 20 & $\rm 1.3\times 10^{-5}$\\
C & 1h & $10^{9}$ & $10^{43}$ & 0.1    & 1  & 20 &   $\rm 5.5 \times 10^{-7}$\\
D & 3h & $10^{9}$ & $10^{43}$ & 0.1    & 1 & 20   &  $    1.1 \times 10^{-6}$\\
\hline                                        
\end{tabular}
\tablecomments{
Columns denote:
Model ID;
Simulation ID; 
Black hole mass;
Total energy flux rate;
Minimal co-moving pitch angle of emission;
Standard inclination;
Resulting jet mass loss rate in terms of the Eddington accretion rate (with a radiative efficiency of $0.1$).  
}
\end{table*}

For the interpretation of the results, an understanding of the qualitative dependence of the observables on our parametrization is helpful and therefor discussed in the following.  
In terms of the physical scaling and equipartition fraction, we have 
\begin{align}
\epsilon_{\nu}     &\propto \epsilon_{\rm B}\ \dot{E}^{7/4}\ M_{\bullet}^{-7/2}\ \nu^{-\alpha}\\
\kappa_{\nu} &\propto \epsilon_{\rm B}\ \dot{E}^{2}\ M_{\bullet}^{-4}\ \nu^{-\alpha-5/2}\\
\frac{d \chi_{\rm F}}{d l} &\propto \dot{E}^{3/2}\ \mbh^{-3}.
\end{align}
The observable quantities then become 
\begin{align}
I_{\nu}^{\rm thin} &\propto \epsilon_{\nu}\times l \propto \epsb\ \dot{E}^{7/4}\ \mbh^{-5/2}\ \nu^{-\alpha}\\
I_{\nu}^{\rm thick} &\propto \epsilon_{\nu}/\kappa_{\nu} \propto \dot{E}^{1/4}\ \mbh^{3/2}\ \nu^{2.5}
\end{align}
and the opacities follow to
\begin{align}
\tau_{\nu} &\propto \alpha_{\nu}\times l \propto \epsb\ \dot{E}^{2}\ \mbh^{-3}\ \nu^{-\alpha-2.5}\\
\chi_{\rm F} &\propto \dot{E}^{3/2} \mbh^{-2}.
\end{align}
In reverse, based on the latter relations, spectrum and Faraday rotation measurements will allow us to constrain the physical parameters.\footnote{In practice this is further complicated by the dependences introduced by the doppler factor which we omitted here as we will compare only dynamically identical models at a given inclination.}  

We deliberately chose model B to feature $\sim 32$ times higher Faraday depth compared to model A by increasing the jet energy $\dot{E}$, while maintaining a similar spectrum with the choice of $\epsilon_{\rm B}$ (although the total flux is thus decreased by a factor of $\sim 1.8$).

\subsection{Core shift}

\cite{lobanov1998} showed how opacity effects in optically thick jet cores provide valuable information that can help to constrain the dynamical jet quantities.
As the bulk of the radiation originates in the photosphere $\tau=1$, the projected distance of the $\tau=1$ surface will result in a specific core offset.  
Due to the frequency dependent opacity of a synchrotron self absorbing radio source, the measured core position varies systematically with the observing frequency.  
In simple model jet where the magnetic field and relativistic particle density is modeled as $B\propto r^{-m}$ and $N\propto r^{-n}$ \citep{konigl1981}, this projected distance becomes

\begin{align}
r_{\rm core} \propto \nu^{-1/k_{r}}
\end{align}

where $k_{r}$ is a combination of the parameters $m,n$ and the spectral index $\alpha$
\begin{align}
k_{\rm r} = \frac{(3+2\alpha)m+2n-2}{5+2\alpha}.
\end{align}

\begin{figure}[htbp]
\centering
\includegraphics[width=0.49\textwidth]{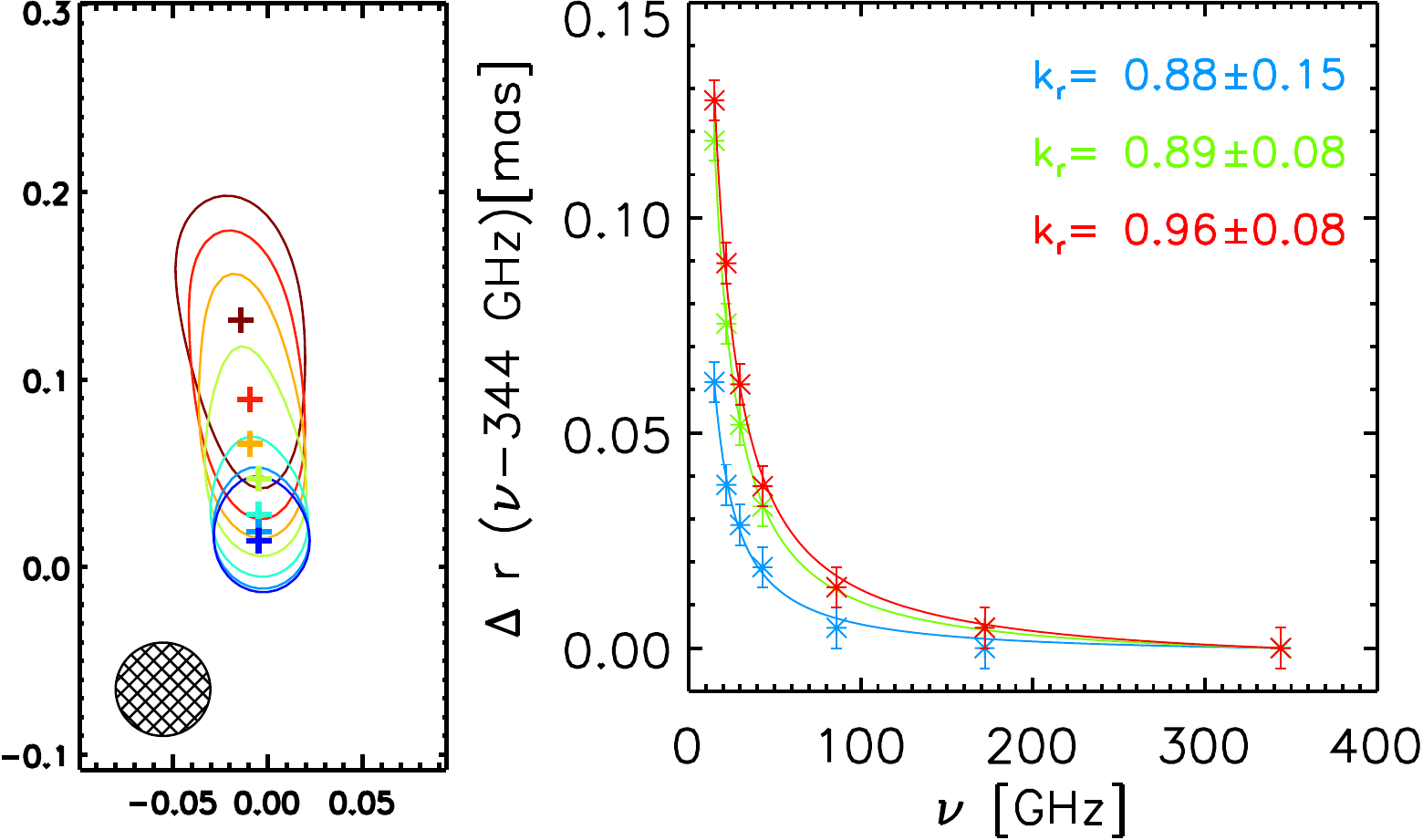}
\caption{\textit{Left:} Half maximum intensity contours and peak positions illustrating the core shift in mas in model A for different frequencies: 344GHz (blue) to 15 GHz (maroon).  The images are aligned with respect to the imaginary black hole line of sight located at the origin.  
\textit{Right:} Fit of the core distances relative to the 344 GHz peak for models C,A,D (top to bottom).  The fit function reads $\Delta r = A (\nu^{-1/k_{\rm r}}-{\rm 344}^{-1/k_{\rm r}})$ with free parameters $A$ and $k_{\rm r}$.  
\label{fig:coreshift}}
\end{figure}

In a conical jet, the (predominantly toroidal) magnetic field strength will follow $m=1$ due to flux conservation.  Conservation of relativistic particles sets for the number density $n=2$.   
It is noteworthy that in this most reasonable case, the exponent $k_{r}$ becomes independent of the spectral index $k_{\rm r}\equiv1$.  Although our jet models are collimating and thus not exactly conical, qualitatively we expect $k_{\rm r}$ close to unity when equipartition particle energy (eq. \ref{eq:minimal e}) and thus $n = 2m$ is assumed.  
The observed core shift is illustrated in figure \ref{fig:coreshift} for the fiducial model A.  Fits of the $k_{\rm r}$ exponent for all runs are shown in the right panel.  

The best fits are slightly steeper than the conical expectation of $k_{\rm r}=1$.  Within the fitting error, models C and D are however still consistent with $k_{\rm r}=1$ while the deviation in model A becomes significant.  
It is tempting to interpret this behavior as a consequence of jet collimation which increases the shift for larger distances from the origin.  As the far regions are probed by lower frequencies, the distance law of a collimating jet is expected to systematically steepen towards the low frequency side.    
The quality of the fits for our model jets suggests that core shifts can provide a robust diagnostic of the AGN jet acceleration region.  
Unfortunately, core shift largely complicates the interpretation of the Faraday rotation maps, as we will show in section \ref{sec:rotationmeasure}.  

\subsection{Depolarization}\label{sec:poldeg}
To understand the polarization signal of the simulated jets, let us briefly review some considerations on the polarization degree in general \citep[e.g.][]{Pacholczyk:1970}, in the presence of Faraday rotation \citep[e.g.][]{burn1966} and when an observing beam is used.  

In the simplified case of a uniform magnetic field and for power-law electron distributions, the polarization degree reads 
\begin{align}
\Pi_{\rm h}=\frac{(p+1)}{(p+7/3)} \to 0.69\ (p=2)
\end{align}
for the optically thin regime and 
\begin{align}
\Pi_{\rm l}=\frac{3}{(6p+13)} \to 0.12\ (p=2)
\end{align}
for optically thick radiation.  
Such high polarization degrees are however rarely observed in AGN cores which go down to the percentage level. Clearly a mechanism for depolarization is required.  

Already \cite{burn1966} considered the admixture of a isotropic random field component $B_{\rm r}$ to an otherwise ordered field $B_{0}$ in the emitting region.  He found the simple relation for depolarization relating the energies of the fields $\Pi_{r} \simeq \Pi_{0} {B_{0}^{2}}({B_{0}^{2}+B_{\rm r}^{2}})^{-1}$ where $\Pi_{\rm r}$ is now the reduced polarization due to the additional random component.  
In result, to obtain significant depolarization the random component needs to be comparable to the ordered field.  
The accompanying dissipation of energy would notably decrease the efficiency of the jet acceleration process, increasing the scales of jet acceleration possibly beyond the parsec scale.  
The occurrence of turbulence and thus randomly oriented fields in the AGN core can serve as an explanation for various multifrequency observations as ventured by \cite{marscher2011}.
Under additional compression, even a completely randomized field structure can account for the high and low observed polarization 
degrees as demonstrated by \cite{1980MNRAS.193..439L}.
However, the complex physics of turbulence \textit{within the jet} can not be incorporated into MHD simulations of relativistic jet formation at this time.
\footnote{Note on the other hand that simulations featuring turbulent slow disk winds serving as Faraday screen were presented by \cite{broderick2010}.  }  
Our current MHD simulation models thus provide highly ordered near force-free fields around which relativistic electrons following an isotropic distribution are assumed to gyrate.  We propose that the most promising site for finding such ordered (helical) fields is in fact the Poynting dominated regime of the jet acceleration region that is simulated here.  

Under the influence of Faraday rotation within the emitting volume, the polarization degree will also depend on the Faraday opacity $\beta_{\rm F}\equiv d\chi_{F}/ds$.  \\
Considering first optically thick radiation where the polarization degree is governed by the ratio of ordinary to Faraday opacity $\delta\equiv\kappa_{\nu}/\beta_{\rm F}$.  
Specifically, with $\kappa_{\nu}\propto\nu^{-p/2-2}$ and $\beta_{\rm F}\propto\nu^{-2}$ it becomes $\delta\propto \nu^{-p/2} \to \nu^{-1}\, (p=2)$.  
Once the photon mean free path is smaller than the correlation length of the field and the mean rotation length $\beta_{\rm F}^{-1}$ in the low frequency limit ($\delta\to\infty$), we expect $\Pi$ to approach the uniform magnetic field case $\Pi_{\rm l}$.  
Accordingly, for a small value of $\delta$, the radiation is depolarized $\Pi\to 0\, (\delta\to0)$.  

For optically thin radiation, the impact of Faraday rotation can be parametrized by $\eta\equiv\beta_{\rm F} s$ describing the angular change during the emission length $s$.  The polarization degree will then decrease and oscillate according to $\Pi=\Pi_{\rm h}|\sin\eta/\eta|$.  
This is known as \textit{differential Faraday rotation} and its influence on $\Pi$ is shown in figure \ref{fig:los} along a line of sight in the simulation.  
Since $\eta\propto \nu^{-2}$, internal Faraday rotation has vanishing influence also in the high frequency limit and $\Pi$ approaches the maximal polarization degree $\Pi_{\rm h}$.  
In a non-uniform field, changes in the emitting geometry lower the maximal polarization degree and $\Pi_{\rm h}$ can only be assumed at the edges of the emission region.  
The direction of preferred emission perpendicular to the projected magnetic field is also the direction of dominant absorption, such that in the uniform field case the optically thick polarization direction is flipped by $90^{\circ}$ with respect to the optically thin polarization.  

Additional depolarizing effects occur when an extended source is observed with finite resolution.  
The radiation is depolarized when the beam encompasses \textit{(1)} intrinsic changes in the emitting geometry, \textit{(2)} optical depth transitions leading to $90^{\circ}$ flips and \textit{(3)} varying Faraday depths, known as (internal or external) \textit{Faraday dispersion} \cite[see also][]{1966ARA&A...4..245G,1998MNRAS.299..189S}.  

For completeness, we should also mention depolarization via ``blending'' - contamination with an unpolarized (thermal) component - for example radiation from the torus in the infrared.  This occurs when the intensity of the contaminant becomes a notable fraction of the total intensity and thus imprints on the spectral energy distribution as well.  Naturally, our results are only valid as long as the jet-synchrotron radiation is dominating the total flux.

\subsubsection{Low Faraday rotation case}

\begin{figure}[htbp]
\centering
\includegraphics[width=.45\textwidth]{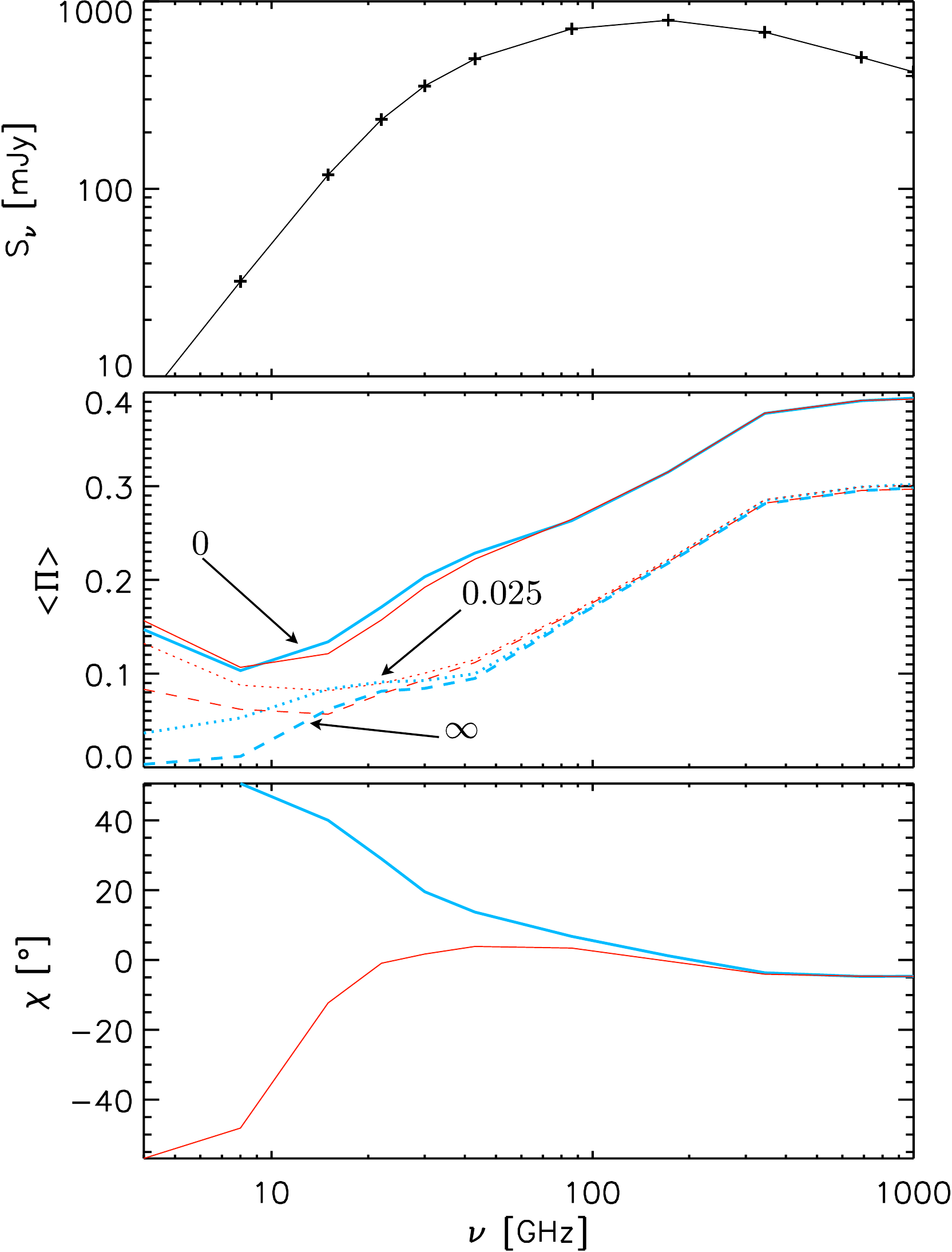}
\caption{
Spectrum, core polarization degree and core polarization direction in model A showing depolarization due to beam- and Faraday effects.  The curves in the middle panel are shown for increasing beam-FWHM in mas as indicated and converge towards the unresolved case.  The unresolved polarization direction is shown in the lower panel.  
Thick blue lines take Faraday rotation into account and thin red lines are calculated in its absence.  
\label{fig:spec}}
\end{figure}

Spectrum and core polarization degree for model A is shown in figure \ref{fig:spec}.  
The observable core polarization degree is defined as 
\begin{align}
\langle \Pi_{\nu} \rangle \equiv \frac{\int d\Omega\ \langle I_{\nu}(\Omega)\rangle\ \Pi(\langle \mathbf{I_{\nu}}(\Omega)\rangle)}{\int d\Omega I_{\nu}(\Omega)}
\end{align}
where the quantities under the integral are themselves subject to beam convolution.  
In absence of the latter, the averaged polarization degree increases monotonically from the expected value of $\sim0.1$ in the low frequency range to $0.4$ in the optically thin case.  Due to the varying emitting geometry, the theoretical maximum of $\Pi_{\rm h}=0.69$ is not realized.  
The influence of differential Faraday rotation seen in the ideal resolution case is in fact small.  Depolarization occurs through ordinary beam depolarization and through Faraday dispersion, once the angle between the (unresolved) emitted polarization direction and the Faraday rotated polarization becomes larger than $\sim 45^{\circ}$ (compare with lower panel of Fig. \ref{fig:spec}).  
We note that also for the unrotated polarization direction, a flip of $\sim90^{\circ}$ between the thick and thin case is not observed.  This is also expected, since the photosphere probes various pitch angles as the frequency is decreased and so the simple uniform field case does not apply.  

With increasing beam size, the polarization degree ultimately converges to the unresolved case.  The convergence is faster for the optically thin regime where the intrinsic emission is less extended and thus quickly masked by the beam.   

\subsubsection{High Faraday rotation case}\label{sec:depol-b}

\begin{figure}[htbp]
\centering
\includegraphics[width=.45\textwidth]{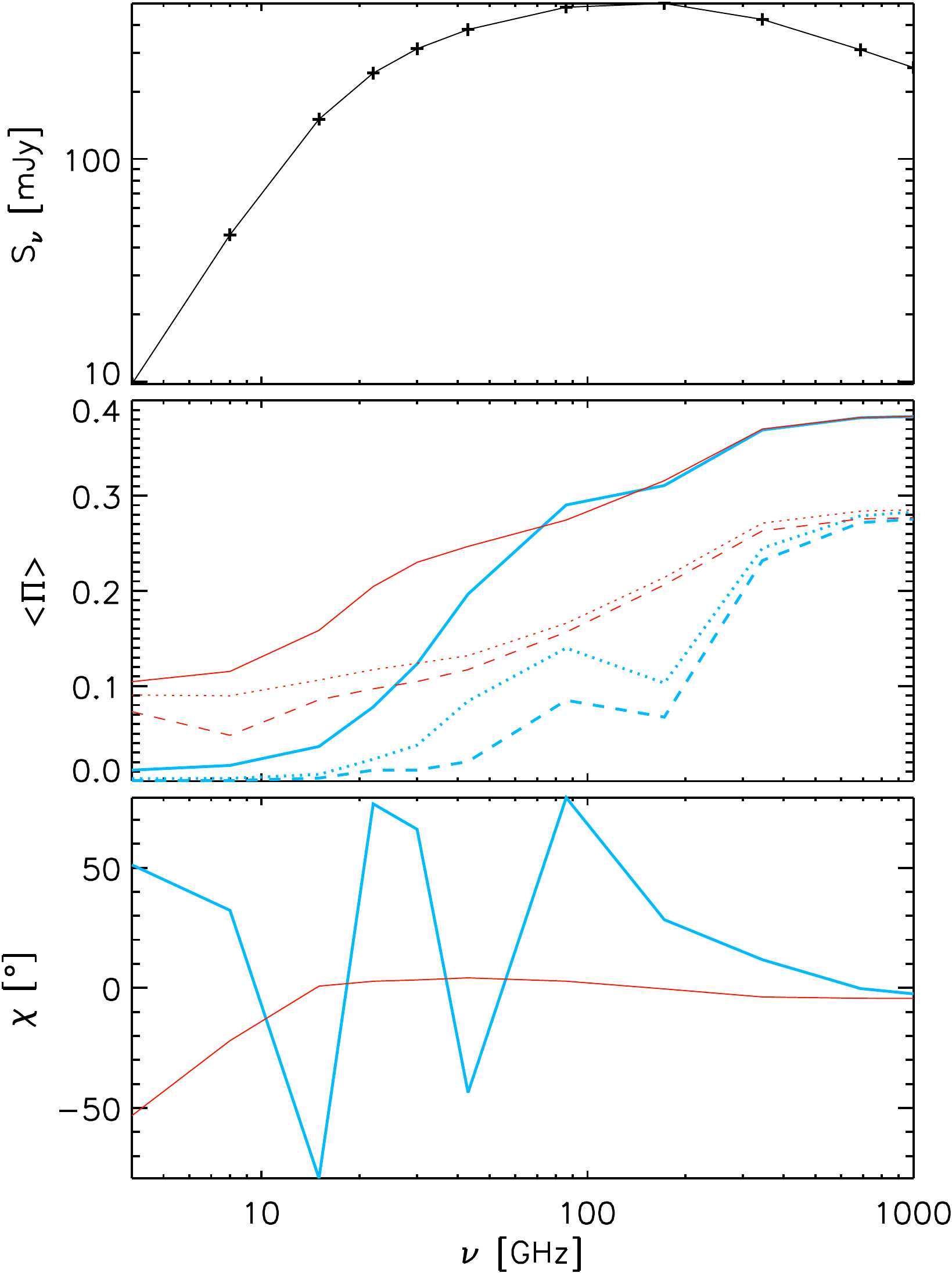}
\caption{
As figure \ref{fig:spec} but for model B with high Faraday rotation.  
\label{fig:specb}}
\end{figure}

To observe the effect of depolarization due to internal Faraday rotation, we perform the same analysis as before for the high Faraday rotation model B.  
Figure \ref{fig:specb} shows the resulting quantities.  
As anticipated, a similar spectrum is obtained but the polarization degree and direction behave differently.  Even when observed with ideal resolution, the polarization degrees of the two cases (Faraday active vs. neglected Faraday rotation) separate clearly as a result of internal depolarization.  
Also the unresolved polarization angles separate at higher frequencies.  Due to multiple rotations, the polarization angle appears to fluctuate below observing frequencies of $\rm 100\, GHz$.  
In principle, the ideal resolution polarization degree is expected to rise again for lower frequencies as $\delta\to \infty$.  However, this did not yet occur at frequencies above $4\rm\, GHz$ that were under investigation.  

\subsection{Rotation measure}\label{sec:rotationmeasure}
A helical field geometry is generally perceived to promote transversal rotation measure (RM) gradients owing to its toroidal field component.  
First evidence for RM gradients was found by \cite{asada2002} and \cite{2005ApJ...626L..73Z} in the jet of $\rm 3C\,273$.  
In several unresolved radio cores, $\lambda^{2}$ law RMs have been detected and are found to follow a monotonic profile for example by \cite{gabuzda2004, 2009MNRAS.393..429O, 2010MNRAS.402..259C}.  \\
\cite{taylor2010} point out the observational requirements of a RM gradient detection as follows:
\textit{1.} At least three resolution elements across the jet.
\textit{2.} A change in the RM by at least three times the typical error.
\textit{3.} An optically thin synchrotron spectrum at the location of the gradient.
\textit{4.} A monotonically smooth change in the RM from side to side (within the errors).

\begin{figure}
\centering
\includegraphics[width=.45\textwidth]{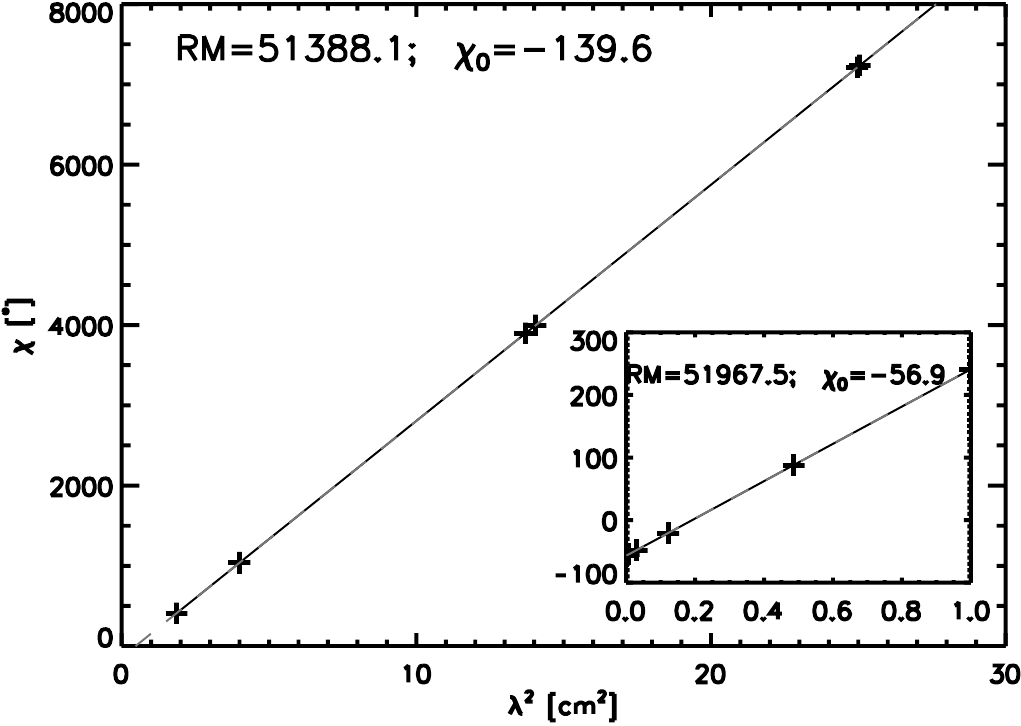}
\caption{
Determination of the rotation measure for an ideal resolution line of sight.  The value of RM (indicated in $\rm rad/m^{2}$) necessitates multiple rotations by $\pi$ for the high $\lambda^{2}$ case.  In the inlay, a different fit for the optically thin regime is shown.  Between the two cases, the effective angle of emission $\chi_{0}$ (indicated in $\rm deg$) is rotated by $83^{\circ}$.  
\label{fig:fitrm}}
\end{figure}

In the following paragraphs we will touch up on each of the aforementioned points.  
Due the flip between the optically thick and thin polarization direction, the measurement of $\lambda^{2}$-law RM around the spectral peak require extra caution.  
Using sufficiently small spacings in $\Delta \lambda^{2}$ to recover $n\pi$ rotations, we can fit the rotation measure law
\begin{align}
\chi(\lambda^{2}) = \chi_{0} + \rm RM \lambda^{2}
\end{align}
to the optically thick and thin cases, where $\chi_{0}$ now denotes the effective angle of emission.  The fits are shown for a particular line of sight in figure \ref{fig:fitrm}.  
Here, $\chi_{0}$ differs by almost $90^{\circ}$, whereas RM is of comparable size.  
However, the latter two findings are not necessarily true for all lines of sight, since the optically thin photons can originate in higher Faraday depths and different emitting geometry, as mentioned previously.  
We stress that with ideal resolution, consistent $\lambda^{2}$-laws are found both for optically thick and thin photons.  Alas, if taken together, we would not be able to fit a linear function for the whole range.  

The two-dimensional RM maps shown in figure \ref{fig:rmfit} demonstrate the effect of beam averaging on the low- and high frequency regime.  
In the high frequency case, the intensity is strongly peaked close to the central object where the Faraday depth is highest, leading to steep radial gradients also in the rotation measure.
As the emission is more extended, the RM is lower and smoother in the low frequency case.
We find an interesting relation between the core shift and the rotation measure:
\begin{figure}\centering
\includegraphics[width=8cm]{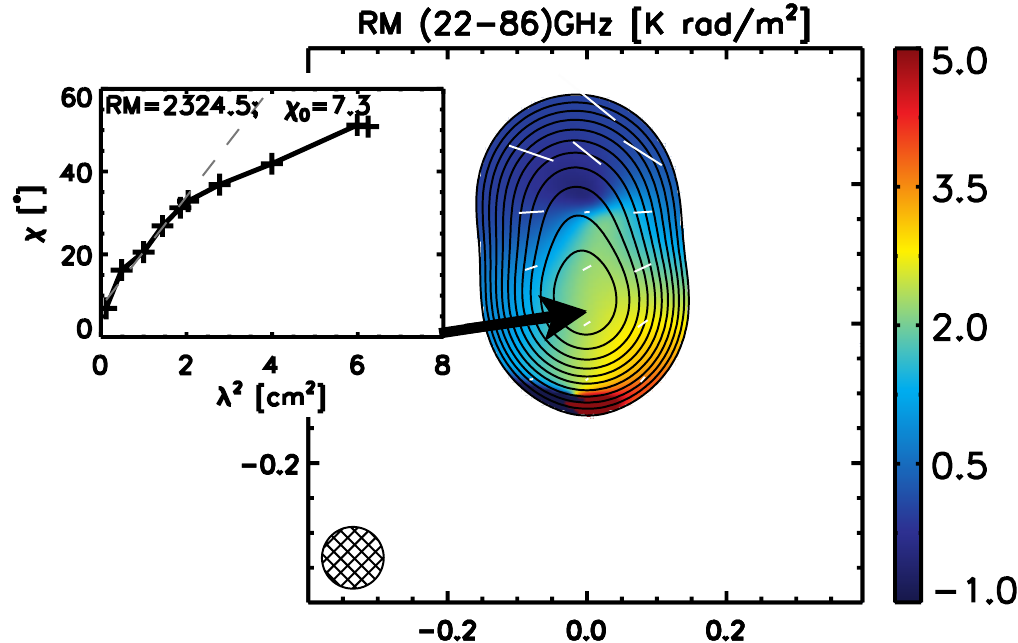}
\includegraphics[width=8cm]{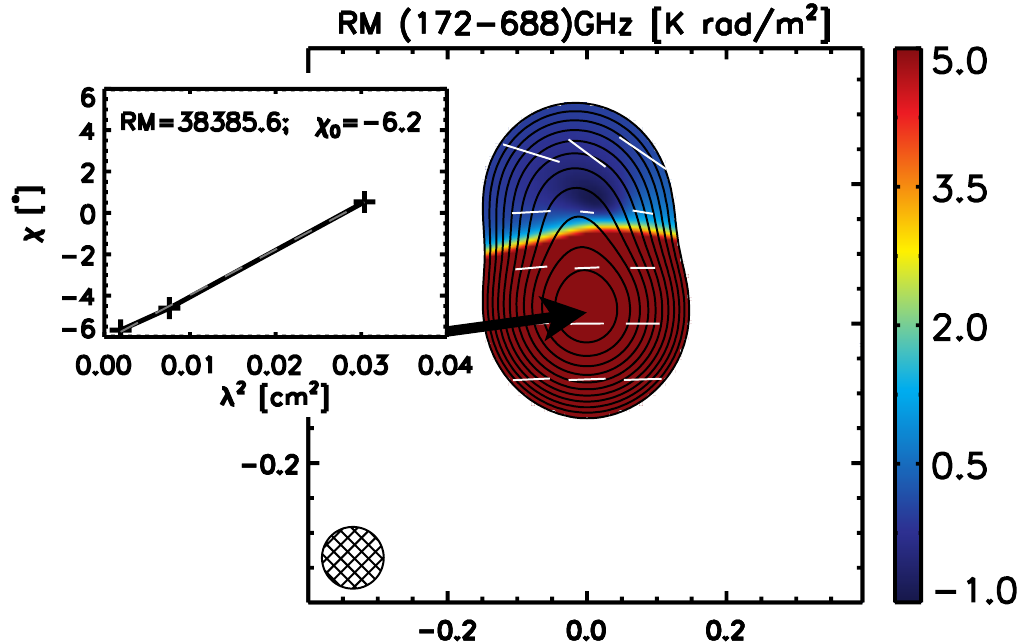}
\caption{
Rotation measure maps ($i=20^{\circ}$) in the optically thick \textit{(top)} and in the optically thin regime \textit{(bottom)}.  White sticks indicate the direction of effective emission, $\chi_{0}$.  In the top plot inlay, only the $22-86 \rm GHz$ region is fitted while the additional points illustrate the $\lambda^{2}$-law breaking due to the core shift (see text).  
\label{fig:rmfit}}
\end{figure}
As the photosphere moves outwards with decreasing frequency, the flux tends to originate further away from the central object leaving systematically less Faraday active material between the observer and the source.
\footnote{As a limitation of our direct ray tracing method, the photosphere can in principle shift to the outer boundary (the ``lid'') of our domain, such that the very optically thick regime below $8\rm GHz$ can not reliably be probed.}
After beam convolution, the polarization angle is biased towards the outward moving photosphere, resulting in a shallower rotation measure for lower frequencies which ultimately breaks the $\lambda^{2}$-law that is valid for each ideal resolution line of sight, (e.g. figure \ref{fig:fitrm}).  

Once the core shift distance is large compared to the scale of typical changes in the Faraday depth or emission angle, a $\lambda^{2}$-law will not be observed when the observations are aligned according to the core position.  Even absolute positioning as done here\footnote{Our images are aligned with respect to the imaginary black hole line of sight. In practice, optically thin features should be used as indicators for an absolute alignment.} can successfully reestablish the $\lambda^{2}$-law only when the beam size is smaller than the core shift.  
In practice, unresolved core shifts could well be the origin of optically thick $\lambda^{2}$-law breakers.

\subsection{Resolution, mm-VLBI}

In order to observe the helical fields of the jet acceleration region via the associated rotation measure, beam sizes able to resolve the dynamics are essential. In addition, the frequency must be sufficiently high to peer through the self-absorption barrier.  With the advance of global mm-VLBI experiments, substantial progress will be made on both of these fronts.  
The theoretical resolution of a $10^{4}\rm km$ mm-observatory (at $300\rm GHz$) evaluates to $10\mu \rm as$, similar to the resolution of space-VLBI at $86\rm GHz$.  Corresponding to a physical scale of $\sim 60~ r_{\rm S}$, our reference object is thus entering the regime of interest for RM studies.  

At present, the record holders in terms of physical and angular resolution are the $1.3\rm mm$ observations of the radio source Sgr A* near the galactic center black hole that were reported by \cite{2008JPhCS.131a2055D} and \cite{fish2011}.  
Coherent structures on scales less than $45 \mu\rm as$ or $\sim 4 r_{\rm S}$ could already be detected\footnote{Given several Jansky flux in the mm-range, imaging the black hole shadow in Sgr A* becomes a mere problem of visibility in the southern hemisphere \citep[e.g.][]{miyoshi2007}.} 
and sub-mm rotation measure magnitudes in excess of $4\times 10^{5} \rm rad/m^{2}$ were discovered and confirmed by \cite{macquart2006} and \cite{marrone2007}.  

For the prominent case of M87, \cite{broderick2009} elaborate that the inner disk and black hole silhouette at $5~ r_{\rm S}$ could be observed with (sub) mm-VLBI.  
In M87, high frequency radio observations are already pushing towards the horizon scale \citep[e.g.][]{krichbaum:2006,Ly:2007}, only the important core polarization signal is still inconclusive \cite[e.g.][]{walker2008}.  
Rotation measures at core distances of $\approx 20 \rm mas$ vary between $-5000$ and $10^{4}$ $\rm rad/m^{2}$ depending on the location in the jet \citep[e.g.][]{zavala2002}.  
It is tempting to extrapolate these values to the $\mu \rm as$ scale, assuming the observed RM values are non-local enhancements due to cold electron over densities.  Following this argument that was put forward by \cite{broderick2009}, the resulting core RM's would be on the order of $10^{8} \rm rad/m^{2}$, enough to account for the low observed polarization degree via Faraday depolarization. 

In the mm wavelength range, large rotation measures are needed to produce detectable deviations in the polarization angle due to the decreasing coverage of $\lambda^{2}$ space.  
Typical calibration errors of $\sim 1^{\circ}$ require $\rm{RM}>18\times 10^{3}\rm rad/m^{2}$ for a $3\sigma$ detection between $172$ and $688$ GHz and $\rm{RM}>4.3\times 10^{3} \rm rad/m^{2}$ when $86$ GHz observations are added.  
The steep spectrum synchrotron flux of the jet rapidly declines when higher frequencies are considered, however, opacity and Faraday depth also decrease to yield a higher contribution of polarized flux from the core (e.g. Figure \ref{fig:spec}).  
Also the deviations introduced by the core-shift as described in the previous section will pose lesser problems in (sub-) mm observations.

\subsubsection{sub-mm Rotation measure maps}

\begin{figure*}[htbp]
   \centering
{\Large RM [$\rm K\, rad/m^{2}$]}\\
\hfill
\includegraphics[width=\textwidth]{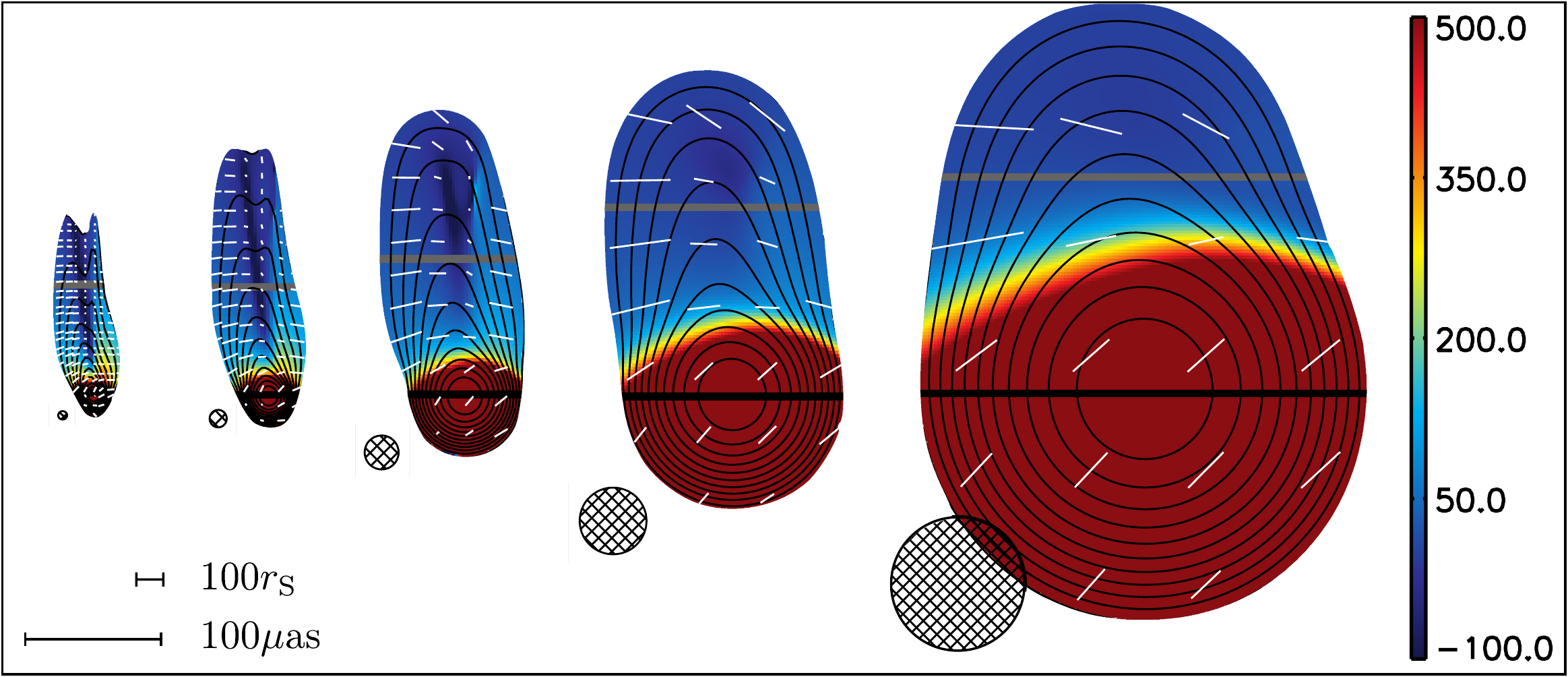}
\hfill
   \caption{RM maps in the optically thin wavelength range ($\rm 344-688\, GHz $) of model B when observed with decreasing resolution from $6.25\rm \mu as$ to $100 \rm \mu as$ doubling the beam size for each image. The effective emission angle $\chi_{0}$ is indicated by white sticks.  }
   \label{fig:bw}
\end{figure*}

To obtain detectable deflections of the polarization angle also at short wavelengths, we now focus on the high Faraday depth model B.  As discussed in section  \ref{sec:depol-b}, at frequencies beyond $172\, \rm GHz$, the unresolved polarization vector exhibits changes below $45^{\circ}$ and internal depolarization is not observed (see also Figure \ref{fig:specb}), such that we find consistent $\lambda^{2}$-law rotation measures in the mm wavelength range.  
For an increasing beam FWHM from $6.25\rm \mu as$ to $100 \rm \mu as$, sub-mm RM maps for the optically thin radiation are shown in figure \ref{fig:bw}.  The corresponding physical resolution is $31 \rs - 500 \rs$.  
Steep gradients of RM across the jet axis and ``spine and sheath'' polarization structures are observable down to a resolution of $125\,\rs$, below which most information is destroyed by the beam.  With increasing beam size, only the central Faraday pit remains detectable.  Here the core rotation measure reaches values as high as $10^{6}\rm rad\, m^{-2}$.

\begin{figure}[htbp]
   \centering
   \includegraphics[width=9cm]{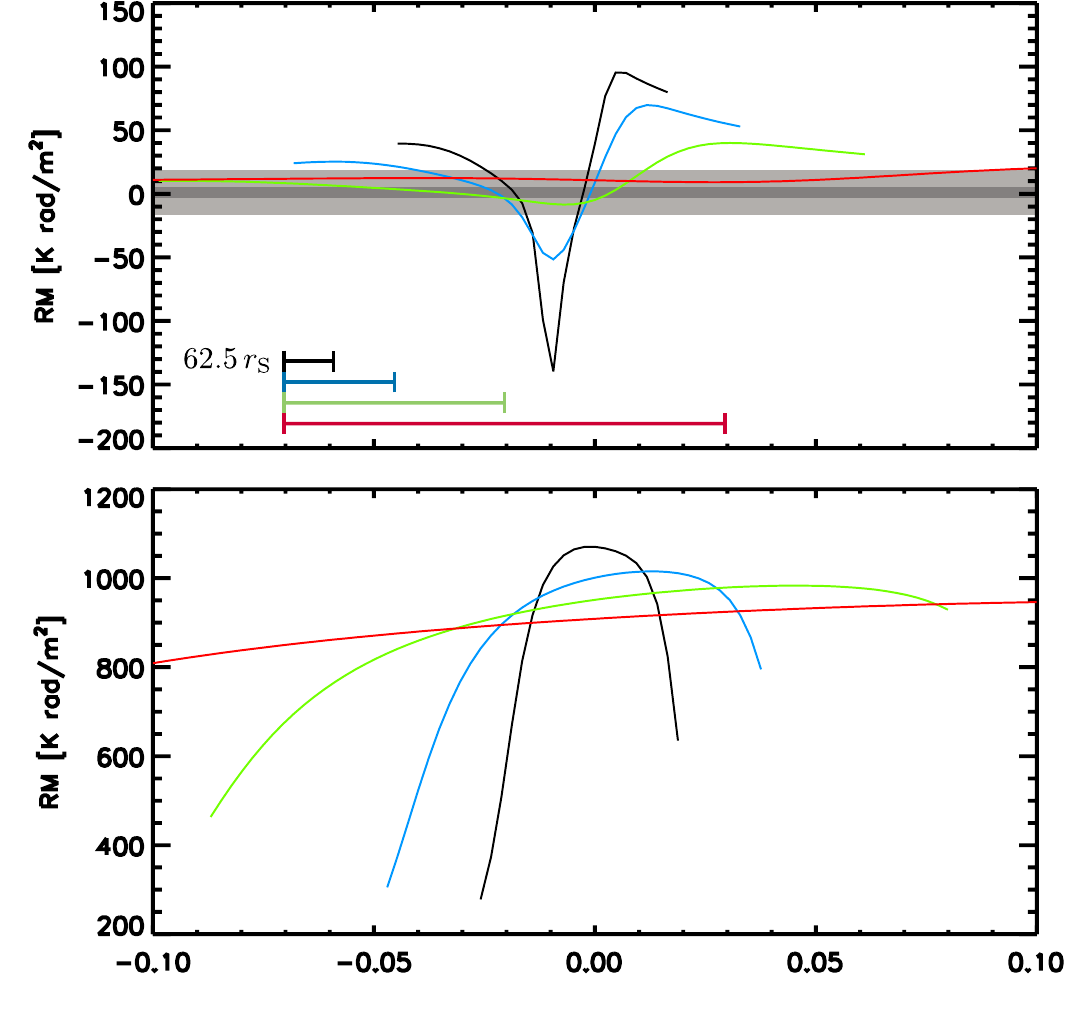}
   \caption{RM cuts along the paths indicated in figure \ref{fig:bw} for various beam sizes.  The physical resolution of the cuts is $62.5 \rs - 500 \rs$ as indicated and $3^{\circ}$ detection limits for $86-688\, \rm GHz$ and $172-688\, \rm GHz$ observations are shaded grey in the top panel.  Curves are cropped where the intensity falls below $5\times 10^{-4} I_{\nu,\rm peak}$ as in figure \ref{fig:bw}.  }
   \label{fig:rm_cut_bw}
\end{figure}

We show the transversal cuts along the core and jet in figure \ref{fig:rm_cut_bw}.  
At a beam size of $500 \rs$ and observing frequencies between $172-688\rm GHz$ the transversal RM gradients of the jet fall below the assumed detection limit of $3^{\circ}$ and become consistent with a constant.  
Interestingly, beam convolution decreases the magnitude of RM not only in the cuts exhibiting a sign change, but also for the cuts across the intensity peak.  
We note that the intrinsic jet RM profiles are non-monotonic.  Only when under-resolved, the RM features the monotonic profiles ``from side to side'' that are typically observed.

\subsection{Viewing angle}
\begin{figure*}[htbp]
\begin{center}
   \includegraphics[width=0.99\textwidth]{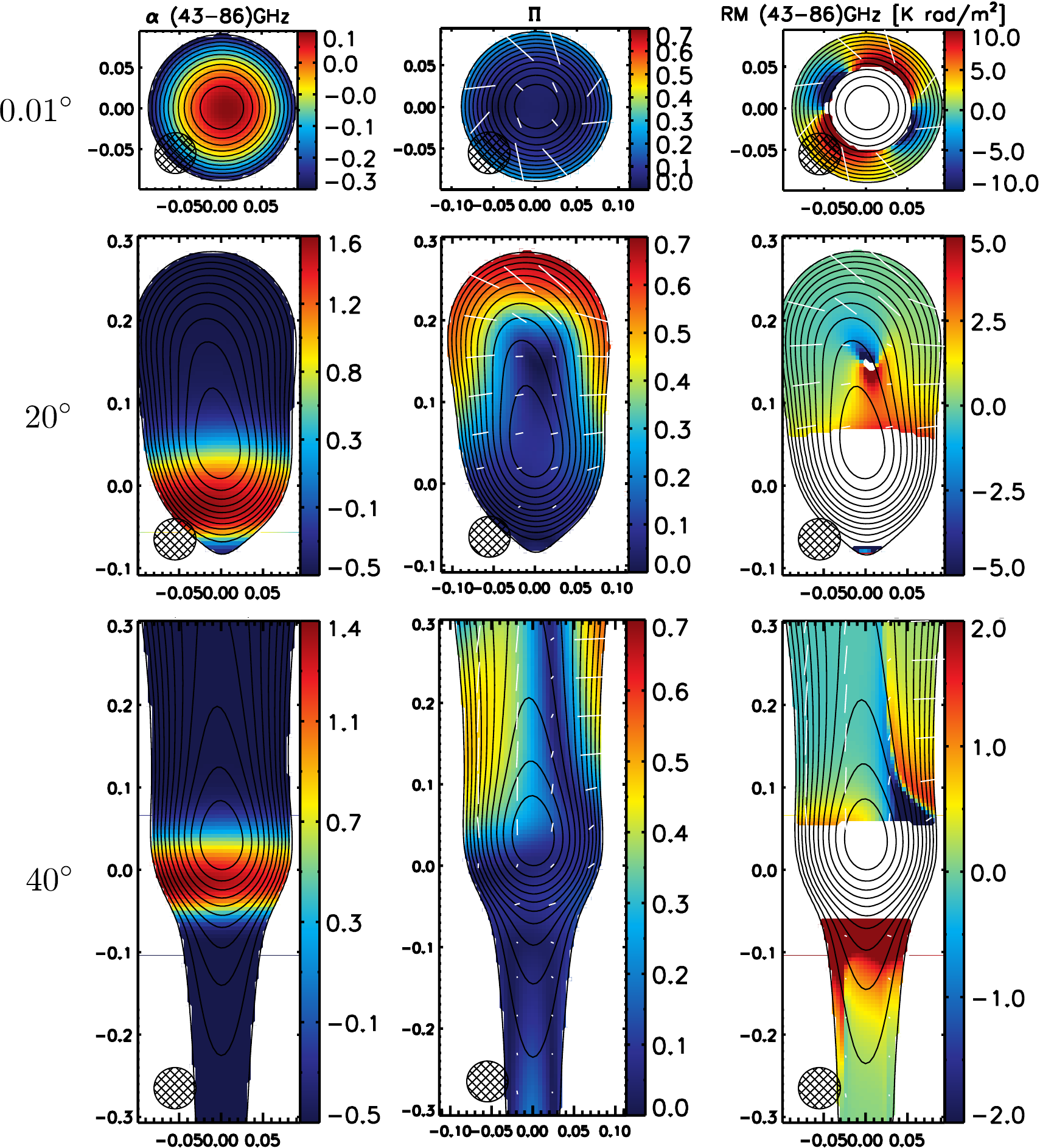}
\caption{Mock observations based on model A at $43 \rm GHz$ and $86 \rm GHz$ for viewing angles $i\in(0.01^{\circ},20^{\circ},40^{\circ})$ with resolution of $50 \mu \rm as$.  Contours according to $43 \rm GHz$ intensity and spaced by factors of two.  From left to right: Spectral index $\alpha$, $43\rm GHz$ polarization degree $\Pi$ and rotation measure in optically thin regions.  }
\label{fig:angle}
\end{center}
\end{figure*}

To investigate the viewing angle dependence on the observations we show radio observables for different inclinations from $0.01^{\circ}$ to $40^{\circ}$ in figure $\ref{fig:angle}$.  With a resolution of $50\mu \rm as$ in the 43-86 GHz frequency range these images preview the next generation space VLBI experiments.  We observe asymmetric features in the spectral index as proposed by \cite{clausen-brown2011}.  Due to the core shift, the maximum of the spectral index appears ``behind'' the low frequency intensity peak.  
When seen right down the jet, the polarization vectors are radially symmetric and show an inclination about the radial direction.  With increasing viewing angle, the predominating polarization direction with respect to the jet flips from perpendicular to parallel orientation.  
Also the transversal polarization structure exhibits asymmetries as mentioned in section \ref{sec:pitchangle}.  
To produce rotation measure maps, we fitted the $\lambda^{2}$ law to observations at $43,85$ and $86$ GHz.  In order to avoid the problems due to optical depth effects (see section \ref{sec:rotationmeasure}) we exclude regions of spectral index between $0<\alpha<2$ in the maps.  
The strong feature in the $20^{\circ}$ RM map at $(0,0.15)\rm mas$ is most likely an artifact from the finite ray-casting domain and corresponds to the region where the axis pierces through the ``lid'' of the domain. 
For $i=40^{\circ}$ this region is excluded from the RM map and for $i=0.01$ the problem does not arise.  
At high inclinations, we observe steep RM gradients that coincide with the spine-sheath flip of polarization.  

\begin{figure}[htbp]
\begin{center}
   \includegraphics[width=0.5\textwidth]{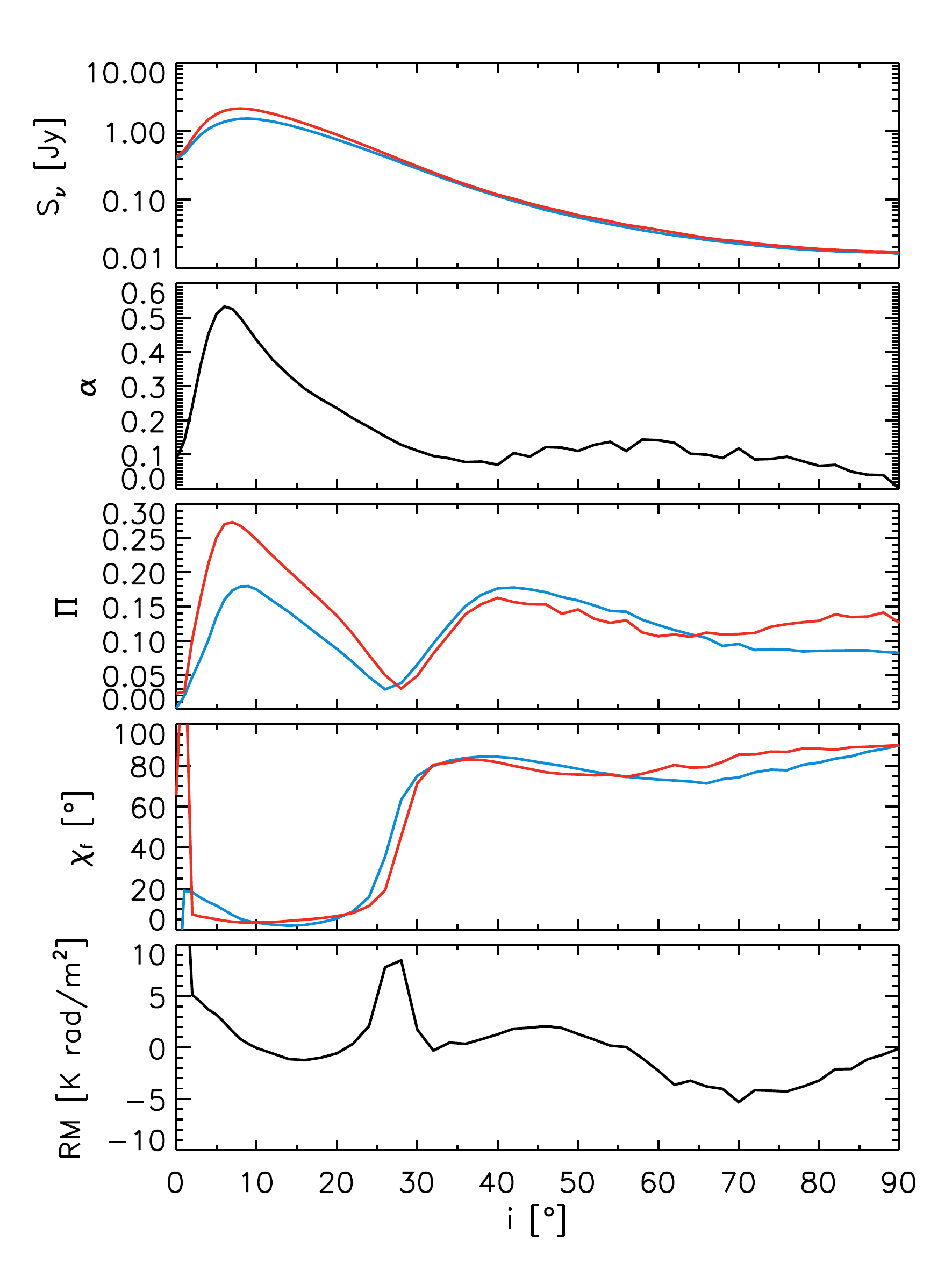}
\caption{The dependence on viewing angle $i$ of unresolved quantities for 86 GHz (red) and 43 GHz (blue). Top to bottom: Beamed flux, spectral index, integral polarization degree, observed polarization vector $\chi_{\rm f}$ and apparent rotation measure derived from $\Delta\chi_{\rm f}$.  }
\label{fig:fluxvsangle}
\end{center}
\end{figure}

As such high resolution data are not readily available, we also quantify the integral values that should reflect unresolved core properties.  The viewing angle dependence of flux, spectral index, unresolved polarization degree, polarization direction and rotation measure are shown in figure \ref{fig:fluxvsangle}.  
The flux peaks at viewing angles of roughly $5^{\circ}$ and not when looking directly down the jet as would be the case for simplified cylindrical flows with toroidal fields.  This reflects the fact that our jet solutions are not perfectly collimated and exhibit a small degree of rotation such that the Doppler factor attains its maximum value of $D\simeq15$ at $i\simeq5^{\circ}$.  
The dominating polarization direction flips from perpendicular ($0^{\circ}$) to parallel ($90^{\circ}$) for viewing angles $i>25^{\circ}$.  At this flip, the polarization degree shows a local minimum through beam depolarization.  
In the framework of the unified model proposed by \cite{urry1995}, this suggests that the core polarization direction in the radio galaxy case (viewed at high inclination) should be clearly distinct from the blazar case (looking down the jet).  
For $i=0^{\circ}$, the polarization degree approaches zero and its direction fluctuates between $0^{\circ}$ and $90^{\circ}$ due to the axial symmetry.

\subsection{Towards modeling actual observations}

Individual sources are modeled by constraining all free parameters (tables \ref{tab:sim} and \ref{tab:radmod}) with observations of spectra, core shift, polarization and rotation measures.  Taken together, we obtain seven interrelated parameters in our radiation models.  \\
Two of these are related to jet dynamics, parametrizing the energy partitioning with $\beta_{1}$ and the current distribution with the parameter $s$.  We find that the current distribution has little influence on the observables and hence this dependence can eventually be dropped, which reduces the number of free parameters but at the same time looses predictive power.\\
Two parameters are related to the physical scaling of the simulations, namely the total energy flux $\dot{E}$ and the black hole mass $\mbh$.  
With the scale free nature of the underlying MHD, a single dynamical simulation can be used to construct a multitude of objects for arbitrary total energy flux and black hole mass. Thus in principle a library of physical jet models can be constructed from a few dynamically distinct simulations (see also: Physical scaling, section \ref{sec:scaling}).  \\
Compared to the simulations, the raytracing is fast and can be used to vary the remaining parameters $\epsilon_{\rm B},\Psi'$ and $i$.

An alternative to modeling individual sources is to compare the statistical properties of a set of models with a sample of AGN cores as observed e.g. by \citet[][]{agudo2010}.  

Such an effort is beyond the scope of the current paper and we leave this open for future investigation.  However, already with the acquired data (e.g. figure \ref{fig:fluxvsangle}), our simulations strongly suggest a bimodal distribution of the polarization angle.

\section{Summary and Conclusions}

We have performed axisymmetric, special relativistic MHD simulations of jet acceleration 
and collimation. 
The resulting dynamical variables are applied to calculate polarized synchrotron radiation 
transport in postprocessing, providing emission maps consistent with the jet dynamical
structure.  

\subsection{Jet acceleration}

Our jets are realistically modeled to consist of two components: an inner thermal 
spine assumed to originate in the a black hole corona, and a surrounding self-collimating
disk jet driven by Poynting flux.  

We follow the flow acceleration for more than 3000 Schwarzschild radii reaching Lorentz factors 
in the disk jet of $\Gamma\sim 8$ within the AGN ``blazar zone'' where we calculate the
synchrotron emission maps.
In application to a $10^{9}\msun$ black hole this translates to a distance of $0.3\rm pc$.  
 
Although the Poynting dominated jet flow becomes super fast-magnetosonic within the domain, 
it has not yet reached equipartition between Poynting and kinetic energy - jet acceleration 
is still ongoing.  
According to the available energy budget, in the case of high energy disk jets terminal 
Lorentz factors of $\Gamma_{\infty}\sim20$ would be aquired asymptotically. 
At the fast magnetosonic point we find the Michel scaling $\Gamma(x_{\rm F})=\Gamma_{\infty}^{1/3}$
to be satisfied within $5\%$.  
We find that the jet acceleration up to a distance $z\sim 3000\rs$ is well described by 
the linear relation $\Gamma\propto r$ as proposed by \cite{2008MNRAS.388..551T} and \cite{2009MNRAS.394.1182K}.  
We do however not reproduce the tight coupling of acceleration and collimation 
$\Gamma \tan \theta_{\rm fl}\simeq 1$ observed in the latter communications but
instead find $\Gamma \tan \theta_{\rm fl}$ to monotonically decrease along the flow.  
The fast jet component in all models considered collimates to half-opening angles of 
$\sim 1^{\circ}$.
We find the causal connection within the flow - its ability to communicate with the axis 
via fast magneto-sonic waves - to be well maintained in our simulations.  

Also the thermal spine acceleration is shown to be efficient with $\Gamma\propto r$ and limited only by the amount of enthalpy available at the sonic point, in our case to $\Gamma_{\infty}<4$.  
 
We have placed the location of the outflow boundaries out of causal contact with the 
propagating jet beam of interest.
Thus, we can be sure that the calculated jet structure is purely self-collimated, and does 
not suffer from spurious boundary effects leading to an artificial collimation.  
We have investigated jets with a variety of poloidal electric current distributions.
We find - somewhat surprisingly - that the topology of the current distribution, 
e.g. closed current circuits in comparison to current-carrying models,
has little influence on the jet collimation.  

\subsection{Jet radiation}

The benefit of having performed relativistic MHD simulations of jet formation is that
we could apply them to produce dynamically consistent emission maps 
to predict VLBI radio and (sub-) mm observations of nearby AGN cores.  
For this purpose, we have developed a special relativistic synchrotron transport code 
fully taking into account self-absorption and internal Faraday rotation.  
Since the acceleration of non-thermal particles can not be followed self-consistently 
within the framework of pure MHD, it remains necessary to resort the particle energy
distribution to simple recipes.    
We have compared three prescriptions of the non-thermal particle energy distribution. 
We found good agreement in the alignment of the polarization structure, but considerable
differences in the intensity maps.  
Thus, the polarization maps derived in this work can be considered as robust, 
while the intensities distribution should be regarded with caution.  

The strict bi-modality of the polarization direction suggested by
\cite{Pariev2003,2005MNRAS.360..869L} and others can be circumvented when the 
structure of a collimating jet is considered.  
However, the efficient collimation to near-cylindrical jet flows in general confirms these
results obtained for optically thin cylindrical flows when the fast jet is considered.  
Thus, depending on the pitch angles of the emission region, also a spine-and-sheath 
polarization structure could be observed.  
The relativistic swing effect skews the polarization compared to the non-relativistic case.  
Our radiation models affirm the finding of \cite{2005MNRAS.360..869L} and \cite{clausen-brown2011} 
that relativistic aberration promotes asymmetries in the polarization (half spine-sheaths) and also
in the spectral index.  
The observational detection of such features would allow to determine the spin direction of the 
jet driver, be it the accretion disk or the central black hole.  

The frequency-dependent core shift in the radiation maps following our jet simulations is 
consistent with analytical estimates of conical jets by \cite{lobanov1998} in two jet models 
and slightly steeper in the third case considered.  
We attribute this discrepancy to fact of jet {\em collimation}.  
The overall good agreement with the analytical estimate suggests that the standard diagnostics 
should provide robust results capable of determining the jet parameters.  
With our radiation models we have confirmed the intuition that unresolved core shifts should 
lead to a breaking of the $\lambda^{2}$ rotation measure law.
Further, we have demonstrated that that law can be restored again as soon as the resolution 
is increased.  
Opacity effects do not allow to obtain a consistent $\lambda^{2}$-law across the spectral peak. 
Once the regimes are separated however, we obtain two valid relations for which the optically 
thin rotation measure is substantially increased over the optically thick case as it peers 
deeper into the Faraday pit.  

The interpretation of observations featuring both internal Faraday rotation and changes in 
opacity is one of the most challenging aspects in polarimetric imaging of jets.  
With our detailed modeling, we were able to disentangle the depolarizing effects of 
opacity transition, differential Faraday rotation, and also beam effects such as ordinary 
beam depolarization or Faraday dispersion for two exemplary jet models.
We find that the unresolved, optically thin mm-wavelength radiation is depolarized due to 
both the changing emission geometry (down to $\sim 40\%$), and the additional beam 
depolarization (down to $\sim30\%$).  
Increasing the Faraday opacity by observing at lower frequencies would lead
to depolarization below the $1\%$ level due to \textit{1.} Faraday dispersion 
and \textit{2.} differential Faraday rotation.  

We have also investigated the influence of resolution on the detectability of rotation measure 
(RM) gradients in the optically thin parsec-scale jet-core previewing mm-VLBI observations.  
To detect such gradients across the jet, a resolution of $\sim 100 \rs$ would be required.  
Increasing the beam size leads to more monotonic transversal RM profiles.  
We find the peak magnitude of the RM to increase with resolution.  
High RMs beyond $10^{4} \rm rad\, m^{-2}$ are required to obtain a noticeable 
deflection in the mm-wavelength range.  
From the sources where high resolution data is already available, namely Sgr A* and M87, 
such high rotation measures are in fact observed, and we predict that many more objects in 
this class will be found at the advent of ALMA and global mm-VLBI.  

Finally, we have presented mock observations of spectral index, polarization degree and 
rotation measure for various inclinations.  
Asymmetries in the spectral index and polarization degree can be observed most clearly
at high inclinations $>30^{\circ}$.  
The necessary resolution for this detection in a fiducial low Faraday rotation case 
(our model A) amounts to $50 \mu\rm as$, 
which could be reached with the next generation space-VLBI.
At $\sim30^{\circ}$, the predominant polarization vector flips from perpendicular alignment 
(with respect to the projected jet direction) for the blazar case to parallel alignment for 
the radio galaxy case at high inclinations.  
The flip in polarization is clearly detectable also from unresolved quantities.  
In summary, these findings suggest a bimodal distribution of the observed polarization 
direction of AGN core jets.  
However, by adding a substantial amount of Faraday rotation (our model B), this signature 
will be scrambled unless the observing frequency is chosen high enough - confirming the 
popular intuition.

In this paper we have focussed on general signatures of the synchrotron radiation in the
large-scale helical fields in the acceleration region of relativistic MHD jets.  
With the developed tool set in hand, further progress can be made when calibrating the 
observational diagnostics with the mock observations that are detailed here.  
We expect a substantial improvement from a more consistent treatment of the non-thermal particles, 
taking into account particle acceleration and cooling.  
Potentially, also modeling of individual sources, or the cumulative statistics of AGN surveys 
applying dynamical simulations could be undertaken in the future.

\begin{acknowledgements}
This work was partly carried out under the HPC-EUROPA2 project (project number: 228398), 
with the support of the European Community - Research Infrastructure Action of the FP7.
Post-processing of the simulations was performed on the THEO cluster of MPIA and fits were obtained with the flexible MPFIT routines provided by \cite{markwardt2009}.  
The authors thank an anonymous referee for comments and suggestions that have helped to improve the presentation of this work.  
O.P. likes to thank Christophe Sauty for discussions and kind hosting during an 
interesting research visit.  
\end{acknowledgements}

\appendix

\section{Stokes transport}\label{sec:stokest}

In the $\{I^{l},I^{r},U^{lr}\}$ basis, the linear transfer equation $d\mathbf{I}/dl = \mathbf{\boldsymbol{\mathcal{E}}-\underline{A}\ I}$ (\ref{eq:radtransp}) has the coefficients
\begin{align}
\left(\mathbf{\underline{A}}\right) = \left(\begin{array}{lll}
a_{11} & 0 & a_{13}\\
0 & a_{22} & a_{23}\\
2a_{23} & 2a_{13} & a_{33}
                            \end{array}\right)
\end{align}
\begin{align}
\left(\mathbf{\boldsymbol{\mathcal{E}}}\right) = D^{2+\alpha}\left(\begin{array}{l}
\epsilon^{(e)}\cos^{2}\chi_{e}+\epsilon^{(b)} \sin^{2}\chi_{e} \\
\epsilon^{(e)}\sin^{2}\chi_{e}+\epsilon^{(b)} \cos^{2}\chi_{e} \\
- \left(\epsilon^{(e)}-\epsilon^{(b)}\right)\sin2\chi_{e}
                             \end{array}\right)
\end{align}
\begin{align}
  a_{11} &= D^{\alpha+1.5}\left[\kappa^{(e)}\cos^{4}\chi_{e}+\kappa^{(b)}\sin^{4}\chi_{e}+\frac{1}{2}\kappa \sin^{2}2\chi_{e}\right]\label{eq:a11} \\
  a_{13} &= -\left[\frac{1}{4}D^{\alpha+1.5}\left(\kappa^{(e)}-\kappa^{(b)}\right)\sin2\chi_{e}+\frac{d\chi_{F}}{dl}\right]\\
  a_{22} &= D^{\alpha+1.5}\left[\kappa^{(e)}\sin^{4}\chi_{e}+\kappa^{(b)}\cos^{4}\chi_{e}+\frac{1}{2}\kappa \sin^{2}2\chi_{e}\right]\label{eq:a22} \\
  a_{23} &= -\left[\frac{1}{4}D^{\alpha+1.5}\left(\kappa^{(e)}-\kappa^{(b)}\right)\sin2\chi_{e}-\frac{d\chi_{F}}{dl}\right]\\
  a_{33} &= D^{\alpha+1.5}\kappa.
\end{align}
Naturally, $\epsilon^{(e,b)}$ and $\kappa^{(e,b)}$ indicate the comoving emissivity/absorptivity in direction $\mathbf{\hat{e}'=\hat{n}'\times\hat{B}'}$ respectively $\mathbf{\hat{b}'=\hat{n}'\times e'}$ given by line of sight $\mathbf{\hat{n}'}$ and magnetic field direction $\mathbf{\hat{B}'}$, while $\kappa=1/2(\kappa^{(e)}+\kappa^{(b)})$ is the average absorption coefficient.  
We apply the standard expressions for $\epsilon^{(e,b)}$ and $\kappa^{(e,b)}$ valid for isotropic power-law particle distributions with index $p=2\alpha+1$ following \cite{Pacholczyk:1970}.  For completeness, the expressions read: 

\begin{eqnarray}
  \epsilon_\nu^{(e,b)} &=& \frac{1}{2} c_5(\alpha) N_0 (B'\sin\vartheta')^{\alpha+1}\left(\frac{\nu}{2c_1} \right)^{-\alpha} 
    \left[1\pm\frac{2\alpha+2}{2\alpha+10/3}\right]\\
  \kappa_\nu^{(e,b)} &=& c_6(\alpha)N_0\left(B'\sin\vartheta' \right)^{\alpha+3/2}\left(\frac{\nu}{2c_1}\right)^{-\alpha-5/2}
    \left[1\pm\frac{2\alpha+3}{2\alpha+13/3}\right]
\end{eqnarray}

depending on the constants 
\begin{align}
  c_1 &= \frac{3e}{4\pi m^3c^5} = 6.27\times 10^{18}\ \  \rm cm^{-7/2} g^{-5/2} s^{4}\\
  c_3 &= \frac{\sqrt{3}}{4\pi}\frac{e^3}{mc^2} = 1.87\times10^{-23}\ \ \rm cm^{5/2} g^{1/2} s^{-1}\\
  c_5(\alpha) &= \frac{1}{4}c_3
    \Gamma\left(\frac{6\alpha+2}{12}\right)\Gamma\left(\frac{6\alpha+10}{12}\right)\Gamma\left(\frac{2\alpha+10/3}{2\alpha+2}\right)\\
    &\to1.37\times10^{-23} \rm cm^{5/2} g^{1/2} s^{-1} \ (\alpha \to 0.5)\\
  c_6(\alpha) &= \frac{1}{32}\left(\frac{c}{c_1}\right)^2 c_3 \left(2\alpha+13/3\right)
    \Gamma\left(\frac{6\alpha+5}{12}\right)\Gamma\left(\frac{6\alpha+13}{12}\right)\\
    &\to8.61 \times10^{-41} \rm cm^{5/2} g^{1/2} s^{-1} \ (\alpha\to0.5)
\end{align}
in Gauss cgs units where $\Gamma(\cdot)$ denotes the Gamma-function.  The upper sign corresponds to the direction of the main polarization axis ($\mathbf{\hat{e}'}$).  

While the transformation of line of sight $\mathbf{\hat{n}}$ leads to the well known relativistic aberration,
\begin{align}
\mathbf{\hat{n}'} = D\mathbf{\hat{n}}-(D+1)\frac{\gamma}{\gamma+1}\boldsymbol{\beta}
\end{align}
which supplies for the angle between co-moving magnetic field and line of sight $B'\sin \vartheta'=|\mathbf{\hat{n}'\times B'}|$, 
the change between $\mathbf{\hat{e}'}$ and its observed counterpart $\mathbf{\hat{e}}$ introduces a relativistic ``swing'' of the polarization \citep{Blandford1979}.
The observer system quantities follow as 
\begin{align}
\mathbf{\hat{e} = \frac{\hat{n}\times q}{\sqrt{q^{2}-(\hat{n}\cdot{q})^{2}}}};
\hspace{1cm} \mathbf{q = \hat{B}+\hat{n} \times (\boldsymbol{\beta} \times \hat{B})} \label{eq:ehat}
\end{align}
according to the short formulation discovered by \cite{Lyutikov:2003}. Hence the local observer-system  angle of the emission $\chi_{e}$ as measured from the image axis $\mathbf{\hat{l}}$ (the projection of jet axis onto the plane of the sky) is given by
\begin{align}
\cos \chi_{e} = \mathbf{\hat{l}\cdot \hat{e}}\ ; \hspace{1cm} \sin\chi_e =
  \mathbf{\hat{n}\cdot \left(\hat{l}\times\hat{e}\right)}
\end{align}
and we can conveniently take advantage of the identity $\sin2\chi_{e}=2\sin\chi_{e}\cos\chi_{e}$ to calculate $\sin2\chi_{e}$ that appears in the absorption matrix.  
The description of Faraday rotation in the observers system, eq. (\ref{eq:Frot}) directly enters into the transfer coefficients $a_{13}$ and $a_{23}$.  

With these considerations we solve the ordinary differential equation (\ref{eq:radtransp}) using a fourth-order Runge-Kutta scheme for a grid of lines of sight.  
In practice this is realized by transforming the $(r,z)$ plane information of the simulations into a coarsened cartesian grid with typically $100\times 100\times 500$ cells by means of a Delaunay triangulation.  
For numerical stability, the adaptive step size is chosen small enough to satisfy $\Delta l < 6 \rs$, $\Delta \tau<0.5$ and $\Delta \chi_{\rm F}<\pi/16$ yielding sufficient convergence of the solution.  
\bibliographystyle{apj} 
\bibliography{astro,extra} 
\end{document}